\documentclass[11pt]{article}
\usepackage{arxiv}
\usepackage[utf8]{inputenc}
\usepackage[T1]{fontenc}
\usepackage{lmodern,microtype,amsmath,amssymb,booktabs,array,tabularx,longtable,enumitem,graphicx,xcolor,xurl}
\usepackage{hyperref}
\usepackage[numbers,sort&compress]{natbib}
\usepackage[nameinlink,noabbrev]{cleveref}
\usepackage{caption,subcaption,float,flafter,needspace}
\usepackage[section]{placeins}
\usepackage{etoolbox}
\makeatletter
\patchcmd{\LT@output}{\vss}{\vfil}{}{}
\patchcmd{\LT@output}{\vss}{\vfil}{}{}
\makeatother
\newcolumntype{P}[1]{>{\raggedright\arraybackslash}p{#1}}
\newcolumntype{Y}{>{\raggedright\arraybackslash}X}

\setlist{itemsep=.25em,topsep=.35em}

\hypersetup{colorlinks=true,linkcolor=blue!55!black,citecolor=blue!55!black,urlcolor=blue!55!black,
pdftitle={Trustworthy Execution of Industrial Agents},
pdfauthor={Zhongjie Wang, Mingyi Liu, Xiang He, Xiao Wang, Shuang Yu, Kai Wang},
pdfsubject={V1.25: complete technical framework with differentiated implementation and evidence scopes},
pdfkeywords={industrial agents, trustworthy execution, engineering toolchains, runtime governance}}
\title{Trustworthy Execution of Industrial Agents}
\author{\parbox{\linewidth}{\centering\mbox{Zhongjie Wang}, \mbox{Mingyi Liu}, \mbox{Xiang He}, \mbox{Xiao Wang}, \mbox{Shuang Yu}, \mbox{Kai Wang}}\\[.5em]
Faculty of Computing, Harbin Institute of Technology\\[.15em]
Harbin 150001, China\\[.25em]
{\small\texttt{\{rainy,liumy,hexiang,wxlxq,yushuang,kai\_wang\}@hit.edu.cn}}}
\date{October 2026}
\makeatletter
\renewcommand{\@maketitle}{%
\begin{center}
{\fontsize{16}{19}\selectfont\bfseries\@title\par}\vspace{12pt}
{\normalsize\@author\par}\vspace{7pt}
{\small\@date\par}
\end{center}\vspace{7pt}}
\makeatother
\begin{document}
\maketitle
\begin{abstract}
Industrial agents increasingly modify persistent engineering objects rather than merely generate answers. Authentication and successful calls cannot establish whether intended changes occurred, evidence supports a conclusion, or historical results remain applicable. We present a framework for trustworthy industrial-agent execution within a single enterprise governance domain. Eight control domains assign responsibilities across authority, planning, composition, tool execution, actual effects, evidence, baseline and use governance, and recovery and evolution. A four-plane architecture and three control chains connect these responsibilities through typed objects and version-bound interfaces. The specification separates tool returns from actual engineering effects, evidence coverage from engineering judgment, and historical lifecycle state from current applicability. Multi-agent handoffs are an architectural extension examined through deterministic interface tests.

Distinct prototypes examine selected mechanisms using conformance tests, ablations, uncertainty stress, CAD/CAE execution, and local persistence. An OpenFOAM fixture produces COMPLETE coverage with FAIL judgment; an expired-job case requires readmission despite completed computation. CAD and recovery counterexamples expose omitted mandatory checks, discarded negative facts, and inconsistent producer--consumer mappings. Targeted repairs support local regression conclusions. The reported 0/29 hazardous commits measure closed-set specification conformance, not a safety probability. We distinguish the complete framework, implemented paths, and empirical evidence, identifying check-inventory completeness and cross-module fact preservation as prerequisites for extending these results.
\end{abstract}
\clearpage

\section{Introduction}\label{sec:section_01}

With tool invocation, planning, memory, and persistent execution, industrial agents are moving from generating text to changing an engineering environment. An agent can modify CAD parameters, generate meshes, start CAE jobs, interpret simulation results, write MBSE verification states, and advance PLM baselines. Risk therefore extends beyond answer accuracy to relationships among models, configurations, evidence, and subsequent activities. An incorrect binding can apply an inlet condition to an outlet face; a report on an old mesh can be attributed to new geometry; a successful tool return can indicate only that a file was written, not that a required effect occurred; and recovery can reintroduce invalid downstream results.

Five considerations motivate this research. First, industrial-agent outputs have persistent side effects: text can often be corrected in the next exchange, whereas an object admitted to a formal baseline is consumed by manufacturing, procurement, testing, and certification. Second, engineering has strong version and configuration semantics. Identical filenames, port names, or commands do not establish identical objects, roles, operating conditions, or rule versions. Third, engineering conclusions carry evidence obligations. Obtaining a result is distinct from supporting a target conclusion; evidence must also satisfy coverage, quality, provenance, independence, and configuration-binding requirements. Fourth, heterogeneous and asynchronous tools produce cumulative effects. CAD, meshing, solvers, postprocessing, and PLM report local successes, while hazards often reside in unrepresented relationships between those signals. Fifth, agent execution evolves over time. Baselines, models, tools, capability claims, and dependencies change, so reusable effects and evidence can become inapplicable.

A stronger model, safer prompts, or approval of tool calls is therefore insufficient by itself to establish trustworthy industrial-agent infrastructure. Model-driven service engineering connects high-level objectives to executable service models and reusable components \citep{r52}. Industrial-agent execution must additionally establish whether those objectives are realized in persistent engineering objects. General agent-security research addresses prompt injection, unauthorized calls, and disclosure. Engineering environments require answers to three questions: does a call target the correct engineering object, what observable effect did the tool actually produce, and can the resulting evidence support subsequent decisions for the current configuration and use? We call these the engineering execution semantics gap.

Recent industrial-agent products incorporate tool use, observability, and governance \citep{r3,r4,r5,r6,r12,r13,r20,r21,r22}. This shift toward persistent execution motivates a complementary question: how should industrial integrations represent actual effects, version-bound evidence, and the authority to change an engineering baseline? Section~4 examines the methods relevant to these interfaces.

Delegated service action requires more than a correct task result: authority, commitments, and evidence must remain inspectable as execution changes. The relationship-centered perspective of Services Computing 3.0 \citep{r55} articulates this requirement at the service level; here, we examine its engineering consequences for tool-produced artifacts, evidence-supported decisions, and the continued applicability of historical results.

Our scope is a single enterprise governance domain. One organization controls identity roots, engineering data, tool permissions, baseline services, and accountability rules. Inter-enterprise collaboration contracts, peer trust negotiation, and cross-organizational allocation of responsibility are outside scope. Even within this boundary, the agent runtime must be connected to MBSE, CAD, CAE, PLM, and evidence governance.

The contributions have three mutually supporting levels. First, we identify eight trust-control domains using control time, controlled object, fact source, and state output to assign concrete responsibilities to engineering risks. Second, we develop a technical framework comprising four software planes, three cross-cutting control chains, and typed objects. It specifies task authority and capability applicability, engineering-obligation reconstruction, action and verification-capability composition, actual effects and role binding, evidence responsibilities, conditional activation, use-constraint propagation, recovery, and applicability-scope evolution. Table~\ref{tab:t3} relates these engineering semantics to established supporting mechanisms. Third, we implement selected paths and report conformance, ablation, stress, actual-tool, and local-persistence experiments, using both positive and negative cases to examine interface designs and implementation conditions.

The technical contribution is an arrangement of non-substitutable engineering decisions and the fact contracts connecting them, rather than a new lease, provenance format, or three-valued logic. A valid CAD postcondition, for example, cannot authorize a conclusion computed on an obsolete mesh: the evidence consumer must also bind the mesh, solver inputs, and intended configuration. Sections~3--4 establish responsibilities and related-work boundaries, Sections~5--8 specify architecture and mechanisms, Sections~9--11 report bounded prototype evidence, and Section~12 analyzes implementation conditions.

The contribution should therefore be read at two linked levels. The first is a deliberately broad research framework: T1--T8, the four planes, and the three chains define a reusable map of trust responsibilities for industrial-agent execution, including responsibilities whose complete implementations remain future work. The second is a set of bounded implementations and counterexamples that make selected interfaces executable and expose where fact binding fails. The framework is not presented as a new primitive in capability security, provenance, runtime verification, or three-valued monitoring; its proposed increment is an engineering-object-centric arrangement in which those primitives cannot silently substitute for one another across versions, configurations, evidence scopes, and lifecycle states. This distinction preserves the full research agenda while keeping prototype evidence proportional to the paths actually exercised.

\section{Running Example: Successful Tools Do Not Establish a Trustworthy Cold-plate Change}\label{sec:section_02}

\subsection{Task and nominal execution chain}

Consider an intra-enterprise battery cold-plate change. Scenario baseline B17 contains geometry G17, Mesh-G17, solver inputs for two flow conditions, and temperature and pressure-drop conclusions. G17 has eight channels, each 1.5 mm wide, 2.0 mm high, and 0.25 m long. A reduced-order proxy used to construct counterexamples produces 43.20 °C and 98.15 kPa at the nominal 8 L/min flow. These are scenario-fixture values, not validated product performance. The task requires Tmax\ensuremath{\leq}40 °C and \ensuremath{\Delta}p\ensuremath{\leq}45 kPa at nominal flow, Tmax\ensuremath{\leq}40 °C and \ensuremath{\Delta}p\ensuremath{\leq}60 kPa at the rated 9 L/min flow, and preservation of port roles and mounting datums.

The agent proposes geometry modification, rebinding, remeshing, two-condition solving, and an update request. Candidate G18 has twelve 1.6 mm channels; the proxy produces 39.44 °C/40.44 kPa at nominal flow and 39.04 °C/50.07 kPa at rated flow. These values meet the deliberately bounded scenario criteria, but a state transition still depends on object versions, preservation conditions, provenance, and the current baseline. B18 denotes a target baseline state in the control scenario, not approval of a real product design on the basis of proxy values.

The deterministic reduced-order values construct a control scenario jointly constrained by thermal, fluid, and version relationships. They do not establish cold-plate physical accuracy. Section 10 separately reports actual OpenFOAM cases for the mandatory scenarios; the two sets of numbers play different evidential roles, and numerical agreement is not a criterion for substrate correctness.

\begin{figure}[H]\centering
\includegraphics[width=\linewidth,height=0.77\textheight,keepaspectratio]{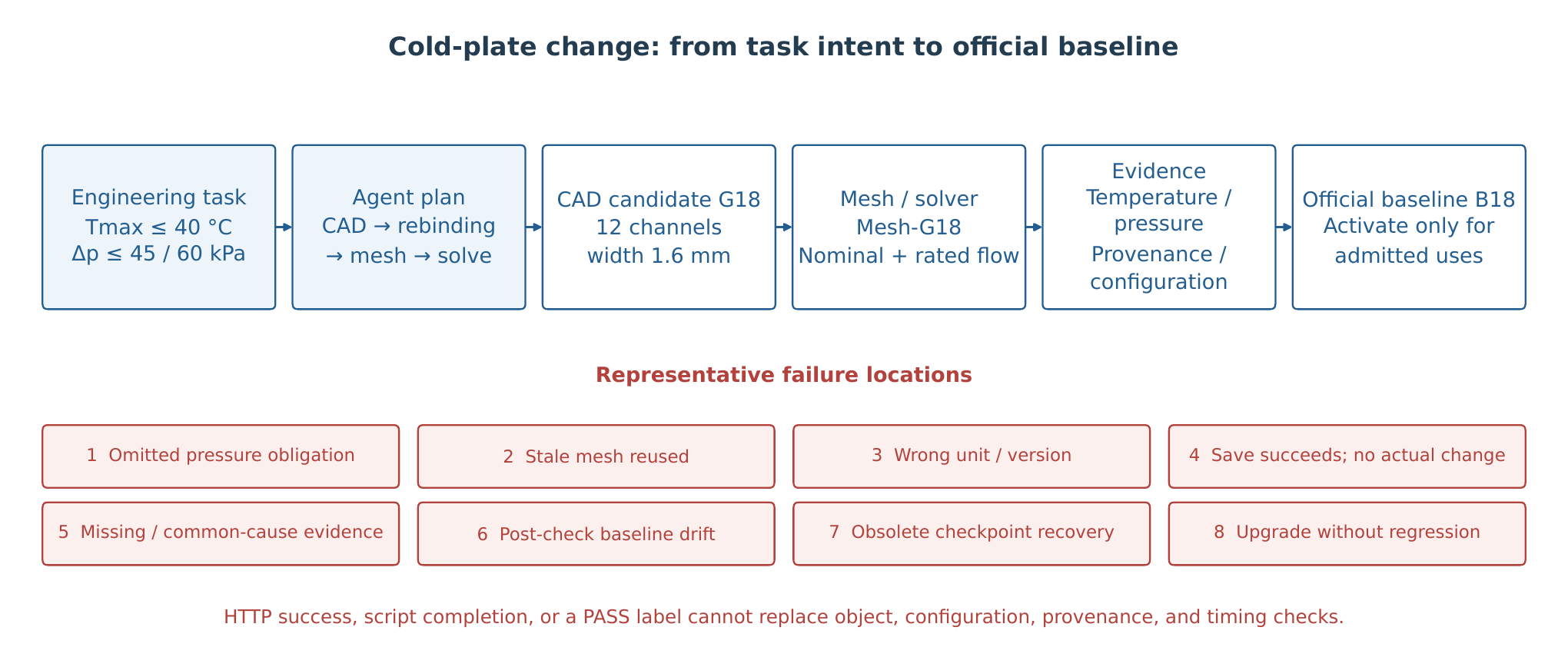}
\caption{Engineering-object chain and representative failure locations. Expired capability belongs primarily to T1, plan omission to T2, stale input or incorrect invocation to T3/T4, absent effects to T5, shared-source evidence to T6, concurrent activation drift to T7, and obsolete-checkpoint reuse to T8. Auxiliary domains may participate without replacing the primary decision.}\label{fig:1}
\end{figure}

\subsection{Eight failures along the same chain}

Numerical threshold satisfaction is only one step. An agent may use an expired solver capability claim; omit pressure drop from its plan; reuse an old mesh after modifying geometry; interpret millimeters as meters; save a CAD file without changing the target dimension; treat two reports based on the same faulty mesh as independent; activate a candidate after another task advances the baseline; or resume from an old-geometry checkpoint. These failures arise at different control times, concern different objects, and have different consequences.

For example, G18 may be written successfully while its target parameter remains 1.5 mm: invocation logs record CAD success, but the required effect is absent. A solver input that still references G17 may converge and produce a well-formed report that does not belong to G18. If B17 advances to B18' after evidence assessment, a previously correct conclusion cannot be written unconditionally into the current baseline. Trustworthy industrial execution therefore requires observation of engineering facts, not merely agent messages and return codes.

\subsection{A design extension to intra-enterprise multi-agent collaboration}

The task can be distributed among design, simulation, verification, and engineering-change agents. The design agent creates geometry candidates and change descriptions; the simulation agent handles meshes, boundary conditions, and jobs; the verification agent evaluates coverage, quality, and provenance; and the change agent requests baseline activation and subsequent asset use. They share an enterprise identity root and fact services but hold separate task authority, tool manifests, contexts, and evidence responsibilities. This is an architectural extension, not an additional experimental result.

A design-to-simulation handoff of G18 should include an immutable artifact reference, version and configuration, actual delta, interface roles to preserve, and unresolved observations. The simulation agent must still verify that the request concerns G18, role bindings are established, and the mesh derives from that geometry, before acquiring authority for its own invocation. An upstream completion message cannot replace these checks, nor can a CAD write lease become solver authorization for another agent.

After solving, the verification agent receives actual jobs, input manifests, operating conditions, artifacts, quality information, and provenance, and evaluates coverage separately from engineering judgment. Reports from different agents can still share a common cause through the same mesh or parameter source. The change agent consumes versioned decisions and their grounds and rechecks B/C/A before activation. If G19 is produced during a G18 solve, the G18 results retain their historical identity but cannot be registered unconditionally as G19 evidence. T8 triggers dependency-based reevaluation; T7 rebases or holds against current references.

\section{Eight Trust-control Domains Across the Execution Lifecycle}\label{sec:section_03}

\subsection{Classification by executable control responsibility}

Rather than classifying general qualities such as security, reliability, privacy, or explainability, we use the tuple control time–\allowbreak{}controlled object–\allowbreak{}observable fact–\allowbreak{}control output. A runtime infrastructure requirement must identify where risk becomes observable, whether judgment concerns an object, capability, evidence, baseline, or historical effect, where non-self-declared facts are obtained, and how the decision changes execution. This avoids a purely principled account of trust and prevents counting a single failure repeatedly under overlapping quality labels.

The eight domains organize responsibilities at execution boundaries. A failure may have a primary control domain and auxiliary ones: T1 first judges capability-version mismatch, whereas T8 triggers regression after change. Controlled objects, required facts, and prohibited substitutions maintain internal consistency. The taxonomy locates checks and interface responsibilities; its coverage limits are discussed in Section 12.

The construction is responsibility-driven, not a Cartesian enumeration of the four dimensions. We start from decisions that change what may be invoked, concluded, committed, reused, or exposed. Checks sharing a controlled object and enforcement consequence are grouped: identity, task authority, and capability enter T1, while evidence applicability, coverage, independence, and support enter T6. Boundaries are retained when a positive result has a different consumer or cannot authorize the next operation, as with T3 admission versus T4 invocation, and T5 effects versus T6 conclusions. T8 groups change-triggered reevaluation rather than adding a final linear stage. Versions, provenance, and uncertainty are cross-cutting properties, not extra domains. Network security, physical safety certification, and cross-organizational governance are excluded by scope. Thus eight is a design decomposition with explicit merge and exclusion rules, not a proof that all possible failures fall into eight mutually exclusive classes.

\subsection{The eight domains and their boundaries}

Figure 2 and Table~\ref{tab:t0} define the domains. T1 includes dynamic tool authority and capability regression; T2 places engineering obligations in intent and planning; T3 covers both repair-action and verification-capability composition; T6 covers anomaly triggering, causal diagnosis, evidence independence, and model-support closure; T7 covers baseline admission, asset use, computation location, field exposure, and output release; and T8 covers interrupted execution, dependency invalidation, and evolution of model credibility and skill-capability scopes.

\begin{figure}[H]\centering
\includegraphics[width=\linewidth,height=0.77\textheight,keepaspectratio]{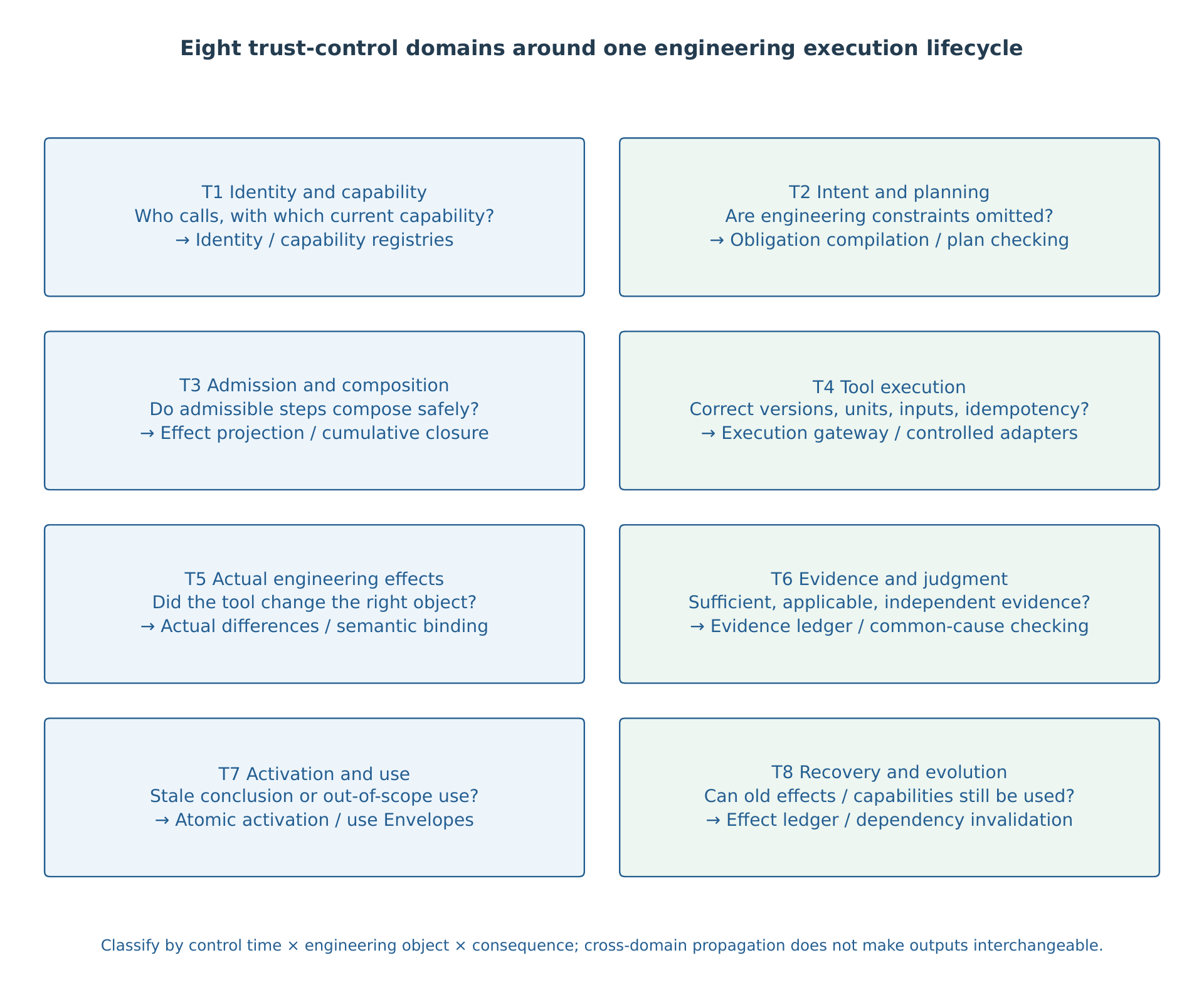}
\caption{Eight trust-control domains and their runtime boundaries.}\label{fig:2}
\end{figure}

\FloatBarrier
\begingroup
\small
\setlength{\parskip}{0pt}\setlength{\LTpre}{8pt}\setlength{\LTpost}{10pt}
\begin{longtable}{@{}P{53pt}P{97pt}P{89pt}P{119pt}P{81.75pt}@{}}
\caption{Primary control responsibilities and prohibited substitutions.}\label{tab:t0}\\
\toprule
\textbf{Domain} & \textbf{Primary control-boundary criterion} & \textbf{Cold-plate counterexample} & \textbf{Typical outcome} & \textbf{Non-substitutable result} \\
\midrule\endfirsthead
\multicolumn{5}{l}{\small\itshape Table \thetable\ (continued)}\\
\toprule
\textbf{Domain} & \textbf{Primary control-boundary criterion} & \textbf{Cold-plate counterexample} & \textbf{Typical outcome} & \textbf{Non-substitutable result} \\
\midrule\endhead
\midrule\multicolumn{5}{r}{\small Continued on next page}\\\endfoot
\bottomrule\endlastfoot
T1 Identity, authority, capability & First boundary deciding who may do what in this task from identity roots, context, capability Envelope, and tool manifest & Valid agent has excessive CAD write authority; old capability claim survives a solver upgrade & LEASE/REVOKE; \mbox{MATCH/}\allowbreak{}\mbox{MISMATCH/}\allowbreak{}\mbox{UNRESOLVED} & Valid identity does not establish task authority or capability applicability \\
T2 Intent, obligations, planning & First boundary compiling language goals into objects, criteria, conditions, invariants, and evidence requirements & Plan omits pressure drop, rated conditions, or mounting datums & Gap; RECONSTRUCT; HOLD & Syntactic completeness does not establish engineering-obligation coverage \\
T3 Candidates, capabilities, composition & First boundary evaluating candidate projections, cumulative action effects, or verification capability closure & Individually admissible actions solve before remeshing; shared output misses Gaps & \mbox{EXECUTABLE/}\allowbreak{}\mbox{REJECTED/}\allowbreak{}\mbox{UNRESOLVED} & Individual admission does not establish sequence executability; task success does not establish per-Gap acceptance \\
T4 Invocation and execution & First boundary controlling calls through actual parameters, references, tokens, idempotency keys, and job instances & Unit scaling error, duplicate asynchronous job, post-admission input substitution & \mbox{EXECUTE/}\allowbreak{}\mbox{DENY/}\allowbreak{}\mbox{HOLD/}\allowbreak{}\mbox{TAKEOVER} & Tool success cannot close T5 effect checking \\
T5 Actual effects and binding & First boundary determining what changed from actual before/after artifacts and downstream references & CAD save without width change; wrong inlet; mesh references old geometry & \mbox{PASS/}\allowbreak{}\mbox{FAIL/}\allowbreak{}\mbox{UNKNOWN}; impact entry & Neither T4 success nor agent declarations replace \mbox{Actual}\allowbreak{}\mbox{Engineering}\allowbreak{}\mbox{Delta} \\
T6 Diagnosis, evidence, judgment & First boundary determining supported conclusions from comparable observations, applicable evidence, provenance, and support & Missing rated condition; evidence shares a faulty mesh; model support breaks & DiagnosticGap; \mbox{COMPLETE/}\allowbreak{}\mbox{PARTIAL}; \mbox{PASS/}\allowbreak{}\mbox{FAIL/}\allowbreak{}\mbox{UNKNOWN} & Evidence existence does not establish coverage; COMPLETE does not trigger T7 acceptance \\
T7 Activation, use, disclosure & First unavoidable boundary for official baseline entry or specific asset use/disclosure & Post-check baseline drift; use outside admitted conditions; raw sensitive data export & \mbox{ACCEPT/}\allowbreak{}\mbox{HOLD/}\allowbreak{}\mbox{\texttt{REBASE\_REQUIRED}}; ALLOW/DENY & Historical ACCEPT/ALLOW cannot prevent T8 invalidation after change \\
T8 Recovery and evolution & First boundary re-evaluating historical effects, support, and capabilities under a new snapshot & Resume an old-mesh checkpoint; silently inherit after CoU expansion or skill upgrade & \mbox{VALID/}\allowbreak{}\mbox{INVALID/}\allowbreak{}\mbox{UNRESOLVED}; \mbox{\texttt{REVALIDATION\_REQUIRED}} & Historical commit, publication, or tool success does not establish current applicability \\
\end{longtable}
\endgroup

\subsection{Propagation and orthogonal relationships}

A forward chain determines who may propose actions, what obligations must be fulfilled, whether actions can execute together, how tools are invoked, what actually changed, what conclusions the facts support, whether conclusions and assets may enter subsequent activities, and which results remain dependable after time and change. These responsibilities correspond to capability and authority, planning, composition admission, tool control, effects, diagnosis and evidence, activation and use, and recovery and evolution.

Three orthogonal dimensions cut across the chain: \mbox{version/}\allowbreak{}\mbox{configuration}, \mbox{provenance/}\allowbreak{}\mbox{responsibility}, and \mbox{time/}\allowbreak{}\mbox{lifecycle.} Every decision binds a version and configuration; evidence and derived assets retain provenance; historical passes, commits, and releases remain reevaluable against current snapshots. T4 success cannot bypass T5 to close an engineering obligation; complete T6 evidence cannot bypass T7 to update a baseline; and historical T7 ACCEPT or ALLOW cannot prevent T8 invalidation after dependency drift.

\subsection{Confirmed violations and required unknowns}

V denotes confirmed per-rule violations supported by reliable engineering observations; U denotes facts required for evaluation but still unknown. Final decisions also incorporate the scope and object-closure evaluation, yielding V* and U* (Section 7.7). A nonempty V* produces the object's negative state while retaining U*. Empty V* with nonempty U* produces a hold or unresolved state. A positive state is permitted only when both are empty and the domain-specific positive closure condition holds. The shared convention is decision priority, not state names: hypotheses use S/R/U, effect ledgers use \mbox{VALID/}\allowbreak{}\mbox{INVALID/}\allowbreak{}\mbox{UNRESOLVED}, model admission distinguishes \mbox{ACCEPT\_CANDIDATE} from ACCEPT, and use decisions employ \mbox{ALLOW/}\allowbreak{}\mbox{DENY/}\allowbreak{}\mbox{UNRESOLVED.}

\section{Related Work and Remaining Interface Gaps}\label{sec:section_04}

\subsection{Tool-agent security evaluates hazards and policy adherence}

AgentDojo, ToolEmu, \ensuremath{\tau}-bench, and SafeToolBench evaluate prompt injection, dangerous tools, policy adherence, and final states \citep{r7,r8,r35,r36,r37}. We adopt the view that task completion alone is insufficient, and examine constraints linking CAD roles, mesh-parent versions, operating conditions, evidence domains, and formal baselines. We do not claim that existing governance lacks all such capabilities; rather, an engineering system must specify its observation sources and state consequences. Actual tool records in Section 11.3 also test whether our own adapters fulfill this responsibility.

\subsection{Runtime verification and assurance address traces and safe switching}

Runtime verification and execution monitoring check properties over event traces; enforcement can suppress, replace, or repair violating actions \citep{r33,r38}. Simplex and black-box extensions combine an unverified advanced controller, a trusted fallback controller, and switching logic \citep{r39}. Adaptive service infrastructures likewise connect monitoring, analysis, planning, and execution to runtime intervention; EPF4M/EI4MS provides such a framework for microservices in cloud--edge environments \citep{r54}. Our work also emphasizes runtime control, but concerns long-lived CAD, mesh, solver, evidence, and PLM artifacts, for which no single safe controller necessarily exists. The question is whether multiversion objects, evidence-support relationships, and historical effects are closed under a current snapshot. Monitoring and switching underpin T4 but cannot substitute for the engineering-fact evaluation in T5–T8.

\subsection{Artifact-centric processes jointly model data and lifecycles}

Artifact-centric approaches such as Guard-Stage-Milestone treat business data and lifecycle state as first-class objects and use guards, stages, and milestones for long-lived activities and event responses \citep{r40,r41}. Their incremental and fixed-point semantics provide a rigorous basis for event-driven propagation. We adopt the idea that object and process state jointly govern execution, but do not equate engineering trust with milestone attainment. Tool completion, artifact existence, or a previously attained milestone cannot replace an \mbox{Actual}\allowbreak{}\mbox{Engineering}\allowbreak{}\mbox{Delta}, evidence-support closure, or current applicability. Beyond how a process artifact evolves, we ask what actually changed, whether it satisfies frozen conditions, and whether its evidence remains usable for a specified conclusion.

\subsection{Scientific workflows and provenance support traceability and reproduction}

W3C PROV, \mbox{Prov}\allowbreak{}\mbox{ONE/}\allowbreak{}\mbox{Prov}\allowbreak{}\mbox{ONE}+, RO-Crate, and software provenance record derivation among data, activities, software, and environments \citep{r23,r42,r43,r44}. They underpin our provenance graphs, events, and unified fact sources. However, provenance does not imply independence; replayability does not imply applicability to the current configuration or sufficiency for an engineering conclusion. We therefore project provenance into common-cause factors, applicability domains, coverage states, and engineering judgments instead of treating a complete lineage as evidence assessment.

\subsection{Digital engineering, change propagation, and model assurance}

NASA-STD-7009B and model-lifecycle V\&V/VVUQ emphasize intended use, evidence, and applicability conditions \citep{r14,r30,r34}. Early MBSE V\&V, engineering change propagation, and model-based assurance cases address design behavior, impact dependencies, and claim–\allowbreak{}argument–\allowbreak{}evidence relations \citep{r45,r46,r47,r48}. ACCESS also links heterogeneous artifacts and supports development- and runtime assurance evaluation \citep{r48}. Version dependencies also guide evolution planning in multi-version microservice systems \citep{r53}; for engineering artifacts, those dependencies must additionally determine whether existing evidence remains applicable after a change. These approaches cannot be dismissed as static documentation. Our focus is a unified interface arrangement with engineering-version and configuration semantics between agent-proposed action, tool-produced artifacts, evidence admission, and historical-effect reevaluation.

\subsection{Contract-based tool execution and verifiable CAD generation}

ToolGate uses typed symbolic states and \mbox{pre/}\allowbreak{}\mbox{postcondition} contracts for tool calls and result commits \citep{r49}. SPADA uses deterministic tests as execution contracts for parametric CAD assembly generation \citep{r50}. ReliCAD links design intent, geometric relations, topological dependencies, and runtime evidence to verification feedback \citep{r51}; the latter is currently a preprint. These works directly concern T2–T5, so postcondition checking, structured design intent, and geometric feedback are not individually claimed as our inventions. Our framework coordinates CAD, meshing, solving, evidence, use, and evolution, emphasizing their dependence on fact mapping. We have not compared these systems under matched conditions and claim no superior performance or safety.

\subsection{Framework contribution and comparison limits}

The incremental question is not whether a framework can express a postcondition, but which engineering obligations must reach that postcondition and its downstream consumers. Consider a required channel-width change followed by simulation. ToolGate can reject an unchanged width if the relevant measured predicate is included in its postcondition; our omitted-width counterexample instead concerns how that predicate enters the mandatory inventory. SPADA's deterministic tests and ReliCAD's geometric feedback address important CAD-level correctness questions. A subsequent claim additionally needs a mesh bound to the accepted geometry, evidence for the requested operating domain, and current baseline correspondence. A provenance graph can identify two analyses sharing a mesh; deciding whether that dependency undermines independent support requires a target-relevant error-propagation predicate. None of these downstream judgments follows merely from a successful CAD test.

Table~\ref{tab:related_interfaces} compares documented emphases and the additional interface questions pursued here. Its right column identifies our research questions, not capabilities presumed absent from the compared systems. Our distinction is the explicit distribution of these responsibilities and their non-substitution contracts; comprehensive enforcement is a specification goal whose present coverage is reported separately in Table~\ref{tab:t2}.

The comparison is consequently about interface composition rather than a claim that the neighboring primitives are inadequate. A conventional postcondition can be correct while its mandatory-check inventory is incomplete; a CAD test can pass while a downstream mesh is bound to another revision; a provenance graph can be complete while two results share a target-relevant failure source; and a release workflow can be orderly while an historical effect is no longer applicable. The framework makes these transitions explicit and assigns each one a separately typed decision and fact contract. Whether this arrangement reduces integration effort or bypass risk relative to a particular industrial product is not measured here; those are follow-on comparative questions, not hidden claims in the present prototype results.

\FloatBarrier
\begingroup\small
\setlength{\parskip}{0pt}
\begin{longtable}{@{}P{85pt}P{174pt}P{193pt}@{}}
\caption{Related methods and the engineering interfaces examined here; a document-level comparison, not a benchmark.}\label{tab:related_interfaces}\\
\toprule
\textbf{Work / basis} & \textbf{Established focus} & \textbf{Additional question in this framework} \\
\midrule\endfirsthead
\multicolumn{3}{l}{\small\itshape Table \thetable\ (continued)}\\
\toprule
\textbf{Work / basis} & \textbf{Established focus} & \textbf{Additional question in this framework} \\
\midrule\endhead
\bottomrule\endlastfoot
ToolGate \citep{r49} & Typed symbolic states; preconditions and postconditions around tool invocation and result commitment. & How are mandatory checks derived from frozen obligations, and how do measured facts survive adapter and consumer transformations? \\
SPADA \citep{r50} & Deterministic execution tests and repair for parametric CAD assemblies. & Which downstream mesh and solver versions support the requested engineering claim? Cross-object support is a separate interface responsibility. \\
ReliCAD \citep{r51} & Design intent, geometric and topological relations, and runtime verification feedback. & How do evidence coverage, common-cause analysis, baseline activation, and later reuse retain distinct decision authority? \\
PROV and model assurance \citep{r23,r14} & Provenance representation and model credibility for intended use. & Which actual dependencies invalidate support, and which current-context evidence authorizes continued consumption? \\
This framework & Eight responsibilities, four planes, three chains; explicit check inventories and fact-consumption contracts. & Selected effect, evidence, activation, and recovery paths are tested. General closure and full lifecycle enforcement remain design requirements. \\
\end{longtable}
\endgroup

\section{Problem Definition, Trust Assumptions, and Fault Model}\label{sec:section_05}

\subsection{System boundary and research object}

We study industrial-agent execution systems that produce persistent effects on engineering objects within one enterprise governance domain. Components include an agent Runtime, task/planning services, industrial-software adapters, \mbox{MBSE/}\allowbreak{}\mbox{CAD/}\allowbreak{}\mbox{CAE/}\allowbreak{}\mbox{test/}\allowbreak{}\mbox{PLM} tools, \mbox{object/}\allowbreak{}\mbox{configuration} services, \mbox{evidence/}\allowbreak{}\mbox{provenance} services, \mbox{rule/}\allowbreak{}\mbox{capability} registries, and human takeover endpoints. The concern is not whether a model response appears plausible, but whether autonomous execution maintains consistency among objects, rules, configurations, evidence, and responsibilities from intent through engineering effects to evidence and lifecycle evolution.

Controlled actions include reading, modification, solving, postprocessing, writeback, baseline activation, purpose-specific use, information release, recovery, and incremental revalidation. Cross-enterprise identity recognition, contract negotiation, and responsibility allocation are excluded. Externally supplied data is treated as an asset with provenance and use constraints once it enters the enterprise. This boundary isolates execution-semantic gaps that remain even under unified governance.

\subsection{Trust assumptions and roots of trust}

We assume organizational identity roots, controlled \mbox{version/}\allowbreak{}\mbox{baseline} services, and managed publication of rules and capability manifests. Plans, return codes, file existence, and matching names are not engineering facts. Placement in the control plane does not make an adapter correct: it is an observation and semantic-transformation component requiring validation of field provenance, coordinates, units, applicable versions, and unknown scope. Section 11.3 shows an adapter emitting MATCH despite an existing negative interface fact, and an audit sharing an erroneous parameter interpretation. Traceable provenance and correct field semantics are separate conditions.

A decision and its grounds must refer to the same scope: object, rule, and capability versions, observation scope, and invalidation conditions belong in the frozen context. Current and reference facts are checked again before the decision takes effect. Unreadable required facts produce HOLD or UNRESOLVED, not assumed satisfaction.

\begin{figure}[H]\centering
\includegraphics[width=\linewidth,height=0.77\textheight,keepaspectratio]{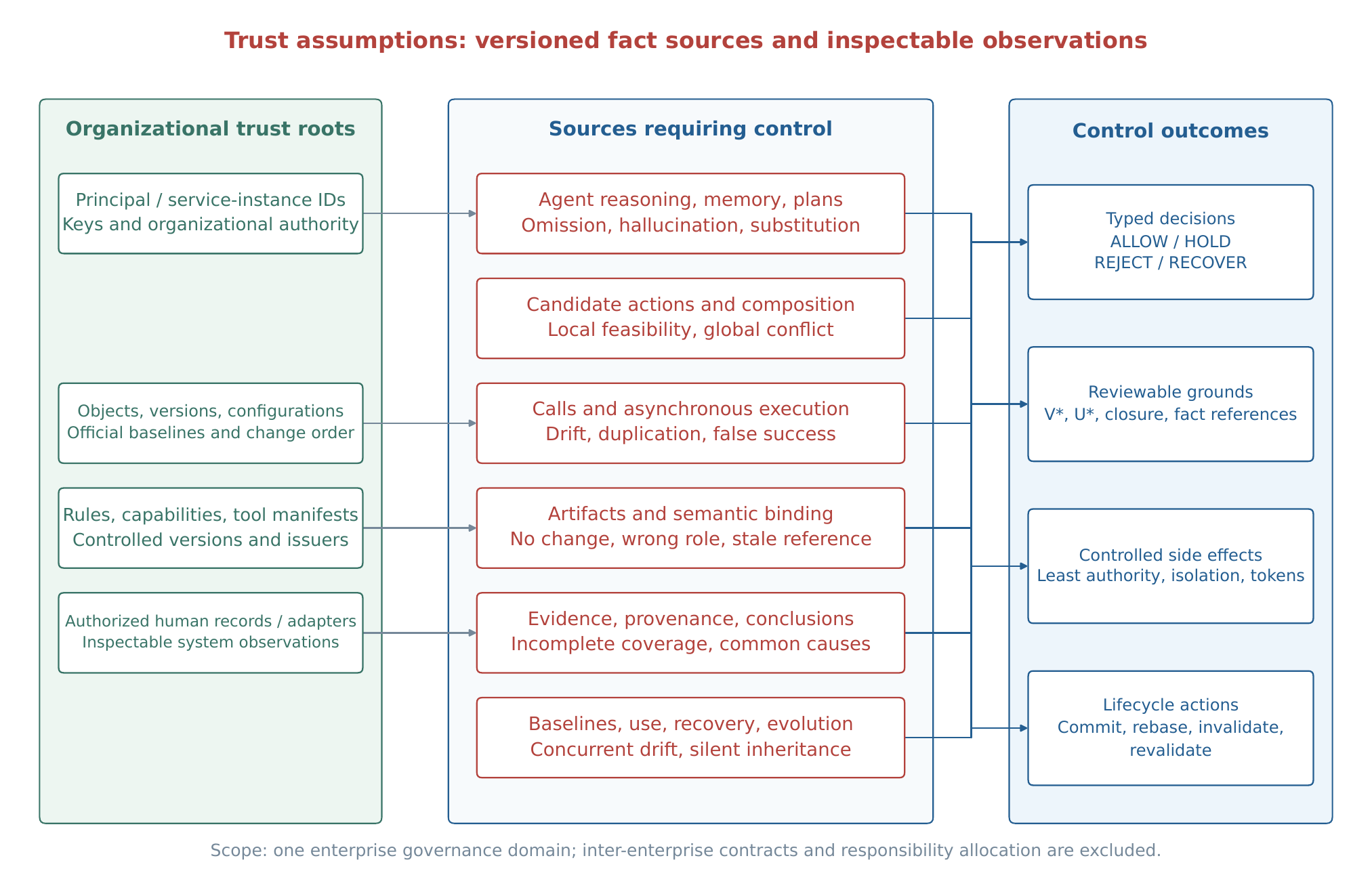}
\caption{Trust assumptions and fault model for intra-enterprise industrial-agent execution.}\label{fig:3}
\end{figure}

\subsection{Fault and adversarial model}

Nonmalicious faults include planning omissions, object/version misbinding, unit and coordinate errors, duplicate asynchronous jobs, no-op tools, stale downstream references, inadequate coverage, common-cause sources, concurrent baseline drift, invalid checkpoints, and \mbox{model/}\allowbreak{}\mbox{skill-}\allowbreak{}\mbox{scope} change. Bounded adversarial behavior includes inducing unauthorized action through prompts or external content, substituting parameters or paths after admission, fabricating tool success, disguising semantic mismatch with identical names, and evading use constraints by renaming, repackaging, or relocating assets.

We do not assume an adversary can compromise organizational identity roots, alter protected clocks, or bypass all adapters to modify databases directly. Such compromise requires lower-level protection. Nor do software rules prove complex physical models absolutely correct. Each engineering decision must instead be traceable to checkable models, observations, and evidence for the current configuration.

\subsection{Research questions and design principles}

RQ-A asks which non-substitutable trust requirements need separate control objects. RQ-B asks how to realize them in interfaces between agent environments, industrial adapters, and fact services. RQ-C asks how mechanisms compose within a frozen context while preserving effects, coverage, judgment, and admission boundaries. RQ-D asks which relationships local experiments support and which observation, adaptation, and evolution conditions they expose. Classification and architecture address the first two; mechanism analysis and bounded evidence jointly address the latter two.

Six design principles guide the framework:
\begin{enumerate}[label=P\arabic*.,leftmargin=2.5em]
\item Agent declarations do not establish engineering facts.
\item Tool success does not establish the required engineering effect.
\item Result existence does not establish sufficient evidence.
\item Sufficient evidence does not imply a passing engineering judgment.
\item Historical commitment does not establish current applicability.
\item Legitimate identity does not authorize every asset use.
\end{enumerate}
These are normative requirements. The remainder of the paper assigns their enforcement to version-bound observations, typed decisions, and reevaluation interfaces, and examines where implementations satisfy or violate them.

\subsection{Scope limitations}

The domains concern engineering execution trust inside an enterprise, not replacement of network or operating-system security, industrial functional-safety certification, or inter-enterprise governance. Confidential, federated, or multiparty computation are pluggable information-release execution modes; cryptographic security proofs are not evaluated. For complex CFD/FEA, we evaluate closure among models, configurations, applicability, and evidence, not engineering certification through lightweight proxies.

\section{Software Architecture Embedded in the Agent Execution Environment}\label{sec:section_06}

\subsection{Infrastructure requirements derived from failures}

Sections 2–5 imply nine requirements beyond ordinary agent orchestration. R1 places control both before side effects and after tool returns, closing admission–\allowbreak{}execution–\allowbreak{}observation–\allowbreak{}acceptance. R2 requires object identity, version, configuration, and role, not merely paths, names, or digests. R3 binds plans, execution, effects, evidence, and commitment to a frozen context. R4 natively preserves confirmed violations V and required unknowns U rather than treating read failures as low-confidence passes. R5 equips evidence with scenarios, configurations, provenance, quality, and coverage and supports common-cause and closure evaluation. R6 detects drift between checking and baseline writes. R7 propagates invalidation along engineering dependencies and reevaluates only the affected closure. R8 separately controls computation location, field exposure, and output granularity because allowed use does not permit raw-data export. R9 narrows, revokes, or reestablishes tool authority and capability as tasks, risk, and dependencies change.

The substrate is not another LLM planner. LLMs retain language understanding, proposal generation, and heuristic search; CAD/CAE tools retain geometric and physical computation; IAM, OPC UA secure channels, and MCP authorization address identity and communication boundaries. The substrate turns high-risk effects into controlled transactions with objects, versions, conditions, evidence, purposes, and lifecycles. OPC UA itself distinguishes communication security from application conditions for enabling mechanisms \citep{r9,r10}. We address engineering-semantic conditions in the latter category.

\subsection{Four planes and trust boundaries}

Figure 4 separates agent orchestration, trust control, industrial execution, and engineering facts. The orchestration plane contains models, planning, memory, multi-agent orchestration, and task state and proposes \mbox{intents/}\allowbreak{}\mbox{candidates.} The control plane hosts eight modules. The execution plane connects MBSE, CAD, meshing, CAE, testing, PLM, and controlled adapters. The fact plane stores versioned objects, rules, observations, evidence, ledgers, scopes, and change events. Requests pass task/action controls; observed artifacts feed evidence and baseline consumption; T8 subscribes to changes and triggers reevaluation.

\begin{figure}[H]\centering
\includegraphics[width=\linewidth,height=0.77\textheight,keepaspectratio]{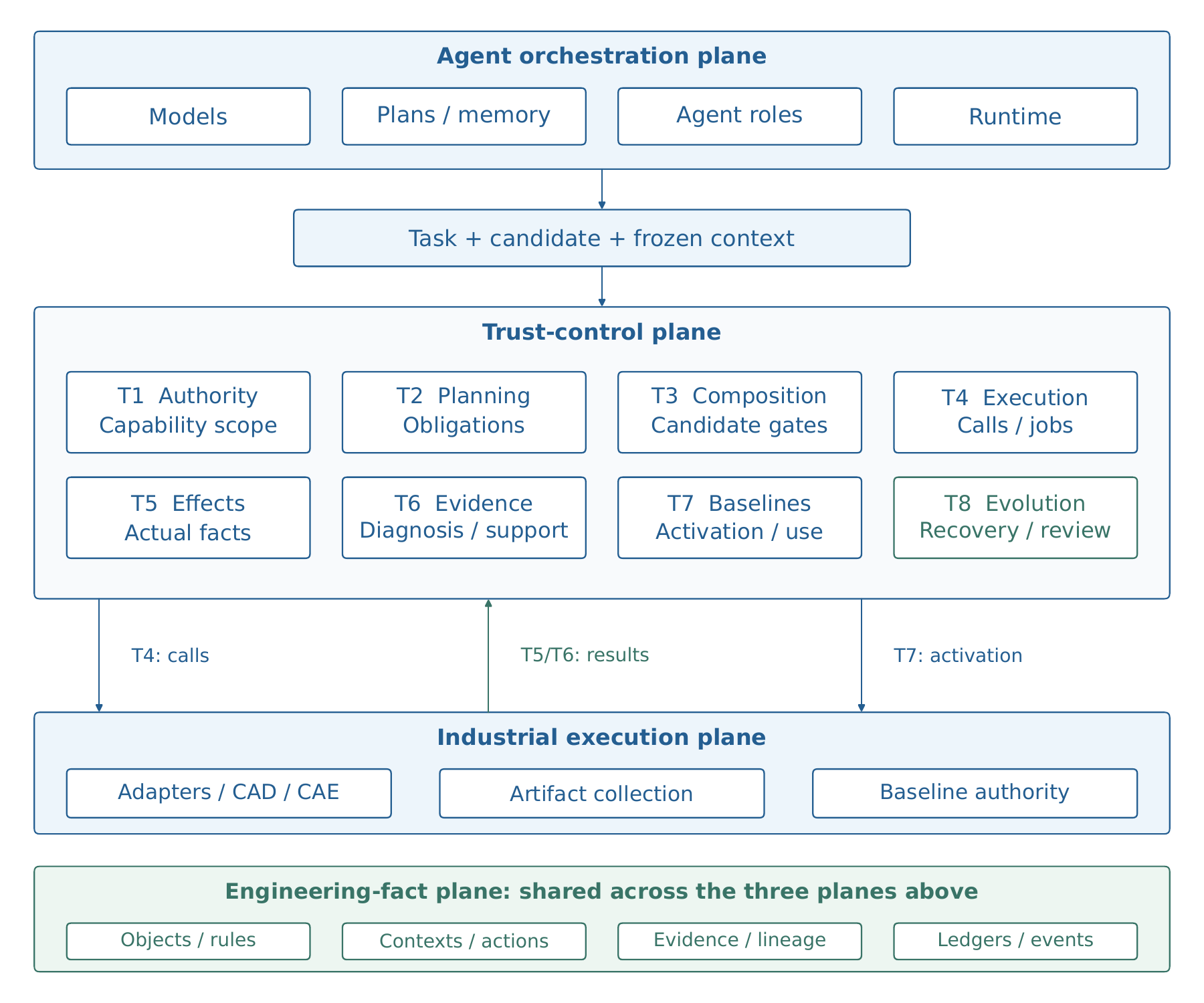}
\caption{Control modules, industrial-execution interfaces, and change feedback across four planes.}\label{fig:4}
\end{figure}

Five boundaries are important: agent to trusted proxy, proxy to adapter, adapter to industrial software, candidate workspace to formal baseline, and evidence/state services to consumers. They use different objects: ActionIntent instead of arbitrary commands; ExecutionToken plus normalized parameters; actual input and job identifiers; \mbox{Admission}\allowbreak{}\mbox{Record} plus conditional writing; and use/disclosure envelopes. A single generic token would conceal these semantic differences. Tokens reference and protect controlled records; they do not replace them.

\subsection{Mechanisms mapped to architectural modules}

The mechanisms in Section 8 reside in the eight modules of Figure 4. T1–T3 determine authority, obligations, and composition before calls; T4 controls invocation; T5 observes changes; T6 carries diagnostic and evidential responsibility; T7 governs activation, use, and release; and T8 maintains current applicability and triggers \mbox{recovery/}\allowbreak{}\mbox{regression.} Table~\ref{tab:t1} maps deployment and I/O, Table~\ref{tab:t2} distinguishes implementation evidence, Table~\ref{tab:t3} describes mechanisms, and Table~\ref{tab:t4} assigns collaborative handoff responsibilities.

\FloatBarrier
\begingroup
\small
\setlength{\parskip}{0pt}\setlength{\LTpre}{8pt}\setlength{\LTpost}{10pt}
\begin{longtable}{@{}P{41.78pt}P{87.95pt}P{87.95pt}P{114.34pt}P{107.74pt}@{}}
\caption{Mechanisms, architectural modules, deployment locations, and interfaces.}\label{tab:t1}\\
\toprule
\textbf{Domain} & \textbf{Logical module in Figure 4} & \textbf{Deployment location} & \textbf{Authoritative facts read directly} & \textbf{Output and enforcement role} \\
\midrule\endfirsthead
\multicolumn{5}{l}{\small\itshape Table \thetable\ (continued)}\\
\toprule
\textbf{Domain} & \textbf{Logical module in Figure 4} & \textbf{Deployment location} & \textbf{Authoritative facts read directly} & \textbf{Output and enforcement role} \\
\midrule\endhead
\midrule\multicolumn{5}{r}{\small Continued on next page}\\\endfoot
\bottomrule\endlastfoot
T1 & Identity–\allowbreak{}authority–\allowbreak{}capability controller & Central \mbox{policy/}\allowbreak{}\mbox{capability} service; tool-gateway intercept & Identity instance, task context, tool manifest, capability Envelope, risk rules & Capability match, least-authority lease, \mbox{revocation/}\allowbreak{}\mbox{takeover} conditions \\
T2 & Obligation compiler and plan checker & After agent runtime, before candidate execution & Target, \mbox{requirements/}\allowbreak{}\mbox{MBSE} semantics, criteria, conditions, invariants, evidence requirements & Typed obligations, activity graph, uncovered Gap/HOLD \\
T3 & Candidate and composition gate & Control service; execution-plan entry & Effect projections, action dependencies, intermediate invariants, verification inputs/outputs and provenance constraints & Individual admission, composite executability, capability closure \\
T4 & Execution gateway and adapter controller & API/MCP gateway; tool-side Sidecar/RPA container & Admitted action, actual paths, parameter schema, idempotency key, tool/adapter version & ExecutionToken, controlled job, observations, \mbox{DENY/}\allowbreak{}\mbox{HOLD/}\allowbreak{}\mbox{TAKEOVER} \\
T5 & Effect observer and binding resolver & Between tool observer and \mbox{object/}\allowbreak{}\mbox{configuration} services & Before/after artifacts, role predicates, \mbox{topology/}\allowbreak{}\mbox{interfaces}, actual downstream references & \mbox{Actual}\allowbreak{}\mbox{Engineering}\allowbreak{}\mbox{Delta}, binding states, effect compliance, impact entry \\
T6 & Diagnosis–\allowbreak{}evidence–\allowbreak{}support controller & Evidence, provenance-graph, and engineering-decision services & Comparable observations, evidence, lineage, failure factors, Claim support & \mbox{Diagnostic/}\allowbreak{}\mbox{independence} gaps, coverage, judgments, revalidation needs \\
T7 & Activation–\allowbreak{}use–\allowbreak{}disclosure controller & \mbox{Candidate/}\allowbreak{}\mbox{baseline} boundary; asset-consumption enforcement point & Candidate content, current baseline, admission conditions, use lineage, sensitive fields, execution domain & \mbox{ACCEPT/}\allowbreak{}\mbox{HOLD/}\allowbreak{}\mbox{REBASE\_}\allowbreak{}\mbox{REQUIRED}; ALLOW/DENY; constrained release \\
T8 & Ledger–\allowbreak{}invalidation–\allowbreak{}evolution manager & Event bus, state service, dependency graph & Effect ledger, engineering dependencies, \mbox{object/}\allowbreak{}\mbox{configuration} changes, \mbox{credibility/}\allowbreak{}\mbox{capability} domains & \mbox{INVALID/}\allowbreak{}\mbox{UNRESOLVED}, recovery plan, incremental revalidation, Envelope update \\
\end{longtable}
\endgroup

A nominal action proceeds through T1\ensuremath{\rightarrow}T2\ensuremath{\rightarrow}T3\ensuremath{\rightarrow}T4 to tools, then through artifact observation at T5, evidence evaluation at T6, and activation/use control at T7. T8 interacts through change events and engineering dependencies and can reopen historical decisions. Modules may share processes or storage but must retain distinct objects and decision conditions. Section 6.12 describes their cross-plane relationships.

\FloatBarrier
\begingroup
\small
\setlength{\parskip}{0pt}\setlength{\LTpre}{8pt}\setlength{\LTpost}{10pt}
\begin{longtable}{@{}P{75.78pt}P{124.81pt}P{115.90pt}P{129.27pt}@{}}
\caption{Full framework and the scope of current implementation evidence.}\label{tab:t2}\\
\toprule
\textbf{Domain} & \textbf{Framework design} & \textbf{\mbox{Implementation/}\allowbreak{}\mbox{evidence} available} & \textbf{Scope not established} \\
\midrule\endfirsthead
\multicolumn{4}{l}{\small\itshape Table \thetable\ (continued)}\\
\toprule
\textbf{Domain} & \textbf{Framework design} & \textbf{\mbox{Implementation/}\allowbreak{}\mbox{evidence} available} & \textbf{Scope not established} \\
\midrule\endhead
\midrule\multicolumn{4}{r}{\small Continued on next page}\\\endfoot
\bottomrule\endlastfoot
T1 & Identity, task authority, capability \mbox{matching/}\allowbreak{}\mbox{regression} & Token/version rules; closed-set capability mismatch & Enterprise IAM integration; complete skill \mbox{regression/}\allowbreak{}\mbox{publication} \\
T2 & Obligations, coverage, activity reconstruction & Typed obligations and plan rules; omission counterexamples & Complete arbitrary-language compilation; enterprise MBSE integration \\
T3 & Individual candidates, cumulative effects, capability composition & \mbox{Candidate/}\allowbreak{}\mbox{action-}\allowbreak{}\mbox{order} rules; per-family ablation & Complete composition for arbitrary engineering actions \\
T4 & Actual inputs, tokens, idempotency, long jobs & Open-source \mbox{CAD/}\allowbreak{}\mbox{mesh/}\allowbreak{}\mbox{structural} calls; E6; local concurrency & Commercial adapters; arbitrary external exactly-once effects \\
T5 & Actual differences and semantic binding & BFM-1.1 replay; bracket/cold-plate CAD paths; separate STEP inspection path; predicate-omission counterexample & Independent full-geometry validation; general mandatory-check closure; arbitrary adapters \\
T6 & Coverage, judgment, common causes, diagnosis, support closure & Closed-set evidence \mbox{rules/}\allowbreak{}\mbox{common-}\allowbreak{}\mbox{cause} ablation; OpenFOAM E2 & Complete \mbox{diagnosis/}\allowbreak{}\mbox{assurance-}\allowbreak{}\mbox{case} service; sufficient product-model V\&V \\
T7 & B/C/A activation, use, disclosure & Local conditional activation, use rules, persistence faults & Production PLM transactions; TEE/MPC; complete disclosure backend \\
T8 & Dual-state ledger, dependency invalidation, \mbox{recovery/}\allowbreak{}\mbox{evolution} & Old checkpoints; unknown-edge propagation; version-bound recovery decisions and local Schema repair & Complete recovery-to-commit service; all malformed states; continuous-evolution validation \\
\end{longtable}
\endgroup

Table~\ref{tab:t2} distinguishes designed mechanisms, \mbox{implementation/}\allowbreak{}\mbox{experimental} records, and unestablished scope. Table~\ref{tab:t3} identifies the engineering semantics contributed by each domain. Table~\ref{tab:t4} specifies which actor transfers which facts and what recipients must check. Together they distinguish maturity, mechanism content, and responsibility handoffs.

\FloatBarrier
\begingroup
\small
\setlength{\parskip}{0pt}\setlength{\LTpre}{8pt}\setlength{\LTpost}{10pt}
\begin{longtable}{@{}P{75.78pt}P{124.81pt}P{115.90pt}P{129.27pt}@{}}
\caption{Technical objects, mechanism arrangements, and supporting infrastructure.}\label{tab:t3}\\[-0.5pt]
\toprule
\textbf{Domain} & \textbf{Technical \mbox{objects/}\allowbreak{}\mbox{methods}} & \textbf{Proposed engineering semantics} & \textbf{Interfaces and enabling mechanisms} \\
\midrule\endfirsthead
\multicolumn{4}{l}{\small\itshape Table \thetable\ (continued)}\\
\toprule
\textbf{Domain} & \textbf{Technical \mbox{objects/}\allowbreak{}\mbox{methods}} & \textbf{Proposed engineering semantics} & \textbf{Interfaces and enabling mechanisms} \\
\midrule\endhead
\midrule\multicolumn{4}{r}{\small Continued on next page}\\\endfoot
\bottomrule\endlastfoot
T1 & Task authority and skill applicability & Separate demonstrated capability from permitted task actions; dependency changes trigger scope regression & Capability claims, Scope, task leases; IAM/leases enable authorization \\
T2 & Typed obligations and activity requirements & Difference objects, criteria, conditions, and invariants; synthesize activities through rules & ObligationDiff \ensuremath{\rightarrow} \mbox{Activity}\allowbreak{}\mbox{Delta}\allowbreak{}\mbox{Graph}; \mbox{names/}\allowbreak{}\mbox{traceability} do not replace semantic differences \\
T3 & Candidate effects and two compositions & Project modifications; evaluate cumulative effects at checkpoints; close verification inputs, quality, and provenance & Separate action executability and capability composition; per-Gap acceptance; graph/rule engines \\
T4 & Actual inputs and asynchronous jobs & Bind admission to actual objects, parameters, jobs; isolate and readmit expired results & ExecutionToken, input manifest, job observations; \mbox{idempotency/}\allowbreak{}\mbox{version} checks \\
T5 & Actual differences and fact mapping & Evaluate \mbox{artifacts/}\allowbreak{}\mbox{downstream} bindings; preserve \mbox{negative/}\allowbreak{}\mbox{unknown} facts; derive checks from obligations and actual effects & \mbox{Actual}\allowbreak{}\mbox{Engineering}\allowbreak{}\mbox{Delta}, role predicates, impact entries; traceable coordinates, units, sources \\
T6 & Diagnosis, evidence, credibility support & Separately evaluate hypotheses, obligation coverage, actual common causes, and Claim domains; retain \mbox{coverage/}\allowbreak{}\mbox{judgment} states & DiagnosticGap, \mbox{Repair}\allowbreak{}\mbox{Obligation}, support closure; provenance graphs and V\&V \\
T7 & Model activation and asset consumption & Separate candidate eligibility from activation; review B/C/A; monotonically narrow use/disclosure along lineage & Acceptance records and use/disclosure Envelopes; conditional writes/access controls \\
T8 & Historical effects and domain evolution & Separate \mbox{lifecycle/}\allowbreak{}\mbox{applicability}; propagate true/unknown dependencies; revalidate model and skill domains separately & Dual-state ledger, conditional propagation, \mbox{Credibility}\allowbreak{}\mbox{Delta/}\allowbreak{}\mbox{Capability}\allowbreak{}\mbox{Delta}; events/graphs \\
\end{longtable}
\endgroup

Table~\ref{tab:t3} organizes the definitions and interfaces proposed here without claiming all supporting mechanisms as new inventions. Table~\ref{tab:t2} and Sections 9–11 provide implementation evidence; Sections 7–8 define methods and evaluation rules. Public, checkable research supplies methodological foundations; unpublished materials do not replace definitions within the paper.

\FloatBarrier
\begingroup
\small
\setlength{\parskip}{0pt}\setlength{\LTpre}{8pt}\setlength{\LTpost}{10pt}
\begin{longtable}{@{}P{75.78pt}P{124.81pt}P{115.90pt}P{129.27pt}@{}}
\caption{Artifact handoffs and control responsibilities in intra-enterprise collaboration.}\label{tab:t4}\\
\toprule
\textbf{Role} & \textbf{Handoff content} & \textbf{Required recipient checks} & \textbf{Control boundary} \\
\midrule\endfirsthead
\multicolumn{4}{l}{\small\itshape Table \thetable\ (continued)}\\
\toprule
\textbf{Role} & \textbf{Handoff content} & \textbf{Required recipient checks} & \textbf{Control boundary} \\
\midrule\endhead
\midrule\multicolumn{4}{r}{\small Continued on next page}\\\endfoot
\bottomrule\endlastfoot
Design agent & Candidate geometry, actual difference, invariants, unknowns & \mbox{Version/}\allowbreak{}\mbox{configuration}, actual artifact, roles, modification scope & T1–T5; write candidates, not official baseline publication \\
Simulation agent & Mesh parent version, actual inputs, job, conditions, raw results & Geometry/mesh lineage, boundaries, tool capability, job lease & T1, T3–T6; produce evaluable results \\
Verification agent & Per-obligation coverage, engineering judgment, \mbox{provenance/}\allowbreak{}\mbox{support} & Target obligations, evidence applicability, quality, common causes & T6; report completion does not establish product PASS \\
Engineering-change agent & Activation request, current references, use request & B/C/A, admission conditions, use, change impacts & T7/T8; baseline authority performs controlled updates \\
\end{longtable}
\endgroup

\subsection{Deployment: centralized policy and execution-local enforcement}

Deployment is logically centralized and execution-local. Enterprise services manage policy, rule versions, object types, and state semantics; action gates, interceptors, and frequent object resolvers reside near agents/tools as sidecars, in-process middleware, or local proxies. OPA recommends placing policy decision points near enforcement \citep{r28}. We add engineering-object resolution and \mbox{dependency/}\allowbreak{}\mbox{evidence} evaluation to their inputs. Table~\ref{tab:t2} and Section 11.6 identify the actual local-service prototype configurations.

Adapters for stable APIs can be MCP servers, REST/gRPC proxies, or vendor-SDK wrappers. MCP exposes structured tool inputs and model-initiated calls \citep{r29}, but CallToolResult.isError=false is only a protocol-level result, not an \mbox{Actual}\allowbreak{}\mbox{Engineering}\allowbreak{}\mbox{Delta}. GUI/RPA-only tools require pre-call snapshots, immutable parameter manifests, post-call artifact rereading, and replayable job/file handles on abnormal exit. Capability manifests declare observable fields; unobservable required fields enter U rather than being filled by an agent.

Fact services can use a small monolith, a plant-level control service with sidecars, or layered services across sites in one governance domain. These configurations share requirements for versioned objects, actual observations, evidence, and durability. References and events synchronize boundary changes. T7 can connect to isolated or confidential-computation backends; implementation and evaluation limits are stated in Table~\ref{tab:t2} and Section 12.

\paragraph{Mapping to a PLM authority.}
A PLM adapter must resolve an immutable object revision and configuration, keep working candidates distinct from released baseline membership, and relate the proposed change to its change-order or approval record. The activation request carries the expected baseline revision, candidate digest, and frozen acceptance conditions; approval of a change order does not by itself establish the candidate's engineering judgment. The authoritative service returns the committed revision and membership record for readback. If it lacks an atomic compare-and-update operation, the adapter must use the system's locking, staging, or reconciliation protocol and expose an unresolved outcome when commitment cannot be confirmed. These are adapter obligations, not claims of tested Teamcenter, Windchill, or Aras integrations.

\subsection{Runtime states and transitions}

Figure 5 presents the main controlled-execution state machine: CREATED, \mbox{CONTEXT\_}\allowbreak{}\mbox{FROZEN}, \mbox{PLAN\_}\allowbreak{}\mbox{EVALUATED}, \mbox{ACTION\_}\allowbreak{}\mbox{ADMITTED}, EXECUTING, \mbox{EFFECT\_}\allowbreak{}\mbox{OBSERVED}, \mbox{EVIDENCE\_}\allowbreak{}\mbox{EVALUATED}, \mbox{ACTIVATION\_}\allowbreak{}\mbox{CANDIDATE}, and COMMITTED. Evaluation can enter HELD or REJECTED; interruption, context drift, or historical-effect invalidation leads to \mbox{RECOVERY\_}\allowbreak{}\mbox{REQUIRED}; mismatched pre-commit references produce \mbox{REBASE\_}\allowbreak{}\mbox{REQUIRED.} HOLD means required facts are missing and the same candidate may continue after observation. REJECT means reliable facts confirm a violation; unrelated evidence cannot automatically turn it into a pass. The distinction must be persisted.

\begin{figure}[H]\centering
\includegraphics[width=\linewidth,height=0.77\textheight,keepaspectratio]{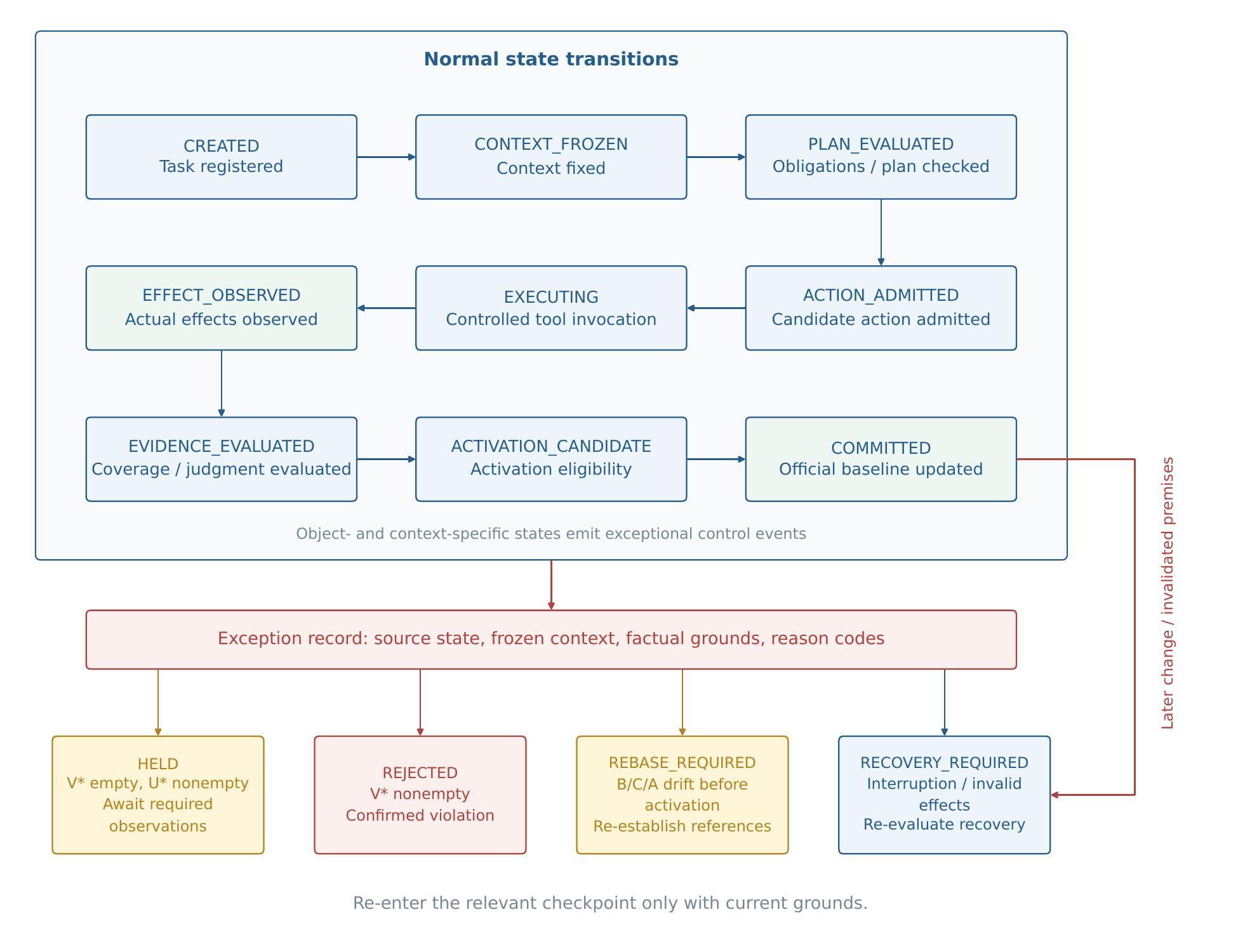}
\caption{State machine for controlled intra-enterprise industrial-agent execution.}\label{fig:5}
\end{figure}

Events and guards govern transitions. \mbox{ACTION\_}\allowbreak{}\mbox{ADMITTED} requires plan coverage, action gates, and authority; \mbox{EFFECT\_}\allowbreak{}\mbox{OBSERVED} requires a readable terminal job state and preserved artifacts; COMMITTED requires current B/C/A to correspond to evaluation references. DecisionRecord stores guard grounds and unknowns. Low-risk read-only queries may take shorter paths, but outputs used as engineering evidence still require T6; T5 must preserve existing negative facts for subsequent evaluation.

\subsection{Command, query, and event contracts}

Commands change controlled state: \mbox{freeze\_}\allowbreak{}\mbox{context}, \mbox{evaluate\_}\allowbreak{}\mbox{plan}, \mbox{admit\_}\allowbreak{}\mbox{action}, \mbox{start\_}\allowbreak{}\mbox{execution}, \mbox{register\_}\allowbreak{}\mbox{actual\_}\allowbreak{}\mbox{delta}, \mbox{evaluate\_}\allowbreak{}\mbox{evidence}, \mbox{request\_}\allowbreak{}\mbox{activation}, \mbox{issue\_}\allowbreak{}\mbox{usage\_}\allowbreak{}\mbox{permit}, invalidate, and \mbox{request\_}\allowbreak{}\mbox{recovery.} Queries read objects, decisions, and dependencies; events propagate facts that have occurred. Every write command carries at least \mbox{request\_}\allowbreak{}\mbox{id}, \mbox{trace\_}\allowbreak{}\mbox{id}, \mbox{actor\_}\allowbreak{}\mbox{ref}, \mbox{target\_}\allowbreak{}\mbox{ref}, \mbox{expected\_}\allowbreak{}\mbox{revision}, \mbox{context\_}\allowbreak{}\mbox{ref}, and \mbox{idempotency\_}\allowbreak{}\mbox{key.}

CloudEvents-style envelopes support transport across buses \citep{r26}: id supports deduplication, source identifies the producing system, type identifies events such as object.changed, evidence.registered, baseline.advanced, and tool.capability.changed, subject references the object, time records observation time, dataschema identifies JSON Schema, and data contains the payload. JSON Schema 2020-12 validates payloads \citep{r24}. OpenTelemetry Context and Propagator carry traceparent, but \mbox{context\_}\allowbreak{}\mbox{ref}, \mbox{baseline\_}\allowbreak{}\mbox{token}, and \mbox{policy\_}\allowbreak{}\mbox{revision} must be persisted business fields, not optional Baggage. OpenTelemetry correlates calls; it is not an engineering-consistency protocol \citep{r25}.

\FloatBarrier
\begingroup
\small
\setlength{\parskip}{0pt}\setlength{\LTpre}{8pt}\setlength{\LTpost}{10pt}
\begin{longtable}{@{}P{115pt}P{110pt}P{105pt}P{115.76pt}@{}}
\caption{Minimal interfaces, inputs, outputs, and blocking conditions.}\label{tab:t5}\\
\toprule
\textbf{Interface} & \textbf{Key input} & \textbf{Output} & \textbf{Typical \mbox{rejection/}\allowbreak{}\mbox{suspension} condition} \\
\midrule\endfirsthead
\multicolumn{4}{l}{\small\itshape Table \thetable\ (continued)}\\
\toprule
\textbf{Interface} & \textbf{Key input} & \textbf{Output} & \textbf{Typical \mbox{rejection/}\allowbreak{}\mbox{suspension} condition} \\
\midrule\endhead
\midrule\multicolumn{4}{r}{\small Continued on next page}\\\endfoot
\bottomrule\endlastfoot
\mbox{\texttt{freeze\_context}} & Task, object, baseline, \mbox{rule/}\allowbreak{}\mbox{capability} versions & \mbox{Evaluation}\allowbreak{}\mbox{Context}\allowbreak{}\mbox{Snapshot} & Unreadable mandatory object version: HOLD \\
\mbox{\texttt{evaluate\_plan}} & Obligations, \mbox{activity/}\allowbreak{}\mbox{action} graph, context & Gap, base V/U, final V*/U*, PlanDecision & Missing mandatory obligation; unclosed dependency \\
\mbox{\texttt{admit\_action}} & \mbox{Candidate}\allowbreak{}\mbox{Effect}\allowbreak{}\mbox{Projection}, lease, snapshot & \mbox{Admission}\allowbreak{}\mbox{Decision}, ExecutionToken & Hard-gate failure or authority exceeded \\
\mbox{\texttt{start\_execution}} & Token, normalized parameters, actual input manifest & \mbox{job\_}\allowbreak{}\mbox{id}, \mbox{Execution}\allowbreak{}\mbox{Observation} & Token drift, idempotency conflict, parameter violation \\
\mbox{\texttt{register\_actual\_delta}} & Before/after snapshots, bindings, artifacts & \mbox{Actual}\allowbreak{}\mbox{Engineering}\allowbreak{}\mbox{Delta}, EffectDecision & Required effect absent; binding unknown \\
\mbox{\texttt{evaluate\_evidence}} & EvidenceRecord, obligations, lineage & Per-Gap acceptance, common-cause/closure results & Incomplete coverage, uncontrolled source, mandatory unknown \\
\mbox{\texttt{request\_activation}} & Candidate, admission decision, three reference classes & ACCEPT or \mbox{REBASE\_}\allowbreak{}\mbox{REQUIRED} & Any \mbox{current/}\allowbreak{}\mbox{reference} mismatch \\
\mbox{\texttt{issue\_usage\_permit}} & Asset, purpose, configuration, lineage, release requirements & \mbox{Usage/}\allowbreak{}\mbox{Disclosure} Permit & Forbidden use, excessive exposure, unknown provenance \\
\mbox{\texttt{invalidate}} & Changed facts, graph version & Affected closure, invalidation records & Unknown edge predicates retain UNRESOLVED \\
\mbox{\texttt{request\_recovery}} & Interruption snapshot, effect ledger & Recovery plan & Unrecognized old effect or ambiguous rebinding \\
\end{longtable}
\endgroup

\subsection{Consistency, idempotency, and concurrency}

Interface order is itself a safety invariant. First compare \mbox{expected\_}\allowbreak{}\mbox{revision} with the current target: drift requires \mbox{REBASE\_}\allowbreak{}\mbox{REQUIRED} or readmission even if the ExecutionToken remains live. Then validate the token's action, object, parameter range, and adapter: prior admission cannot override drift. Finally deduplicate by \mbox{idempotency\_}\allowbreak{}\mbox{key} + \mbox{target\_}\allowbreak{}\mbox{ref} + \mbox{context\_}\allowbreak{}\mbox{ref.} Idempotency prevents duplication of an otherwise permitted action; it does not legitimize a forbidden one. Current-version correspondence therefore takes precedence over token scope, which takes precedence over idempotent reuse.

Long-running CAE uses leases, heartbeats, terminal artifact preservation, and result readmission. Token expiry does not necessarily kill an already submitted job, which could destroy unrecoverable intermediate results. The job may finish, but outputs remain isolated until evaluated under current versions and rules for evidence registration or baseline admission. A network timeout must not cause blind resubmission; the caller first queries the existing job by \mbox{job\_}\allowbreak{}\mbox{id} and idempotency key.

We do not assume cross-system ACID support in CAD, CAE, or PLM. Candidate writing and formal activation are separate, and a baseline service is authoritative for membership updates after B/C/A checks. Preparation records, conditional writes, and compensation markers for legacy PLM are adaptation requirements, not proof of cross-system atomic commit. HTTP If-Match can carry conditional resource writes \citep{r27}, but external effects do not automatically roll back with a local transaction. Uncertain completion requires querying, reconciliation, or holding rather than blind replay.

\subsection{Policy evaluation and the V/U kernel}

Policies have four layers: static schemas, engineering predicates, dependency closure, and state transitions. Schema rules validate fields/types; domain resolvers evaluate geometry roles, units, physical domains, configurations, and scopes; closure propagates obligations, evidence, or invalidation; state rules aggregate base V/U, closure outcomes, and object conditions into V*/U* and domain states. Generic engines can execute structured rules, but domain plugins must supply engineering predicates: preserving an inlet face cannot become string comparison.

Each rule emits RuleEvaluation=<\mbox{rule\_}\allowbreak{}\mbox{id}, \mbox{target\_}\allowbreak{}\mbox{ref}, status, \mbox{observation\_}\allowbreak{}\mbox{refs}, \mbox{reason\_}\allowbreak{}\mbox{code}>, with SATISFIED, VIOLATED, or UNKNOWN. Dependency-aware aggregation places required violations in V and required unknowns in U. An optional unknown need not block unless it is the only available support path. Confirmed violation takes priority; one object's pass cannot compensate for another required object's violation. The kernel can serve effect conditions, repair-candidate gates, model admission, and capability regression, but each module uses its own RuleSet and positive closure definition.

\subsection{Persistence, indexing, and audit structures}

A hybrid store combines relational records, graph relations, object storage, and an append-only event log. Relations include \mbox{DEPENDS\_}\allowbreak{}\mbox{ON}, \mbox{BOUND\_}\allowbreak{}\mbox{TO}, \mbox{DERIVED\_}\allowbreak{}\mbox{FROM}, SUPPORTS, INVALIDATES, and \mbox{CONSTRAINS\_}\allowbreak{}\mbox{USE.} CAD/CAE artifacts, raw fields, and reports reside in object storage. PROV-O Entity, Activity, Agent, used, wasGeneratedBy, and wasDerivedFrom provide a provenance-exchange foundation \citep{r23}, extended with engineering versions, configurations, roles, contexts of use (CoU), quality, and failure factors.

Control objects have immutable revisions; updates create a new revision linked by supersedes. Relationships are also versioned because evidence support varies by configuration. Minimum indexes cover (\mbox{object\_}\allowbreak{}\mbox{id}, revision, configuration), (\mbox{claim\_}\allowbreak{}\mbox{id}, \mbox{domain\_}\allowbreak{}\mbox{cell}), (\mbox{evidence\_}\allowbreak{}\mbox{id}, \mbox{source\_}\allowbreak{}\mbox{factor}), (\mbox{change\_}\allowbreak{}\mbox{event}, \mbox{dependency\_}\allowbreak{}\mbox{closure}), and (\mbox{decision\_}\allowbreak{}\mbox{id}, \mbox{context\_}\allowbreak{}\mbox{ref}). Audits preserve what a decision relied on at the time, not merely the latest value of an object that subsequently changed.

\subsection{Observability, performance, and degradation}

Operational tracing records RPCs, calls, queueing, and rule time; engineering audits record objects, rules, V/U, evidence, and transitions. They share \mbox{trace\_}\allowbreak{}\mbox{id} but have different lifecycles. Metrics include rule matches, UNKNOWN sources, cache hits, closure size, and admission drift, not only API latency.

Optimization must preserve semantics. Object parsing and pure rules can cache by \mbox{object\_}\allowbreak{}\mbox{revision} + \mbox{rule\_}\allowbreak{}\mbox{revision}; dependency closure can be updated incrementally; large evidence enters hot paths through immutable references and digests. High-risk writes fail closed. Low-risk reads may enter logging-only mode during central-service outages, but their outputs cannot become formal evidence. Offline edge nodes may use expiring policy bundles and snapshots; actions needing current baselines or use constraints enter HOLD.

\subsection{Minimal deployable configuration}

A minimal configuration includes versioned objects, frozen contexts, \mbox{action/}\allowbreak{}\mbox{execution} gates, actual-observation channels, \mbox{evidence/}\allowbreak{}\mbox{judgment} services, conditional activation, and event-driven applicability reevaluation. Selected prototype paths integrate open-source tools, SQLite, and PostgreSQL. A full substrate extends these with enterprise MBSE/PLM adapters, common-cause/support services, use/disclosure controls, and skill evolution through Section 6.6 interfaces.

The minimal configuration preserves the three separations and provides typed extension points for other domains. Deployment must route missing required observations to U, preserve negative facts, and notify consumers of state drift. Section 7.13 specifies the first two requirements, Section 11.3 reports identified gaps, and Section 12 states implementation boundaries.

\subsection{Three cross-cutting control chains}

Figure 6 organizes the domains into three connected chains. The action chain starts with T1 authority, proceeds through T2 \mbox{obligations/}\allowbreak{}\mbox{planning} and T3 \mbox{candidate/}\allowbreak{}\mbox{composition} checks, enters T4 calls and T5 observation, and reaches T7 baseline/use control. T6 supplies evidence evaluation: an action-produced result is a handoff input that still needs evaluation against target obligations and current configuration.

The evidence chain starts with actual objects and observations from T5. T6 evaluates applicability, per-gap acceptance, actual provenance, common causes, coverage, and engineering judgment, producing claim-support domains for T7. Anomalous observations take another T6 path through multiple hypotheses and repair obligations, returning to T2/T3 for checks or modification candidates. The paths share fact sources but maintain separate conclusion-support and anomaly-handling responsibilities.

The evolution chain starts with changes to versions, configurations, rules, purposes, or capabilities. T8 reads historical effects and current snapshots, computes confirmed and possible impacts, and triggers recovery, model revalidation, or skill regression. Updated credibility and capability scopes constrain subsequent action and evidence use. The chains share frozen contexts, objects, dependencies, and provenance. Arrows denote controlled information and responsibility links, not a requirement to execute every module sequentially for every task.

\begin{figure}[H]\centering
\includegraphics[width=\linewidth,height=0.77\textheight,keepaspectratio]{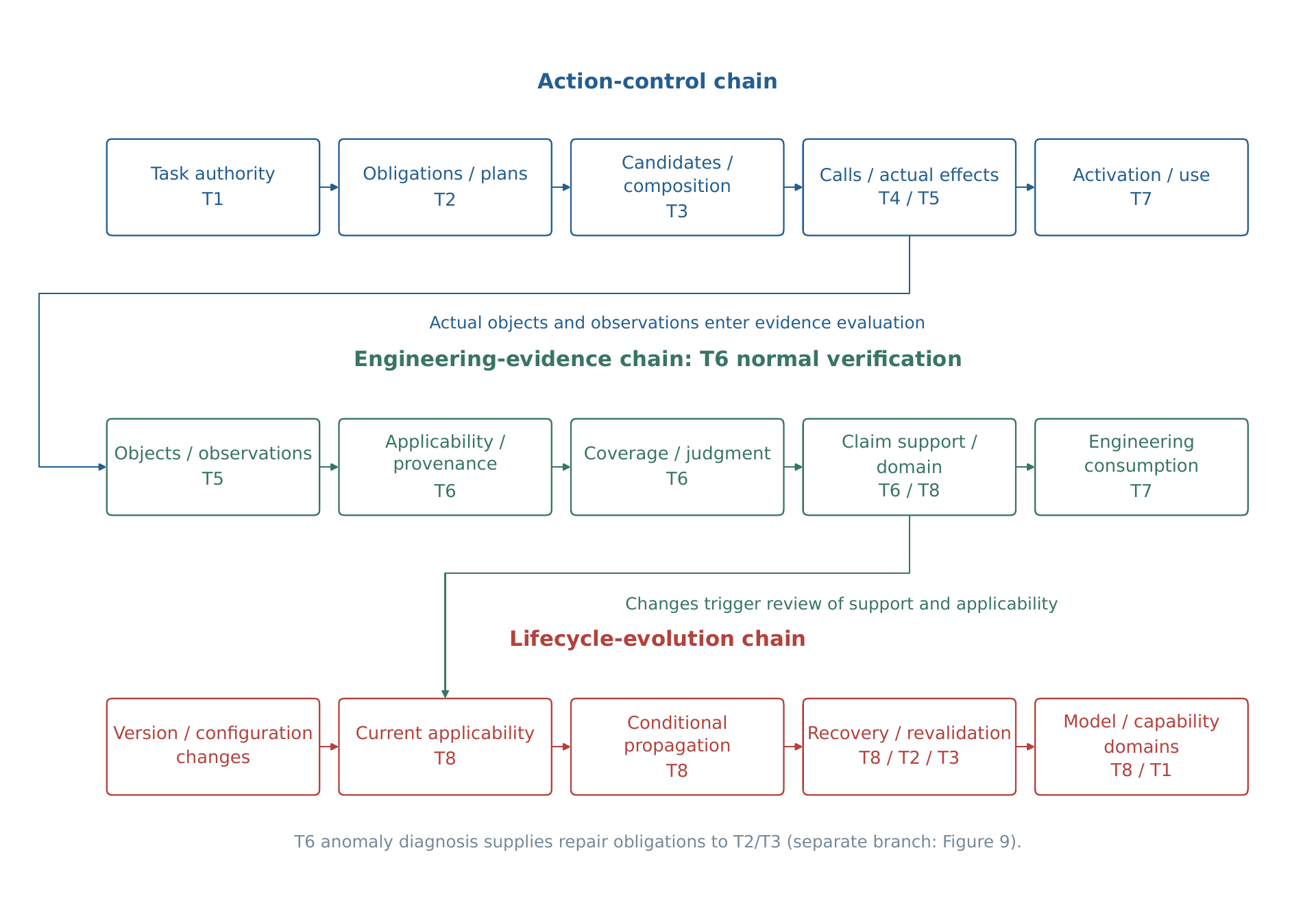}
\caption{Handoffs among eight domains through action, evidence, and evolution chains.}\label{fig:6}
\end{figure}

\subsection{Multi-agent handoff interfaces and collaboration}

The roles in Table~\ref{tab:t4} map to orchestration task instances and control-plane handoff interfaces. A handoff record contains \mbox{producer\_}\allowbreak{}\mbox{actor}, \mbox{consumer\_}\allowbreak{}\mbox{role}, \mbox{task\_}\allowbreak{}\mbox{ref}, \mbox{context\_}\allowbreak{}\mbox{ref}, \mbox{artifact\_}\allowbreak{}\mbox{refs}, \mbox{obligation\_}\allowbreak{}\mbox{refs}, \mbox{actual\_}\allowbreak{}\mbox{delta\_}\allowbreak{}\mbox{ref}, \mbox{evidence\_}\allowbreak{}\mbox{refs}, \mbox{decision\_}\allowbreak{}\mbox{refs}, and \mbox{unresolved\_}\allowbreak{}\mbox{items.} These reference immutable fact-service objects. Messages route references and responsibility; they cannot rewrite registered facts or replace recipient evaluation.

Recipients resolve actual \mbox{objects/}\allowbreak{}\mbox{versions}, check obligations and scope, obtain their own task authority, and recheck evidence at execution or consumption. Upstream decisions are reusable only when objects, conditions, scopes, and versions correspond. Drift requires a new context, reevaluation, or hold. A solver agent cannot inherit a design agent's write authority; a verification agent cannot turn solver exit status into engineering PASS; and unanimous agents cannot bypass activation guards.

T3 dependencies and write conditions, T4 object/job binding, and T7 conditional activation jointly control concurrent conflicts. T8 notifies affected tasks about changes during or after handoff. The orchestrator schedules collaboration, control services \mbox{register/}\allowbreak{}\mbox{evaluate}, and baseline services update controlled membership. This supports agents with different capabilities within the organizational boundary in Section 5.

\section{Data Model and Implementation Constraints of the Trust-Control Kernel}\label{sec:section_07}

\subsection{Why the Data Model Is a Runtime Mechanism Rather Than a Documentation Template}

Data objects in the substrate do not merely enrich logs: every control decision requires minimal facts that existing tool return codes cannot express. Without versioned role bindings, identically named inlets cannot be distinguished. Without a frozen evaluation context, the system cannot detect rule drift after checking. Without evidence applicability domains, it cannot distinguish evidence existence from coverage of the current Claim. Without recording historical effects' applicability relative to the current snapshot, it cannot recover safely. The data model therefore directly determines whether the control algorithms can be implemented.

Figure 7 presents four object layers. Identity and context establish which object is evaluated, when, and under which rules. Obligations and execution describe intended and actual actions. Evidence and decisions identify the grounds and conclusions. Lifecycle and use specify when conclusions remain dependable and within which scope they may be used.

\begin{figure}[!tbp]\centering
\includegraphics[width=\linewidth,height=0.77\textheight,keepaspectratio]{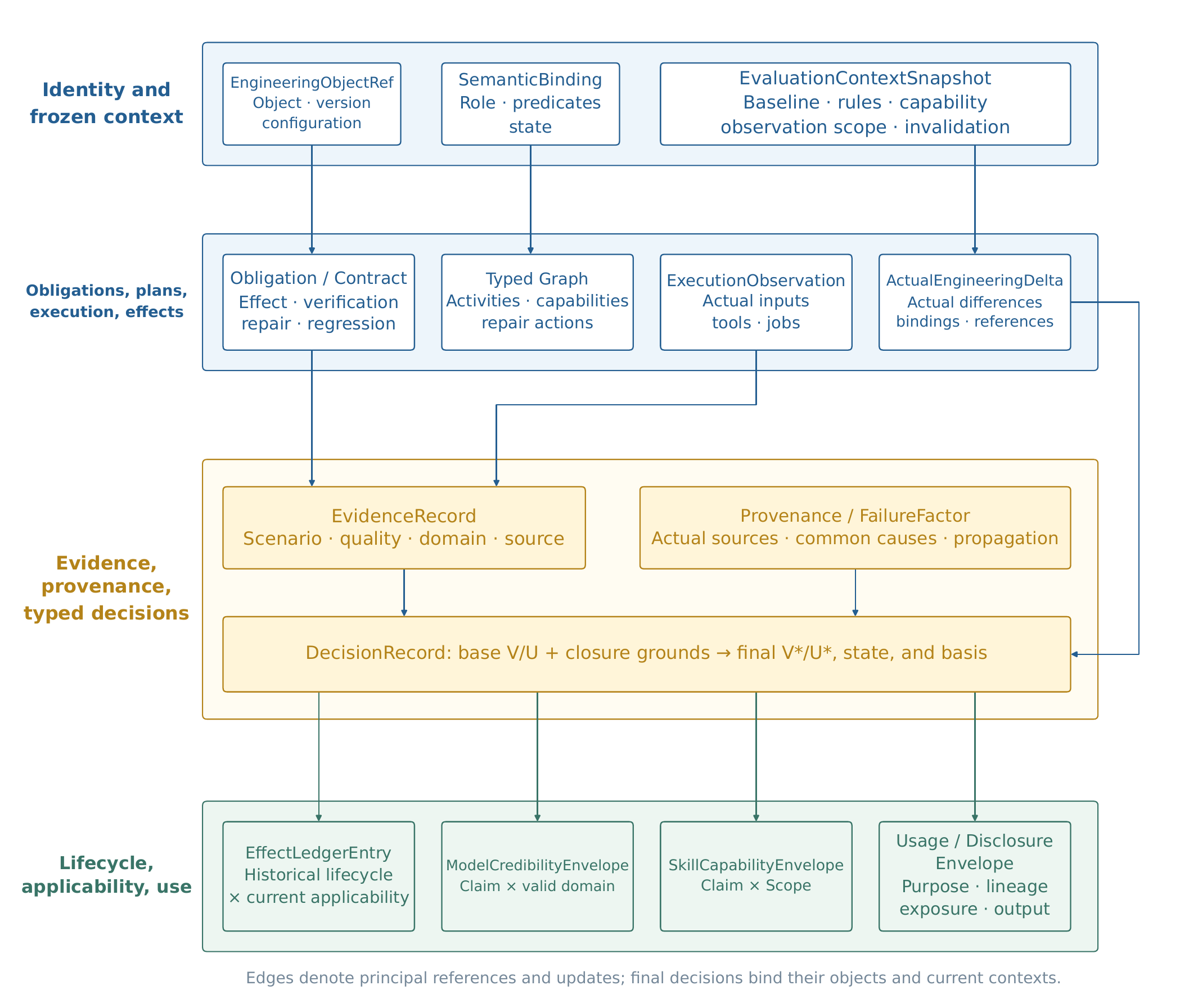}
\caption{Four object layers and their principal relationships in the trust-control kernel.}\label{fig:7}
\end{figure}

\subsection{Versioned Engineering Object References and Actual Semantic Binding}

EngineeringObjectRef separates stable identity, immutable revision, configuration, and artifact location (Table~\ref{tab:t7}). The locator may change without redefining the object; a content digest does not replace engineering identity. CAD roles whose topology changes are represented by a SemanticBinding containing the role, candidate entities, required predicates, selected entity, state, and grounds.

Binding evaluation has three steps. First, a resolver derives candidates from the current artifact, such as all exterior surfaces adjacent to a fluid domain. Second, it filters them using predicates for direction, connectivity, medium, boundary relations, units, and configuration. Third, it returns MATCH, MISMATCH, or UNRESOLVED according to uniqueness. No candidate, together with confirmed absence of the required object, yields MISMATCH; multiple candidates or unreadable required attributes yield UNRESOLVED. MATCH requires exactly one candidate satisfying every mandatory predicate. Names, historical IDs, and geometry hashes may generate candidates but cannot independently establish MATCH.

\subsection{Frozen Evaluation Context}

EvaluationContextSnapshot fixes the decision's object, task, baseline, configuration, rule bundle, capability and policy revisions, observation manifest, and invalidation conditions. It is the reference root for subsequent evaluations, not a mutable cache of the latest values. Freezing can bind immutable references rather than copy every artifact.

Contexts can become invalid through explicit revocation, such as withdrawal of a rule bundle by an administrator; reference drift, such as advancement of the official baseline; or capability drift, such as a tool or adapter version change. Invalidation does not erase historical decisions, but prevents their use for new submissions. A context is valid only within its declared control scope: a snapshot for one cold-plate CAD task cannot be reused for another product configuration.

\paragraph{Consequences of context invalidation.}
Invalidation is an event about a particular authority or premise, not automatic deletion of every dependent artifact. A pending candidate loses permission to submit under the old context. A committed object retains its historical record while T8 reevaluates current applicability along relevant, version-bound dependencies. A running job may continue in isolation when safe and useful, but its output cannot be consumed under the invalid context; a hazardous or unauthorized operation instead requires the execution controller's stop or takeover policy. Downstream agents receive the context identifier, changed premise, affected scope, and reason, and must recheck validity when consuming a result even if event delivery is delayed. Section~7.9 propagates these events through true and unknown dependency predicates.

Job lifetime and decision authority therefore have different clocks. Lease duration should reflect authorization policy, expected runtime, and the consequences of stale premises. A renewal requires fresh authority and reference checks; it must not silently extend permission merely because a solver is still running. Completion after expiry triggers current-context readmission. The experiments test selected expiry and race cases, not an adaptive lease optimizer or all four invalidation paths.

\subsection{Obligations, Contracts, and Applicability Scopes}

A common \mbox{Obligation}\allowbreak{}\mbox{Header} carries identity, object, configuration, scope, provenance, and version, while each obligation retains domain-specific fields. EffectContract contains \mbox{required\_}\allowbreak{}\mbox{effects}, \mbox{allowed\_}\allowbreak{}\mbox{effects}, \mbox{forbidden\_}\allowbreak{}\mbox{effects}, invariants, and \mbox{observation\_}\allowbreak{}\mbox{scope.} \mbox{Verification}\allowbreak{}\mbox{Obligation} identifies objects, criteria, scenarios, conventions, quality, and evidence sources. \mbox{Repair}\allowbreak{}\mbox{Obligation} specifies required and forbidden effects, invariants, evidence requirements, and release conditions. \mbox{Regression}\allowbreak{}\mbox{Obligation} concerns skill capability claims and Scope; \mbox{Revalidation}\allowbreak{}\mbox{Obligation} concerns model \mbox{Credibility}\allowbreak{}\mbox{Claims} and valid domains. The common header supports routing and audit; domain payloads prevent unlike obligations from being treated as the same task.

Scopes are expressed as intersectable predicates rather than free text. Continuous operating conditions may use intervals or multidimensional regions, discrete configurations may use sets, and engineering roles may use typed predicates. Scope comparison supports at least subset, overlap, disjoint, and unknown. Evidence covers only the intersection of its applicability domain with the Claim scope; overlap does not imply complete coverage. Unknown boundary units or coordinate systems yield unknown.

\subsection{Plans, Actions, Executions, and Actual Effects}

The planning layer includes at least \mbox{Activity}\allowbreak{}\mbox{Delta}\allowbreak{}\mbox{Graph}, \mbox{Verification}\allowbreak{}\mbox{Capability}\allowbreak{}\mbox{Fragment}, and \mbox{Repair}\allowbreak{}\mbox{Action}\allowbreak{}\mbox{Graph}. They share graph headers and directed edges but not node semantics. Activity graphs describe required engineering activities; verification capability graphs describe which evidence follows from specified input and transformation semantics; repair action graphs describe how multiple candidate modifications preserve cumulative constraints. Implementations must distinguish them through \mbox{graph\_}\allowbreak{}\mbox{type} and node schemas, rather than collapsing them into a semantically untyped workflow DAG.

\mbox{Candidate}\allowbreak{}\mbox{Effect}\allowbreak{}\mbox{Projection} records predicted effects before execution. \mbox{Execution}\allowbreak{}\mbox{Observation} records actual tools, input parameters, jobs, and return states. \mbox{Actual}\allowbreak{}\mbox{Engineering}\allowbreak{}\mbox{Delta} should record before/after object differences, role bindings, downstream references, and observation gaps. Their separation is an interface requirement, not a guarantee supplied by type names. Difference fields must additionally identify whether they come from artifact readback, generation metadata, or parameter-based derivation; the latter two must not be presented as independent geometric measurement. Tool success is only a process observation, and binding MATCH requires grounds in the relevant role predicates.

\paragraph{A producer--consumer fact contract.}
Each decisive fact has a canonical predicate identifier, typed value and unit, object revision and configuration, artifact digest, observation method, producer Schema revision, and grounds. The observation method distinguishes artifact readback, generation metadata, and parameter-based derivation. A consumer declares the predicate and binding it requires; field names alone do not establish correspondence. Table~\ref{tab:fact_contract} illustrates the normalized contract for an installation interface. This is a specification example; historical payloads are interpreted according to their actual Schema, not retroactively converted into evidence that this full contract was implemented.

\FloatBarrier
\begingroup\small
\setlength{\parskip}{0pt}
\begin{longtable}{@{}P{132pt}P{179pt}P{141pt}@{}}
\caption{Example fact-consumption contract at an effect or recovery boundary.}\label{tab:fact_contract}\\
\toprule
\textbf{Producer information} & \textbf{Canonical meaning / consumer obligation} & \textbf{Failure handling} \\
\midrule\endfirsthead
\multicolumn{3}{l}{\small\itshape Table \thetable\ (continued)}\\
\toprule
\textbf{Producer information} & \textbf{Canonical meaning / consumer obligation} & \textbf{Failure handling} \\
\midrule\endhead
\bottomrule\endlastfoot
\texttt{measured\_facts.}\allowbreak\texttt{mount\_interface.}\allowbreak\texttt{binding\_status} & Status of the installation-interface predicate for the referenced artifact; a declared Schema adapter may accept a documented flat representation. & Missing required field or unsupported nonempty enum contributes U; it is never treated as absence of a violation. \\
\texttt{mounting\_interface\_}\allowbreak\texttt{preserved=false} & Retain the negative predicate and its source grounds when the object and revision match the consumer. & An applicable confirmed violation contributes V; a separate unknown does not erase it. \\
Object revision, configuration, digest & Require correspondence to the consumed artifact; a favorable statement about another revision cannot replace it. & Mismatched or unestablished binding prevents a positive decision. A stale violation is not silently attributed to a new revision. \\
Coordinates, units, extraction method & Interpret relative translation separately from absolute edge distance; retain whether a value is measured or derived. & Ambiguous conversion becomes unknown. Metadata-derived distance is not independent STEP measurement. \\
\end{longtable}
\endgroup

Normalization precedes domain evaluation. A recognized negative fact with valid current binding survives normalization even when another required observation is missing, yielding both V and U and a negative decision. Two useful contract tests are therefore an unrecognized nonempty status and a same-version confirmed violation combined with unavailable auxiliary facts. These are specified acceptance criteria; the recovery paths discussed in Section~12.4 have not been shown to satisfy them universally.

\subsection{Evidence Records, Provenance Graphs, and Common-Cause Factors}

EvidenceRecord binds a target obligation or Claim to object revisions, configuration, scenario, quantities and units, raw artifacts, derivation, quality, provenance, and applicability domain (Table~\ref{tab:t7}). Raw observations, derived evidence, and conclusions remain distinct. PROV-O provides Entity, Activity, and Agent relations \citep{r23}; engineering consumers additionally need role, configuration, valid-domain, and error-propagation semantics.

Common-cause analysis does not equate shared provenance with failure. A FailureFactor = <\mbox{factor\_}\allowbreak{}\mbox{id}, \mbox{source\_}\allowbreak{}\mbox{ref}, \mbox{failure\_}\allowbreak{}\mbox{mode}, \mbox{propagation\_}\allowbreak{}\mbox{condition}, \mbox{affected\_}\allowbreak{}\mbox{quantity}, \mbox{control\_}\allowbreak{}\mbox{state}> is relevant only when an error can propagate to the target quantity or acceptance criterion under its propagation condition. \mbox{Evidence}\allowbreak{}\mbox{Records} sharing a RELEVANT and UNCONTROLLED FailureFactor cannot constitute independent support. If planned tool diversity actually resolves to the same cache or mesh, actual-provenance review takes precedence over planned provenance.

\subsection{Base Observations, Final Decisions, and Positive Closure Conditions}

DecisionRecord stores three related layers: base sets V/U from individual rule observations, the conditions and factual grounds of closure evaluation, and final sets \mbox{V\_}\allowbreak{}\mbox{star/}\allowbreak{}\mbox{U\_}\allowbreak{}\mbox{star}, denoted V*/U* in this paper. Its fields are \mbox{decision\_}\allowbreak{}\mbox{id}, \mbox{decision\_}\allowbreak{}\mbox{type}, \mbox{target\_}\allowbreak{}\mbox{ref}, \mbox{context\_}\allowbreak{}\mbox{ref}, \mbox{rule\_}\allowbreak{}\mbox{bundle\_}\allowbreak{}\mbox{revision}, V, U, \mbox{positive\_}\allowbreak{}\mbox{conditions}, \mbox{closure\_}\allowbreak{}\mbox{evaluation}, \mbox{closure\_}\allowbreak{}\mbox{status}, \mbox{V\_}\allowbreak{}\mbox{star}, \mbox{U\_}\allowbreak{}\mbox{star}, \mbox{result\_}\allowbreak{}\mbox{state}, \mbox{reason\_}\allowbreak{}\mbox{codes}, \mbox{observation\_}\allowbreak{}\mbox{refs}, \mbox{created\_}\allowbreak{}\mbox{at}, and \mbox{superseded\_}\allowbreak{}\mbox{by.} The \mbox{closure\_}\allowbreak{}\mbox{evaluation} contains mandatory-condition identifiers, per-condition outcomes, confirmed violations \mbox{V\_}\allowbreak{}\mbox{c}, mandatory unknowns \mbox{U\_}\allowbreak{}\mbox{c}, and associated fact references. The \mbox{closure\_}\allowbreak{}\mbox{status} stores aggregated Closure(x) \ensuremath{\in} \{true, false, unknown\}. Mandatory rules and closure conditions are determined before evaluation. A sufficiently supported failure of a mandatory closure condition enters \mbox{V\_}\allowbreak{}\mbox{c}; missing facts, undefined conditions, or an unclear checking scope enter \mbox{U\_}\allowbreak{}\mbox{c.} The final sets are \mbox{V\_}\allowbreak{}\mbox{star} = V \ensuremath{\cup} \mbox{V\_}\allowbreak{}\mbox{c} and \mbox{U\_}\allowbreak{}\mbox{star} = U \ensuremath{\cup} \mbox{U\_}\allowbreak{}\mbox{c}, with confirmed violations retaining priority over unknowns. These fields specify the complete substrate; historical implementations remain subject to the distinctions below.

The three-way outcome is not a new three-valued logic. Runtime verification already distinguishes satisfaction, violation, and inconclusive observation \citep{r33,r38}. Here the interface discipline records which unknowns are mandatory for this object, retains their factual grounds separately from confirmed violations, and evaluates the closure of the checking scope itself. Optional observations may improve explanation or selection but cannot discharge a mandatory unknown. Object-specific closure determines the final state; Table~\ref{tab:state_map} prevents unrelated domain states from being promoted into one another.

\begin{equation}
D(x)=\begin{cases}
\text{negative state}, & V^{*}(x)\neq\varnothing,\\
\text{unresolved state}, & V^{*}(x)=\varnothing,\ U^{*}(x)\neq\varnothing,\\
\text{positive state}, & V^{*}(x)=U^{*}(x)=\varnothing,\ \operatorname{Closure}(x)=\mathrm{true}.
\end{cases}
\end{equation}

This definition also handles empty base V and U when closure has not been established. Confirmed non-closure must produce a grounded \mbox{V\_}\allowbreak{}\mbox{c} rather than a positive decision; indeterminacy produces \mbox{U\_}\allowbreak{}\mbox{c.} An unestablished checking scope cannot obtain a positive outcome through vacuous truth over an empty set. Closure is object-specific: plans require obligation coverage and dependency closure; repair combinations require cumulative mandatory effects and intermediate invariants; evidence requires per-obligation coverage; model credibility support requires all mandatory supports and at least one valid path in each alternative-support group. Aggregate scores or successful optional checks cannot replace these mandatory conditions.

Here, \mbox{negative\_}\allowbreak{}\mbox{state} means that the relevant control object's positive conditions are not satisfied; its name follows the domain state table. A confirmed missing mandatory action may make a plan REJECTED. Evidence known to cover only part of the operating domain yields PARTIAL coverage. Confirmed failure of mandatory support can make a support relation BROKEN. Incomplete evidence coverage alone cannot generate an engineering FAIL for a product: engineering judgment must evaluate applicable evidence against actual criteria independently, and may remain UNKNOWN. U* is retained for subsequent evidence completion and diagnosis.

For example, when a repair sequence is known to contain no action realizing a mandatory effect, terminal closure checking must record that omission even if the original field scan produced neither V nor U. If an action's effect cannot be determined because required parameters are unreadable, the result is unknown. DecisionRecord retains condition identifiers and fact references for these derived findings, enabling state traceability beyond a final label. This is an interface requirement of the complete substrate; Table~\ref{tab:t2} identifies current prototype coverage.

Throughout the paper, V/U alone denote base sets obtained from individual rule observations; V*/U* include closure-evaluation results. Final positive decisions concerning admissibility, executability, acceptance, validity, or permission consistently use V*/U* and cannot follow merely from empty base V/U. Domain sets such as causal-hypothesis S(h)/R(h)/U(h) retain their own definitions and enter final decisions through their object-specific rules. Raw experimental V/U fields retain their historical implementation meaning. They are not retrospectively renamed V*/U*, nor taken to establish that historical prototypes performed every closure check.

\FloatBarrier
\begingroup\small
\setlength{\parskip}{0pt}
\begin{longtable}{@{}P{108pt}P{168pt}P{176pt}@{}}
\caption{State semantics by controlled object; outputs on different rows are not interchangeable.}\label{tab:state_map}\\
\toprule
\textbf{Object} & \textbf{States or records} & \textbf{Meaning / prohibited promotion} \\
\midrule\endfirsthead
\multicolumn{3}{l}{\small\itshape Table \thetable\ (continued)}\\
\toprule
\textbf{Object} & \textbf{States or records} & \textbf{Meaning / prohibited promotion} \\
\midrule\endhead
\bottomrule\endlastfoot
Role binding & MATCH / MISMATCH / UNRESOLVED & Predicate binding for this object, not general engineering acceptance. \\
Repair candidate / action sequence & ADMISSIBLE / REJECTED / UNRESOLVED; EXECUTABLE for a plan & Single-candidate admissibility does not establish cumulative plan executability. \\
Evidence coverage & COMPLETE / PARTIAL / UNRESOLVED & Coverage of the obligation scope; COMPLETE does not imply PASS. \\
Engineering judgment & PASS / FAIL / UNKNOWN & Criteria evaluated from applicable evidence; missing evidence does not alone establish product failure. \\
Causal hypothesis & Support S, refutation R, unresolved U & Hypothesis-specific sets, not final acceptance states. \\
Candidate / official membership & ACCEPT\_CANDIDATE / ACCEPT & Eligibility precedes authoritative activation and readback. \\
Historical effect / current applicability & Lifecycle record; VALID / INVALID / UNRESOLVED & Current evaluation does not overwrite the historical commit. \\
Asset use / control action & ALLOW / DENY / UNRESOLVED; HOLD, REBASE\_REQUIRED, REVALIDATION\_REQUIRED & Use permission and instructions to wait or reevaluate do not establish artifact validity. \\
\end{longtable}
\endgroup

\FloatBarrier
\begingroup
\small
\setlength{\parskip}{0pt}\setlength{\LTpre}{8pt}\setlength{\LTpost}{10pt}
\begin{longtable}{@{}P{116pt}P{103pt}P{111pt}P{115.76pt}@{}}
\caption{Object boundaries of easily confused terms.}\label{tab:t6}\\
\toprule
\textbf{Term} & \textbf{Controlled object} & \textbf{Definition} & \textbf{Does not mean} \\
\midrule\endfirsthead
\multicolumn{4}{l}{\small\itshape Table \thetable\ (continued)}\\
\toprule
\textbf{Term} & \textbf{Controlled object} & \textbf{Definition} & \textbf{Does not mean} \\
\midrule\endhead
\midrule\multicolumn{4}{r}{\small Continued on next page}\\\endfoot
\bottomrule\endlastfoot
\mbox{ACCEPT\_CANDIDATE} & Isolated candidate model & Admission checks passed; eligible for atomic-activation review & Official baseline already updated \\
ACCEPT & Official baseline membership & B/C/A review passed and baseline authority completed activation & Product performance necessarily PASS \\
Lease & Temporary task-specific tool authority & Least authorization for principal, task, object, action, and duration & Long-term capability claim or asset-use permission \\
Envelope & Model, skill, or asset scope & \mbox{Applicability/}\allowbreak{}\mbox{use} boundaries maintained with versions, evidence, and dependencies & A single-call token \\
Gap & Computable shortfall against target requirements & Missing facts, evidence, or activities & An already assigned engineering obligation \\
Obligation & Mandatory evaluable engineering responsibility & Object, criteria, scenarios, invariants, or evidence conditions & Generic task advice or workflow node \\
\end{longtable}
\endgroup

\subsection{Lifecycle, Credibility, Capability, Use, and Disclosure Objects}

\mbox{Effect}\allowbreak{}\mbox{Ledger}\allowbreak{}\mbox{Entry} maintains two independent axes: historical lifecycle and current applicability. The historical axis records whether an effect was generated, committed, or superseded; the current axis evaluates it as VALID, INVALID, or UNRESOLVED under the current snapshot. \mbox{Model}\allowbreak{}\mbox{Credibility}\allowbreak{}\mbox{Envelope} maintains model support through \mbox{Credibility}\allowbreak{}\mbox{Claims} and valid domains; \mbox{Skill}\allowbreak{}\mbox{Capability}\allowbreak{}\mbox{Envelope} maintains demonstrated skill capabilities through \mbox{Capability}\allowbreak{}\mbox{Claims} and Scope. Neither is a scalar score. Changes respectively produce \mbox{Credibility}\allowbreak{}\mbox{Delta} and \mbox{Capability}\allowbreak{}\mbox{Delta}.

\mbox{Usage}\allowbreak{}\mbox{Constraint}\allowbreak{}\mbox{Envelope} binds asset lineage, permitted and forbidden purposes, configuration, derivation rules, and version. \mbox{Disclosure}\allowbreak{}\mbox{Envelope} additionally specifies sensitive fields, permitted computation locations, raw-asset visibility, output granularity, de-\mbox{identification/}\allowbreak{}\mbox{aggregation} rules, and proof requirements. With multiple constrained parents, forbidden purposes are unioned and the permitted scope cannot be broader than that of any parent. Output release is limited to fields or statistics explicitly permitted by \mbox{Disclosure}\allowbreak{}\mbox{Envelope}. Valid identity, successful download, tool success, or local publication by a recipient cannot independently establish ALLOW.

\subsection{Engineering Dependency Graph and Edge Predicates}

Nodes of the common dependency graph G = (N, E) include engineering objects, obligations, activities, executions, evidence, claims, capabilities, and decisions. Edges distinguish at least \mbox{DEPENDS\_}\allowbreak{}\mbox{ON}, \mbox{GENERATED\_}\allowbreak{}\mbox{BY}, \mbox{BOUND\_}\allowbreak{}\mbox{TO}, SUPPORTS, \mbox{DERIVED\_}\allowbreak{}\mbox{FROM}, AFFECTS, INVALIDATES, and \mbox{CONSTRAINS\_}\allowbreak{}\mbox{USE.} Each edge carries \mbox{edge\_}\allowbreak{}\mbox{revision}, \mbox{configuration\_}\allowbreak{}\mbox{scope}, \mbox{propagation\_}\allowbreak{}\mbox{predicate}, \mbox{required\_}\allowbreak{}\mbox{or\_}\allowbreak{}\mbox{optional}, and \mbox{observation\_}\allowbreak{}\mbox{basis.}

For a frozen graph G and change seeds S, compute the confirmed affected set D = \mbox{Reach\_}\allowbreak{}\mbox{true}(S) and possibly affected set P = \mbox{Reach\_}\allowbreak{}\mbox{true\_}\allowbreak{}\mbox{or\_}\allowbreak{}\mbox{unknown}(S), with unresolved set \mbox{U\_}\allowbreak{}\mbox{dep} = P \ensuremath{-} D. Both reachability computations continue to a fixed point. If A reaches B through an unknown edge and B reaches C through a true edge, both B and C enter P; without an all-true path, both remain unresolved. Known irrelevant edges stop propagation; a node with another all-true path enters D. The current control object determines mandatory dependencies and edge applicability scopes. The algorithm does not prove graph completeness, predicate correctness, or eventual resolution of unknowns. Claims of unaffected objects outside these sets are relative to the given graph and frozen semantics.

\subsection{Schemas, Primary Keys, and Validation}

External writes first undergo JSON Schema checks for types and mandatory fields \citep{r24}, then domain validation for cross-field invariants. A recommended primary key is <namespace, \mbox{object\_}\allowbreak{}\mbox{id}, revision, \mbox{configuration\_}\allowbreak{}\mbox{id}>, not a file digest; digests serve integrity and content deduplication. DecisionRecord, EvidenceRecord, and \mbox{Context}\allowbreak{}\mbox{Snapshot} use UUIDs or enterprise-wide identifiers with business references as foreign keys. Quantities use value + unit + \mbox{quantity\_}\allowbreak{}\mbox{kind}; bare numbers must not cross interfaces. Both event time and system receipt time are recorded.

\FloatBarrier
\begingroup
\small
\setlength{\parskip}{0pt}\setlength{\LTpre}{8pt}\setlength{\LTpost}{10pt}
\begin{longtable}{@{}P{112.94pt}P{167.15pt}P{171.67pt}@{}}
\caption{Minimal fields and indispensable invariants of key objects.}\label{tab:t7}\\
\toprule
\textbf{Object} & \textbf{Minimal fields} & \textbf{Invariant} \\
\midrule\endfirsthead
\multicolumn{3}{l}{\small\itshape Table \thetable\ (continued)}\\
\toprule
\textbf{Object} & \textbf{Minimal fields} & \textbf{Invariant} \\
\midrule\endhead
\midrule\multicolumn{3}{r}{\small Continued on next page}\\\endfoot
\bottomrule\endlastfoot
\mbox{Engineering}\allowbreak{}\mbox{Object}\allowbreak{}\mbox{Ref} & id, revision, configuration, type, locator & Locator changes do not redefine stable identity; revision references immutable content \\
\mbox{Evaluation}\allowbreak{}\mbox{Context}\allowbreak{}\mbox{Snapshot} & Baseline, rules, capability, observation manifest & No in-place mutation after freezing; drift creates a new context \\
ExecutionToken & Context, action, target, adapter, expiry, nonce & Only declared \mbox{actions/}\allowbreak{}\mbox{objects}; no cross-context reuse \\
\mbox{Actual}\allowbreak{}\mbox{Engineering}\allowbreak{}\mbox{Delta} & Before/after references, changed facts, binding states & Actual artifact readback required; agent self-report is not a substitute \\
EvidenceRecord & Target, scenario, configuration, source, quality, domain & Existence is not coverage; derived evidence retains original provenance \\
DecisionRecord & Target, context, base V/U, \mbox{closure\_}\allowbreak{}\mbox{evaluation} with \mbox{V\_}\allowbreak{}\mbox{c/}\allowbreak{}\mbox{U\_}\allowbreak{}\mbox{c} and grounds, final \mbox{V\_}\allowbreak{}\mbox{star/}\allowbreak{}\mbox{U\_}\allowbreak{}\mbox{star}, \mbox{closure\_}\allowbreak{}\mbox{status}, \mbox{result\_}\allowbreak{}\mbox{state} & Store base observations, closure, and final decision separately; final sets are reconstructible, not labels alone \\
\mbox{Effect}\allowbreak{}\mbox{Ledger}\allowbreak{}\mbox{Entry} & Source action, artifact, dependencies, lifecycle, applicability & Historical COMMITTED does not imply current VALID \\
\mbox{Capability/}\allowbreak{}\mbox{Credibility} Envelope & Claims, scope/domain, \mbox{support/}\allowbreak{}\mbox{dependency} references, revision & New versions do not overwrite old ones; expansion needs new evidence \\
\mbox{Usage/}\allowbreak{}\mbox{Disclosure} Envelope & Lineage, purpose, configuration, \mbox{exposure/}\allowbreak{}\mbox{output} rules & Derivation cannot relax parent constraints; permission does not imply raw-data visibility \\
\end{longtable}
\endgroup

\subsection{Machine-Checkable Data Invariants}

The complete substrate specifies twelve invariants: I1 decisions reference frozen contexts; I2 evidence binds actual object versions and configurations; I3 tool success does not directly produce effect PASS; I4 mandatory unknowns do not become satisfied through defaults; I5 confirmed violations are not overridden by scores or unknowns; I6 candidates do not become official baseline members before ACCEPT; I7 current B/C/A correspond to their references at commit; I8 derived use and disclosure scopes do not relax parent constraints; I9 historical COMMITTED does not imply current VALID; I10 applicability expansion requires new evidence; I11 recovery does not migrate invalidated premises; and I12 negative and unknown states retain their rule and observation grounds. These define module responsibilities, not a claim that the current prototype enforces every invariant for every object. The bracket adapter counterexample directly exposes implementation gaps in I3/I4 at the fact-input layer.

\subsection{Physical Storage and Transaction Boundaries}

Transaction boundaries have three layers. At the database layer, \mbox{register\_}\allowbreak{}\mbox{evidence} can atomically register records, provenance edges, and events; \mbox{request\_}\allowbreak{}\mbox{activation} can atomically update a baseline pointer and acceptance record. Neither covers the external CAE filesystem. Controlled endpoints can handle repeated requests using durable idempotency records and a defined side-effect model; Section 11.6 states the measured scope. External CAD/PLM operations may complete a side effect while losing its acknowledgment, requiring external job queries, stable operation identifiers, or compensating reconciliation. Without those capabilities, UNKNOWN/HOLD is retained. This paper does not establish end-to-end exactly-once execution for arbitrary external tools.

\subsection{Completeness of Mandatory Checks and Observation Mapping}

The mandatory inventory is constructed before evaluation, through the following auditable workflow:
\begin{enumerate}[leftmargin=2em]
\item An authorized requirement owner identifies versioned requirements, invariant sources, target objects, and intended uses. Agent proposals may add candidates but cannot remove these sources.
\item The obligation compiler maps each requirement to one or more predicate obligations, including dependencies and checkpoint or final-state scope. Each requirement retains its mapping grounds.
\item The checker registry binds each obligation to a checker version and the facts it requires. Unsupported obligations remain explicit inventory entries, not absent checks.
\item An observation audit identifies the permitted sources, extraction methods, units, and revision bindings of every required fact. Missing extraction capability produces a closure unknown.
\item The review authority freezes the requirement-to-obligation-to-checker manifest with the evaluation context. Runtime records reconcile required items against actually executed checks and consumed facts.
\end{enumerate}
An inventory entry minimally stores requirement ID and source revision, object and use scope, predicate ID, required fact references and source classes, checker revision, checkpoint, and unknown-handling rule. Closure succeeds only when every mandatory item in that frozen scope has applicable grounds and its object-specific conditions hold. This supports auditable completeness relative to an explicit requirement set; it does not prove that the set exhausts all real engineering obligations. Requirement discovery and the correctness of its decomposition remain assurance tasks.

For the cold-plate example, channel width and installation-hole diameter must occupy separate inventory entries if both occur in the frozen requirements. Installing a width predicate closes that entry, not the diameter entry. If no diameter checker or observation path is installed, the latter contributes a closure unknown and prevents an unqualified positive decision. V4 demonstrates the consequences of omitting this reconciliation; its repaired implementation does not yet realize the general inventory protocol. The fact-consumption contract in Section~7.5 governs how each executed check receives its grounds.

A positive decision requires an established mandatory-check scope, currently applicable grounds for every mandatory item, empty V* and U*, and satisfied object-specific closure conditions. Together with Section 7.7, these conditions constrain interface implementations. The determination of the checking set, fact mapping, and closure judgment must retain testable or independently inspectable grounds. Table~\ref{tab:t2} identifies aspects not yet implemented. These are normative conditions for the complete substrate; the positive states recorded by a particular experimental prototype establish only its installed checks and observed paths. Section 11.7 distinguishes that implementation scope from general mandatory-check closure.

The inventory protocol and producer--consumer contract address different failure surfaces: a missing obligation cannot be repaired by better reading of the checks that remain, while a complete inventory cannot compensate for a consumer discarding an observed negative fact. Both must preserve their own audit trail.

\section{Eight Trust Mechanisms and Their Implementation Conditions}\label{sec:section_08}

\setcounter{subsection}{-1}

\subsection{Common Evaluation Framework}

The eight mechanisms share a skeleton: resolve context, establish the checking scope, obtain observations, construct V*/U*, evaluate positive conditions, and emit domain states. They differ in controlled objects and closure conditions. Figure 8 presents this evaluation cycle. The following subsections specify their inputs, rules, and outputs; Table~\ref{tab:t1} identifies software hosts, Table~\ref{tab:t3} their technical arrangements, and Table~\ref{tab:t2} current implementation scope.

\begin{figure}[H]\centering
\includegraphics[width=\linewidth,height=0.77\textheight,keepaspectratio]{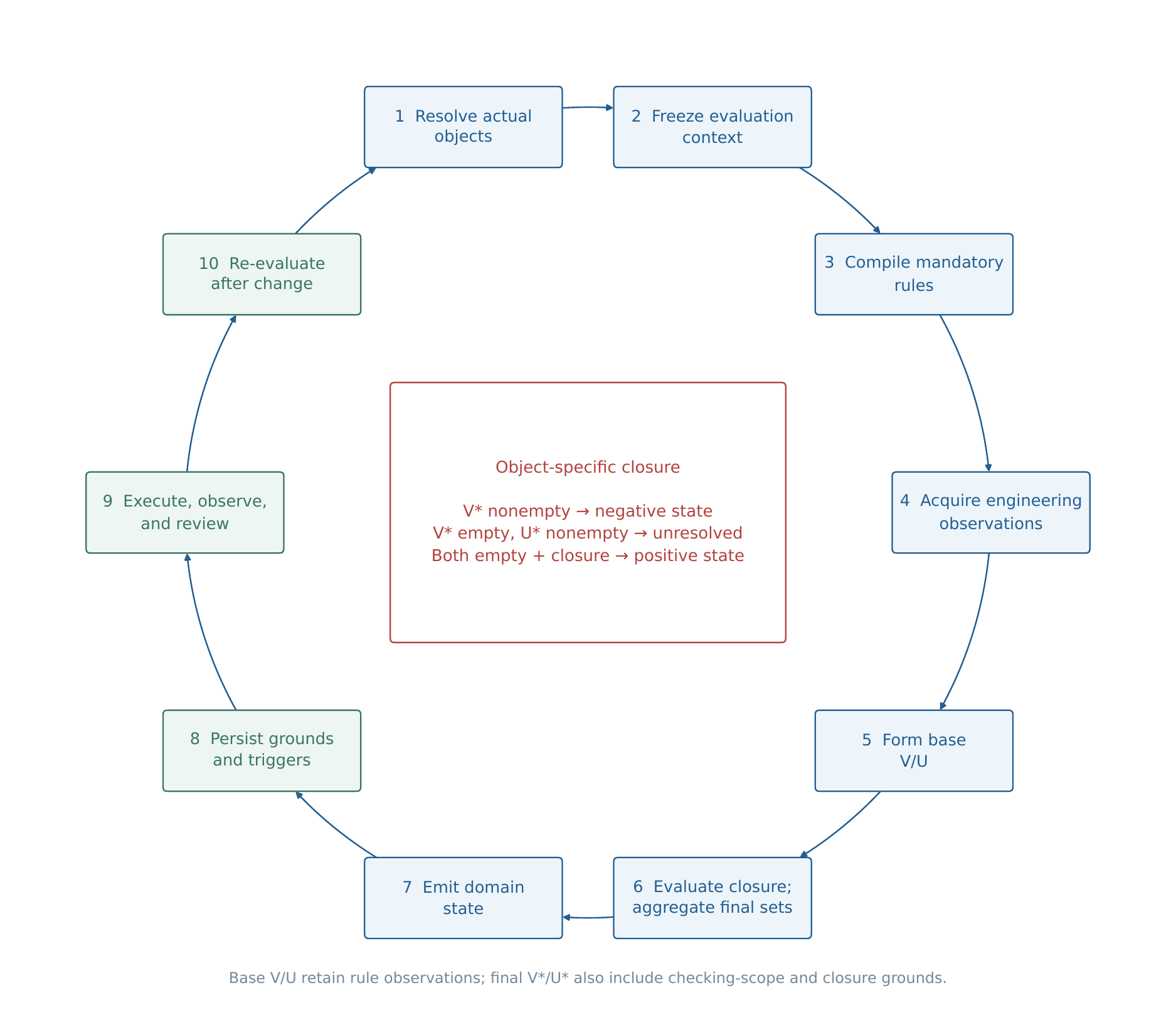}
\caption{Dependency-aware evaluation cycle shared by the eight control mechanisms.}\label{fig:8}
\end{figure}

Evaluation freezes the context, determines mandatory rules from obligations and object semantics, obtains provenance-bearing observations, and evaluates each rule as satisfied, violated, or unknown. It then incorporates closure grounds into V*/U* and persists the domain decision. Scanning only agent-registered fields may omit invariants, while a fixed MATCH may conceal existing negative facts. Section 7.13 therefore makes checking scope and fact transformation part of the control interface; Section 11.3 provides implementation counterexamples.

\subsection{T1: Identity, Authority, and Capability Trust}

\subsubsection{Inputs and Objects}

Inputs include agent identity and instance, current \mbox{task/}\allowbreak{}\mbox{obligation}, \mbox{Tool}\allowbreak{}\mbox{Capability}\allowbreak{}\mbox{Manifest}, \mbox{Skill}\allowbreak{}\mbox{Capability}\allowbreak{}\mbox{Envelope}, actual tool and adapter versions, target engineering objects, risk level, and human approval. T1 produces two distinct objects: \mbox{Agent}\allowbreak{}\mbox{Capability}\allowbreak{}\mbox{Lease} controls the minimal tool authority for this task, while \mbox{Capability}\allowbreak{}\mbox{Match}\allowbreak{}\mbox{Result} determines whether demonstrated capabilities cover the current engineering request. Authentication identifies the caller; capability claims describe what it can do; an authority lease specifies what it may do in this task.

\subsubsection{Capability Matching and Least-Authority Algorithm}

Normalize a request as \mbox{Capability}\allowbreak{}\mbox{Requirement} = <\mbox{task\_}\allowbreak{}\mbox{type}, \mbox{object\_}\allowbreak{}\mbox{scope}, operation, preconditions, \mbox{expected\_}\allowbreak{}\mbox{effect}, invariants, \mbox{tool\_}\allowbreak{}\mbox{scope}, \mbox{evidence\_}\allowbreak{}\mbox{requirement}>. Evaluate scope inclusion for every field. Confirmed version mismatch, unsupported physical domains, operations exceeding \mbox{Tool}\allowbreak{}\mbox{Capability}\allowbreak{}\mbox{Ceiling}, or objects outside Scope enter V. Unreadable mandatory versions, units, or adapter observation capabilities enter U. Only when the capability Envelope covers all mandatory fields and current risk rules permit the request may the tool manifest be restricted to the minimum operation set for a Lease. Before issuance, capability coverage and risk-rule closure are incorporated into V*/U* under Section 7.7. A task lease requires empty final sets and established closure.

A lease binds \mbox{agent\_}\allowbreak{}\mbox{instance} + \mbox{task\_}\allowbreak{}\mbox{ref} + \mbox{obligation\_}\allowbreak{}\mbox{ref} + \mbox{object\_}\allowbreak{}\mbox{scope} + \mbox{operation\_}\allowbreak{}\mbox{set} + \mbox{context\_}\allowbreak{}\mbox{ref} + expiry + \mbox{takeover\_}\allowbreak{}\mbox{condition.} Task changes, object switches, risk escalation, approval revocation, or capability Envelope updates invalidate it. Human takeover may be triggered by irreversible writes, abrupt V/U changes, repeated retries, or high-risk parameter violations. Following takeover, the agent must not retain its previous write lease.

\subsubsection{Capability Changes and Regression}

After changes to tools, foundation models, prompts, knowledge sources, Tool Schemas, adapters, or engineering rules, the capability regression component projects SkillChangeSet onto \mbox{Capability}\allowbreak{}\mbox{Impact}\allowbreak{}\mbox{Set}. Unaffected Claims whose old evidence remains applicable are candidates for retention. Affected Claims, unknown impacts, and proposed capability expansions generate \mbox{Regression}\allowbreak{}\mbox{Obligations}. Re-evaluated regression evidence yields \mbox{Capability}\allowbreak{}\mbox{Delta}. Only PRESERVED, REVALIDATED, and appropriately scoped \mbox{NEWLY\_}\allowbreak{}\mbox{VERIFIED} capabilities enter \mbox{Updated}\allowbreak{}\mbox{Skill}\allowbreak{}\mbox{Capability}\allowbreak{}\mbox{Envelope}. NARROWED reduces Scope; WITHDRAWN and UNRESOLVED do not enter the official capability set.

\subsection{T2: Intent, Obligations, and Planning Trust}

\subsubsection{From Natural Language to Typed Obligations}

A planner must not directly translate “reduce temperature” into a tool sequence. An obligation compiler first identifies the target object, engineering quantities, criteria, scenarios, conventions, quality, provenance, and invariants, producing \mbox{Effect}\allowbreak{}\mbox{Obligation}, \mbox{Verification}\allowbreak{}\mbox{Obligation}, or \mbox{Repair}\allowbreak{}\mbox{Obligation}. For MBSE changes, obligation differencing computes ObligationDiff over objects, criteria, operating conditions, metrics, quality, provenance, and configuration, rather than marking every traceability successor as affected.

\subsubsection{Coverage and Dependency Closure}

Let O be the obligation set, A the candidate activity set, and Cover(a) the expected coverage of activity a. Minimal plan coverage requires \ensuremath{\forall}o \ensuremath{\in} \mbox{O\_}\allowbreak{}\mbox{required}, \ensuremath{\exists}a \ensuremath{\in} A: o \ensuremath{\in} Cover(a). This is necessary but insufficient: activity prerequisites, configuration tokens, write conflicts, and feedback closure must also be satisfied. Each Gap is mapped to activity requirements through \mbox{Activity}\allowbreak{}\mbox{Synthesis}\allowbreak{}\mbox{Rule} = <GapType, \mbox{Applicability}\allowbreak{}\mbox{Condition}, RequiredInput, ActionTemplate, \mbox{Expected}\allowbreak{}\mbox{Coverage}>, preventing \mbox{reconstruct\_}\allowbreak{}\mbox{from\_}\allowbreak{}\mbox{gap} from becoming an unexplained black box.

\subsubsection{Failure Conditions and Outputs}

Confirmed omissions of mandatory criteria, operating conditions, or invariants enter V; unresolved obligation sources, bindings, or coverage relations enter U. Outputs are a checked \mbox{Activity}\allowbreak{}\mbox{Delta}\allowbreak{}\mbox{Graph}, uncovered Gaps, and PlanDecision, not selection of concrete verification capabilities. In the cold-plate case, a nominal-flow Tmax activity without rated-flow or \ensuremath{\Delta}p activities is a confirmed Gap; unreadable inlet-role mapping is U. The obligation compiler connects to \mbox{MBSE/}\allowbreak{}\mbox{requirements} repositories, while the activity checker sits after the planner and before the action gate. A checked \mbox{Activity}\allowbreak{}\mbox{Delta}\allowbreak{}\mbox{Graph} additionally requires empty V*/U* and closure of mandatory coverage, dependencies, and feedback conditions. Base V/U retain the individual checking grounds.

\subsection{T3: Trust in Candidate Actions, Verification Capabilities, and Multistep Composition}

\subsubsection{Hard Gates for Individual Repair Candidates}

The individual-candidate gate transforms \mbox{Repair}\allowbreak{}\mbox{Candidate} into \mbox{Candidate}\allowbreak{}\mbox{Effect}\allowbreak{}\mbox{Projection}, then evaluates five hard constraints: required effects, forbidden effects, invariants, configuration consistency, and evidence availability, yielding base V/U. Evidence availability means that its acquisition path is identifiable, bindable, and executable, not that verification is complete. After mandatory scope and effect closure are included in the final sets, nonempty V* yields REJECTED; empty V* with nonempty U* yields UNRESOLVED. ADMISSIBLE requires empty V*/U* and satisfied positive conditions. Verification levels are determined only for ADMISSIBLE candidates.

\subsubsection{Cumulative Closure of Multiple Repair Operations}

For a candidate sequence \ensuremath{\pi} = (a1, …, ak), recursively apply action projections \ensuremath{\delta}i to initial state S0: Si = Apply(Si\ensuremath{-}1, \ensuremath{\delta}i). Evaluate continuous invariants, checkpoint invariants, and forbidden effects at every prescribed checkpoint, and all required effects at the terminal state. Confirmed inability to realize a mandatory effect, or destruction of its earlier realization by a later action, enters V. Missing mandatory engineering facts that prevent judgment enter U. Checking-scope and cumulative-closure grounds are aggregated under Section 7.7. EXECUTABLE requires empty V*/U*, satisfied constraints at all prescribed checkpoints, and terminal closure of required effects. Plans are selected only from EXECUTABLE candidates; cost optimization cannot promote an ineligible sequence.

\subsubsection{Safe Composition of Verification Capabilities}

The verification orchestrator maps EvidenceGap to \mbox{Verification}\allowbreak{}\mbox{Capability}\allowbreak{}\mbox{Requirement} and selects VCFs. A VCF is not a task name but <input ports, input constraints, transformation semantics, output evidence, quality, provenance, dependencies>. Composition closure recursively propagates outputs to downstream inputs while checking configuration, quality, and provenance; it is not a union of capability names. After execution, outputs are accepted separately for each Gap. Acceptance of a temperature field for Tmax does not imply acceptance for hotspot location or cycle life. Provenance and common-cause modules supply independence constraints; the orchestrator cannot independently declare evidence independence. Each Gap uses its own V*/U* and coverage closure; a positive result for one Gap does not replace evaluation of another.

\subsection{T4: Tool-Invocation and Execution-Process Trust}

\subsubsection{Controlled Invocation Chain}

The invocation chain is \mbox{Action}\allowbreak{}\mbox{Admission} \ensuremath{\rightarrow} ExecutionToken \ensuremath{\rightarrow} Adapter \ensuremath{\rightarrow} Tool Job \ensuremath{\rightarrow} \mbox{Execution}\allowbreak{}\mbox{Observation}. ExecutionToken references an admitted action, object, context, and permitted parameter range, and includes a nonce, validity period, and idempotency key. Before calling a tool, the adapter re-resolves actual input objects to prevent post-admission path substitution, normalizes units and coordinates, creates an immutable input manifest, starts the job, and records \mbox{job\_}\allowbreak{}\mbox{id.} \mbox{Action}\allowbreak{}\mbox{Admission} is an object decision based on final V*/U*. ExecutionToken carries that decision and scope; it does not replace execution-point review of current inputs and context.

\subsubsection{Asynchrony, Retries, and Human Takeover}

Asynchronous job states are SUBMITTED, RUNNING, SUCCEEDED, FAILED, CANCELLED, and UNKNOWN. A network timeout makes the invocation observation UNKNOWN rather than automatically resubmitting; the caller first queries by idempotency key. Irreversible writes may require a human challenge. Risk escalation, out-of-range parameters, repeated failures, or output paths escaping the candidate area trigger TAKEOVER. Human approvals also bind context and action and cannot serve as indefinite authorization.

\subsubsection{Legacy Software and Isolation}

RPA adapters should run on controlled desktops or virtual machines with restricted directories, plugins, and network access; each task receives an isolated candidate workspace. Exit codes, screen messages, and file existence are only process observations. T4 can block unit errors, obsolete CAD inputs, and duplicate jobs. CAD no-ops and incorrect role bindings require T5 observation of actual effects, preventing unsupported claims about tool-gateway coverage.

\subsection{T5: Trust in Actual Engineering Effects and Object Binding}

\subsubsection{Deriving Actual Differences}

The effect observer compares pre- and post-operation objects through structural correspondence and role evaluation, producing added, removed, modified, preserved, \mbox{binding\_}\allowbreak{}\mbox{changes}, \mbox{downstream\_}\allowbreak{}\mbox{reference\_}\allowbreak{}\mbox{changes}, and \mbox{observation\_}\allowbreak{}\mbox{unknowns.} CAD observations concern parameters, topological roles, and mounting relations; CAE observations concern mesh parent versions, operating conditions, and boundaries. Each observation records its extraction source, units, coordinates, and applicable version. Existing negative facts enter corresponding violations; missing decisive observations enter unknowns; binding MATCH references role-predicate grounds. Section 11.3 reports the historical bracket adapter's fixed-MATCH defect and the BFM-1.1 local regression. That regression does not establish general geometric observation or end-to-end assurance.

\subsubsection{Evaluating Effect Contracts}

Required effects are evaluated independently and cannot be replaced by allowed effects or by the fact that a file changed. Confirmed forbidden effects enter V. Invariants are evaluated over their continuous, checkpoint, or terminal scope. Changes outside the observation scope generate impact entries but are not automatically violations. After required-effect, observation-scope, and role-preservation closure enters V*/U*, confirmed violations yield \mbox{Effect}\allowbreak{}\mbox{Compliance} = FAIL; mandatory unknowns without confirmed violations yield UNKNOWN. PASS requires empty V*/U*, all required effects realized, no forbidden effects, and satisfied invariants.

\subsection{T6: Diagnosis, Evidence, and Engineering-Decision Trust}

T6 connects normal verification and anomaly handling through evidential responsibility for engineering conclusions, as shown in Figure 9. The normal path determines which obligations and Claims current evidence supports. The anomaly path determines which surviving hypotheses explain deviations in comparable observations and what checks or repair constraints are needed next. Diagnosis, \mbox{Engineering}\allowbreak{}\mbox{Decision}, and \mbox{Credibility}\allowbreak{}\mbox{Claim} are distinct objects. Survival of a hypothesis does not establish product acceptance, and passing common-cause analysis does not automatically satisfy coverage or physical criteria.

\begin{figure}[H]\centering
\includegraphics[width=\linewidth,height=0.77\textheight,keepaspectratio]{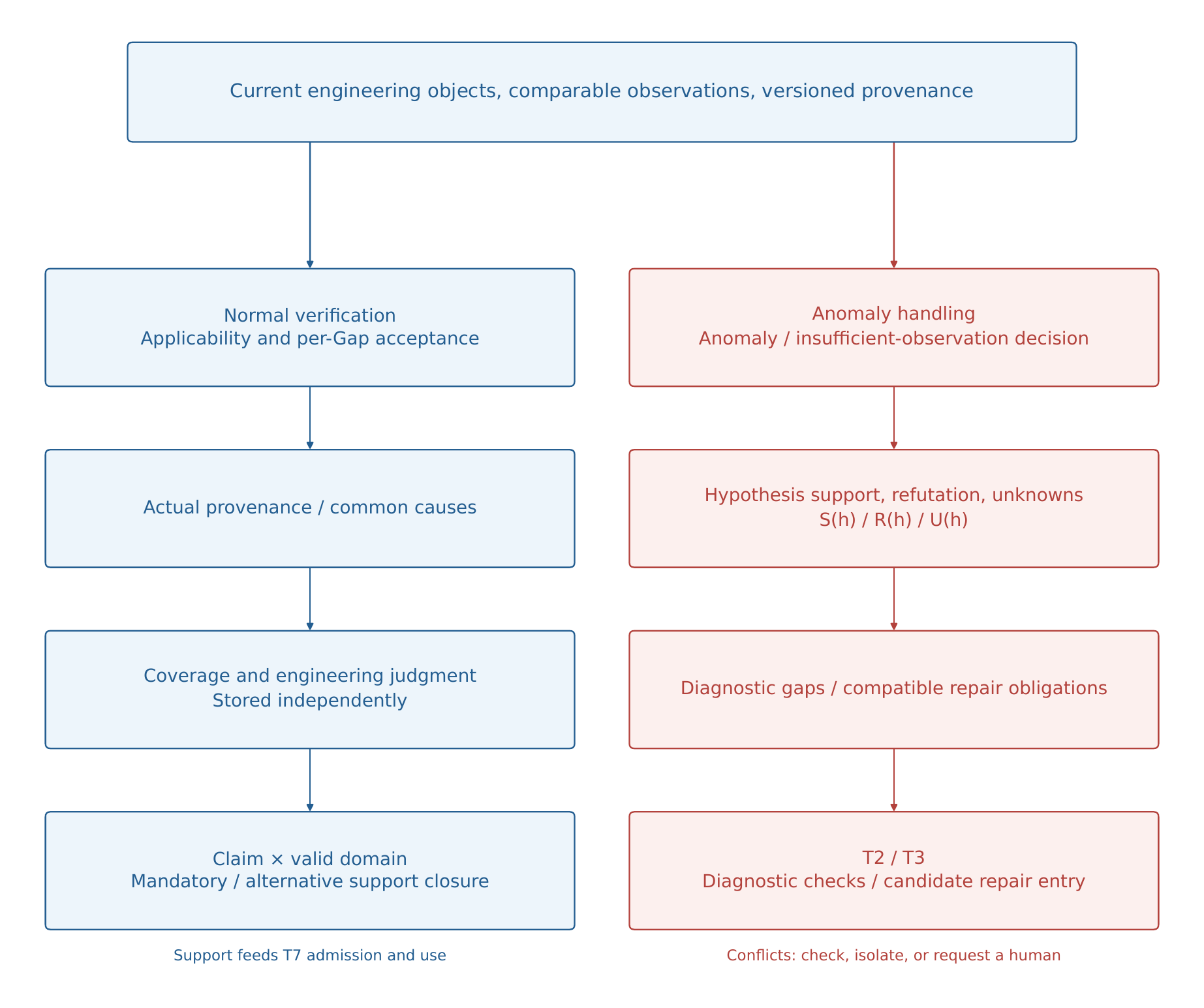}
\caption{Normal verification and anomaly-handling paths in T6 and their outputs to adjacent modules.}\label{fig:9}
\end{figure}

Figure 9 shows responsibility relationships; evidence checks may iterate or run in parallel. Common-cause checks constrain independent support; coverage and engineering judgment remain separate; Claim closure is evaluated against the current model, configuration, and context of use (CoU). The anomaly path forms hypotheses only from comparable observations. Open diagnostic gaps or conflicting modification directions require checking, isolation, or human confirmation first. Both paths persist decisions and grounds in the engineering-fact plane; T8 triggers re-evaluation after changes.

\subsubsection{Anomaly Triggers and Observation Comparability}

The anomaly trigger first creates \mbox{Observation}\allowbreak{}\mbox{Comparability}\allowbreak{}\mbox{Record} by checking \mbox{measured/}\allowbreak{}\mbox{simulated} objects, configurations, operating conditions, units, time windows, and quality. Only comparable observations enter deviation and trend evaluation. A persistent confirmed deviation relevant to engineering constraints generates AnomalyEvent. Incomparability or inadequate coverage generates \mbox{Observation}\allowbreak{}\mbox{Insufficient}. A degradation trend that reaches its trigger without a confirmed anomaly produces \mbox{Trust}\allowbreak{}\mbox{Degradation}\allowbreak{}\mbox{Record} and triggers revalidation. Thus, absence of an observed anomaly is not equated with normality.

\subsubsection{Multiple-Hypothesis Diagnosis and Repair Obligations}

The diagnostician separately maintains support S(h), refutation R(h), and unknowns U(h) for each causal hypothesis h. Reliable observations contradicting mandatory propagation predicates establish priority refutation. Unknowns are not promoted to support, and no unique root cause is forced. For surviving distinguishable hypotheses, DiagnosticGap identifies checks that resolve key predicates or distinguish candidates. Only modification boundaries jointly supported by compatible surviving hypotheses can be compiled into \mbox{Repair}\allowbreak{}\mbox{Obligation}. Conflicting directions first generate checks, isolation, or human confirmation.

\subsubsection{Layered Evidence Applicability and Per-Gap Acceptance}

\mbox{Applicability}\allowbreak{}\mbox{Decision} = <State, Scope, Reason> is evaluated separately for engineering models, raw results, derived results, and engineering conclusions. RequiredAction separately records KEEP, REINTERPRET, REPOSTPROCESS, UPDATE, RERUN, REBUILD, and related actions. Applicability at one layer does not imply applicability at the next. EvidenceGap is the target obligation's required coverage minus coverage supplied by currently applicable evidence. Every task output is accepted separately for each Gap.

\subsubsection{Common Causes and Independent Support}

FailureFactors are constructed for shared objects in actual provenance graphs and evaluated for target relevance and control state. A factor is a common cause only if the same error can propagate to the target quantity or criterion under known conditions. Actual provenance is reread after executing an independent-path plan. If caches, inherited configurations, or shared meshes merge the paths again, Independent Support must not be recorded.

\subsubsection{Model Credibility Support Closure}

Model support is evaluated per \mbox{Credibility}\allowbreak{}\mbox{Claim} and valid domain. Every mandatory support must hold for the current model, configuration, and CoU. Each alternative-support group requires at least one correctly bound path covering the Claim's asserted domain. Confirmed breaks enter \mbox{Broken}\allowbreak{}\mbox{Support}\allowbreak{}\mbox{Set}; insufficient mandatory facts enter \mbox{Required}\allowbreak{}\mbox{Unknown}\allowbreak{}\mbox{Set}; revalidation obligations are generated only for these sets. New evidence restores only the subdomain actually covered: evidence for 8–12 L/min cannot be extrapolated to 12–14 L/min. These sets retain domain-specific grounds, but final closure additionally evaluates mandatory support scope, alternative groups, and applicability domains. Empty base sets alone do not establish positive support.

\subsubsection{Separating Coverage from Engineering Conclusions}

\mbox{Verification}\allowbreak{}\mbox{Coverage}\allowbreak{}\mbox{State} = \{COMPLETE, PARTIAL, UNRESOLVED\} answers whether evidence is sufficient; \mbox{Engineering}\allowbreak{}\mbox{Decision} = \{PASS, FAIL, UNKNOWN\} answers whether an engineering object meets its criteria. With sufficient evidence and Tmax = 43 °C, coverage can be COMPLETE while judgment is FAIL. This separation applies across verification capability composition, obligation differencing, evidence independence, multiple-hypothesis diagnosis, and model support closure, preventing “verification completed” from being misreported as “product passed.”

\subsection{T7: Baseline Activation, Asset Use, and Disclosure Trust}

\subsubsection{Atomic Admission of Candidate Models}

The model admission component evaluates actual semantic binding, at least one class of physical consistency, and executability for the prescribed purpose within the candidate area, including mandatory checking scope and admission closure in V*/U*. Nonempty V* yields REJECT; empty V* with nonempty U* yields HOLD. Empty V*/U* and established closure yield only \mbox{ACCEPT\_CANDIDATE}. Formal ACCEPT additionally requires commit-time correspondence of B (current engineering objects and actual bindings), C (candidate content and external dependencies), and A (role predicates, physical rules, and prescribed scenarios) with their references. Any mismatch yields \mbox{REBASE\_REQUIRED}. This compares three classes of engineering state, not a single hash transaction.

\subsubsection{Monotonic Propagation of Use Constraints}

\mbox{Usage}\allowbreak{}\mbox{Constraint}\allowbreak{}\mbox{Envelope} propagates along actual derivation lineage. For a derived asset $d$ with parent set $\operatorname{Parents}(d)$,
\[
\begin{aligned}
\operatorname{Forbidden}(d)
&\supseteq \bigcup_{p\in\operatorname{Parents}(d)}\operatorname{Forbidden}(p),\\
\operatorname{Allowed}(d)
&\subseteq \bigcap_{p\in\operatorname{Parents}(d)}\operatorname{Allowed}(p).
\end{aligned}
\]
Additional constraints may narrow these scopes further. Renaming, format conversion, repackaging, storage migration, or organizational transfer does not establish a new lineage root. Calls are evaluated against purpose, configuration, derivation context, and expected output. Identity, ACL permission, download, matching digests, publication, or tool success cannot independently establish ALLOW. Final positive use and disclosure decisions likewise require empty respective V*/U* and closure of lineage, configuration, and permitted scope.

\subsubsection{Information Exposure and Output Release}

Disclosure control distinguishes use admissibility from information visibility. \mbox{Disclosure}\allowbreak{}\mbox{Envelope} specifies sensitive fields, execution location, output granularity, and proof requirements. \mbox{Confidential}\allowbreak{}\mbox{Compute}\allowbreak{}\mbox{Plan} references external capabilities such as local isolation, trusted execution environments (TEEs), multiparty computation (MPC), or federated execution. The definition of confidential computing as protection of data in use is given in \citep{r31}. T7 supplies constrained inputs to such backends, obtains execution evidence, and evaluates releasable content against the Envelope. Remote attestation and engineering-result checking retain separate grounds. Table~\ref{tab:t2} identifies current backend implementation scope.

For example, if a material card permits in-domain thermal-fluid computation but forbids exporting raw parameters, the system first establishes computational purpose and execution domain, then restricts returned outputs to Tmax, \ensuremath{\Delta}p, or prescribed error bounds. Use permission, input binding, backend execution records, and release decisions are persisted separately. Changing the backend does not relax parent-asset use or field-exposure constraints. This illustrates T7's disclosure interface; concrete backends depend on deployment configuration.

\subsection{T8: Recovery, Dependency Invalidation, and Capability Evolution Trust}

\subsubsection{Dual-State Effect Ledger and Recovery}

Each \mbox{Effect}\allowbreak{}\mbox{Ledger}\allowbreak{}\mbox{Entry} separately maintains historical lifecycle and current applicability. Recovery first reads the current \mbox{Engineering}\allowbreak{}\mbox{State}\allowbreak{}\mbox{Snapshot}, checks whether historical effects are recognized by the current baseline or consumed by qualified downstream objects, and propagates invalidation through engineering dependencies. Meshes, solver inputs, and evidence tied to obsolete CAD become INVALID; non-unique bindings become UNRESOLVED. Recovery plans select continuation, compensation, rollback, reconstruction, or suspension, but cannot silently migrate invalid premises to a new baseline. Final applicability uses the record's V*/U* and engineering-recognition conditions; historical states do not override mandatory unknowns.

\subsubsection{Incremental Invalidation Propagation}

Propagation uses the two traversals in Section 7.9: one follows true edges to form D; the other follows true and unknown edges to form P. Both enqueue newly reached nodes until a fixed point, then compute \mbox{U\_}\allowbreak{}\mbox{dep} = P \ensuremath{-} D. Successors of mandatory unknown edges continue propagating. Confirmed impacts are invalidated under domain rules; possible impacts enter candidate re-evaluation and prevent unsupported reuse. Changes to graph versions or edge grounds invalidate caches. Decision records identify the frozen graph and observation scope used in that evaluation.

\subsubsection{Evolution of Model Credibility and Skill Capability Domains}

The model credibility manager projects changes in models, configurations, CoU, and supporting evidence onto \mbox{Credibility}\allowbreak{}\mbox{Claim} subdomains to produce \mbox{Credibility}\allowbreak{}\mbox{Delta}. The skill capability manager projects SkillChangeSet onto \mbox{Capability}\allowbreak{}\mbox{Claim} Scope to produce \mbox{Capability}\allowbreak{}\mbox{Delta}. Both follow local change \ensuremath{\rightarrow} impact projection \ensuremath{\rightarrow} revalidation only for \mbox{broken/}\allowbreak{}\mbox{affected} scope \ensuremath{\rightarrow} Envelope update, but their controlled objects differ. The former asks where model results remain credible for particular uses; the latter asks under which objects, operations, and tools a skill remains demonstrably capable. Distinct schemas and state tables prevent a single Trust Score from obscuring local facts.

\subsection{End-to-End Control Path for the Cold-Plate Case}

For the case in Section~2, the normative handoff is compactly expressed as:
\begin{quote}
task authority and frozen obligations (T1--T2) $\rightarrow$ admissible actions and plan (T3) $\rightarrow$ actual invocation and observed differences (T4--T5) $\rightarrow$ separately evaluated coverage and judgment (T6) $\rightarrow$ current-reference activation (T7).
\end{quote}
Each arrow transfers an object and its grounds, not permission inherited from an upstream success label. T8 can reenter any affected boundary after a version, dependency, or capability change.

The three summary tables serve different purposes: Table~\ref{tab:t0} defines responsibility boundaries, Table~\ref{tab:t3} specifies technical objects and transformations, and Table~\ref{tab:t8} locates their host modules and outputs. The empirical sections use these references rather than repeating the full cold-plate control path.

\FloatBarrier
\begingroup
\small
\setlength{\parskip}{0pt}\setlength{\LTpre}{8pt}\setlength{\LTpost}{10pt}
\begin{longtable}{@{}P{75.78pt}P{124.81pt}P{115.90pt}P{129.27pt}@{}}
\caption{Implementation modules, principal objects, and failure outputs of the eight mechanisms.}\label{tab:t8}\\
\toprule
\textbf{Domain} & \textbf{Host module} & \textbf{Principal object and positive closure} & \textbf{\mbox{Failure/}\allowbreak{}\mbox{unknown} output} \\
\midrule\endfirsthead
\multicolumn{4}{l}{\small\itshape Table \thetable\ (continued)}\\
\toprule
\textbf{Domain} & \textbf{Host module} & \textbf{Principal object and positive closure} & \textbf{\mbox{Failure/}\allowbreak{}\mbox{unknown} output} \\
\midrule\endhead
\midrule\multicolumn{4}{r}{\small Continued on next page}\\\endfoot
\bottomrule\endlastfoot
T1 & Identity–\allowbreak{}authority–\allowbreak{}capability controller & \mbox{Lease/}\allowbreak{}\mbox{Capability}\allowbreak{}\mbox{Envelope} covers current task & REVOKE, MISMATCH, HOLD \\
T2 & Obligation compiler/plan checker & Mandatory obligations, dependencies, feedback closed & Gap, RECONSTRUCT, HOLD \\
T3 & \mbox{Candidate/}\allowbreak{}\mbox{composition} gate & Hard gates, cumulative effects, capability coverage closed & REJECTED, UNRESOLVED \\
T4 & Execution \mbox{gateway/}\allowbreak{}\mbox{adapters} & Consistent parameters, objects, versions, token, idempotency & DENY, UNKNOWN, TAKEOVER \\
T5 & Effect \mbox{observer/}\allowbreak{}\mbox{binding} resolver & \mbox{Actual}\allowbreak{}\mbox{Engineering}\allowbreak{}\mbox{Delta} meets effect conditions & FAIL, UNKNOWN, impact entry \\
T6 & Diagnosis–\allowbreak{}evidence–\allowbreak{}support controller & Support, coverage, provenance, applicability closed & BROKEN, PARTIAL, FAIL, UNRESOLVED \\
T7 & Activation–\allowbreak{}use–\allowbreak{}disclosure controller & B/C/A correspondence; permitted use/exposure & HOLD, \mbox{REBASE\_}\allowbreak{}\mbox{REQUIRED}, DENY \\
T8 & Ledger–\allowbreak{}invalidation–\allowbreak{}evolution manager & Current dependencies, support, capability closed & INVALID, \mbox{REVALIDATION\_}\allowbreak{}\mbox{REQUIRED} \\
\end{longtable}
\endgroup

\section{Prototype and Control-Plane Experiments}\label{sec:section_09}

\subsection{Evaluation Questions and Evidential Scope}

The experiments address six local questions: RQ1, discrimination among internal gateway and engineering-rule configurations; RQ2, prescribed states for violations, unknowns, and compliant trajectories; RQ3, responsibility-family failures after mechanism removal; RQ4, isolation and reevaluation costs under observation loss; RQ5, solver-based separation of coverage, judgment, completion, and readmission; and RQ6, consistency among Agent calls, completion scripts, commit records, and artifact facts.

Three hard failure conditions govern the specification tests: E1, missing or violating mandatory actual effects must not produce PASS or activation; E5, a B/C/A mismatch must prevent COMMITTED; E6, expired results require current-context readmission. Registered assertions cover their frozen observation scope. An artifact contradiction remains an implementation failure even when a runner reports no failure.

\paragraph{Common interpretation of the evidence.}
All experimental sections use the following rules. Frozen suites and post-repair reruns measure agreement with stated specifications, not unseen-fault detection. B0/B1/B1+/B-Std/B2 are internally assembled configurations, not replicas of commercial gateways or PLM systems. Counts retain their case, event, or family denominator and are not pooled into a safety probability. Tool execution, artifact inspection, engineering judgment, and baseline readback are distinct evidence levels. A repaired local prototype does not revise historical submissions or establish the same behavior on an unexecuted path. Each subsection below reports only its additional, experiment-specific qualifications; Table~\ref{tab:t32} consolidates the boundaries.

\subsection{Implementation, Toolchain, and Common Fact Sources}

The main control plane runs on Windows 11 and Python 3.12. CadQuery 2.8.0 generates eight-channel G17 and twelve-channel G18 candidates and provides parameter readback, role predicates, topology signatures, and no-op hooks. Gmsh 4.15.2 generates meshes and parent-version lineage. The deterministic reduced-order thermal proxy MOCK-RO serves the 29 control-plane cases; OpenFOAM v1912 reruns only the E2 and E6 hard-requirement scenarios. PyVista and NumPy independently re-extract temperatures, pressure drops, and conservation quantities. git-lfs, SQLite conditional writes, and content-addressed storage emulate candidate storage, evidence repositories, and baseline authority. Environment details, script digests, rule versions, and execution records are stored in manifests and JSONL event streams.

To avoid mixing experimental rounds, reproduction reads only four V2 fact sources: \mbox{records\_}\allowbreak{}\mbox{v2.}\allowbreak{}\mbox{jsonl}, \mbox{ablation\_}\allowbreak{}\mbox{report\_}\allowbreak{}\mbox{v2.}\allowbreak{}\mbox{json}, \mbox{pressure\_}\allowbreak{}\mbox{report\_}\allowbreak{}\mbox{v2.}\allowbreak{}\mbox{json}, and \mbox{perf\_}\allowbreak{}\mbox{report.}\allowbreak{}\mbox{json.} The compliant-equivalence suite retains its unaffected frozen records. The \mbox{extract\_}\allowbreak{}\mbox{detail.}\allowbreak{}\mbox{py} script reconstructs detailed tabulations from the same sources. The \mbox{verify\_}\allowbreak{}\mbox{repro.}\allowbreak{}\mbox{py} script checks the 29 principal records, B2 hazardous commits of 0/29, terminal counts of 22/4/3, A5 = 2/4, E/F/G stress properties, and \mbox{n\_}\allowbreak{}\mbox{ops} = 200 at all three service-level concurrency settings. Original V1 records and the first E6 artifacts are retained but excluded from these aggregates.

\FloatBarrier
\begingroup
\small
\setlength{\parskip}{0pt}\setlength{\LTpre}{8pt}\setlength{\LTpost}{10pt}
\begin{longtable}{@{}P{75.78pt}P{124.81pt}P{115.90pt}P{129.27pt}@{}}
\caption{Prototype toolchain and the evidential role of each component.}\label{tab:t9}\\
\toprule
\textbf{Stage} & \textbf{Implementation} & \textbf{Observable engineering facts} & \textbf{Experimental role} \\
\midrule\endfirsthead
\multicolumn{4}{l}{\small\itshape Table \thetable\ (continued)}\\
\toprule
\textbf{Stage} & \textbf{Implementation} & \textbf{Observable engineering facts} & \textbf{Experimental role} \\
\midrule\endhead
\midrule\multicolumn{4}{r}{\small Continued on next page}\\\endfoot
\bottomrule\endlastfoot
CAD & CadQuery 2.8.0 & Parameter readback, roles, topology signatures, artifact digests & Generate and reread real candidates \\
Meshing & Gmsh 4.15.2 & Geometry groups, mesh digest, parent version & Check stale meshes and lineage binding \\
Control-plane solver & MOCK-RO 0.1.0 & Repeatable Tmax, \ensuremath{\Delta}p, operating conditions & Execute 29 closed-set specification cases \\
Real solver & OpenFOAM v1912 & Native logs, \mbox{T/}\allowbreak{}\mbox{p\_}\allowbreak{}\mbox{rgh/}\allowbreak{}\mbox{U} fields, job duration & Execute E2/E6 hard-requirement scenarios \\
Post-processing & PyVista 0.49.0 and NumPy & Independent extraction, energy balance, decision inputs & Avoid direct reliance on solver summaries \\
\mbox{Baseline/}\allowbreak{}\mbox{evidence} & git-lfs, SQLite CAS & Revisions, candidate states, evidence lineage & Isolation, conditional activation, replay \\
\end{longtable}
\endgroup

\begin{figure}[H]\centering
\includegraphics[width=\linewidth,height=0.77\textheight,keepaspectratio]{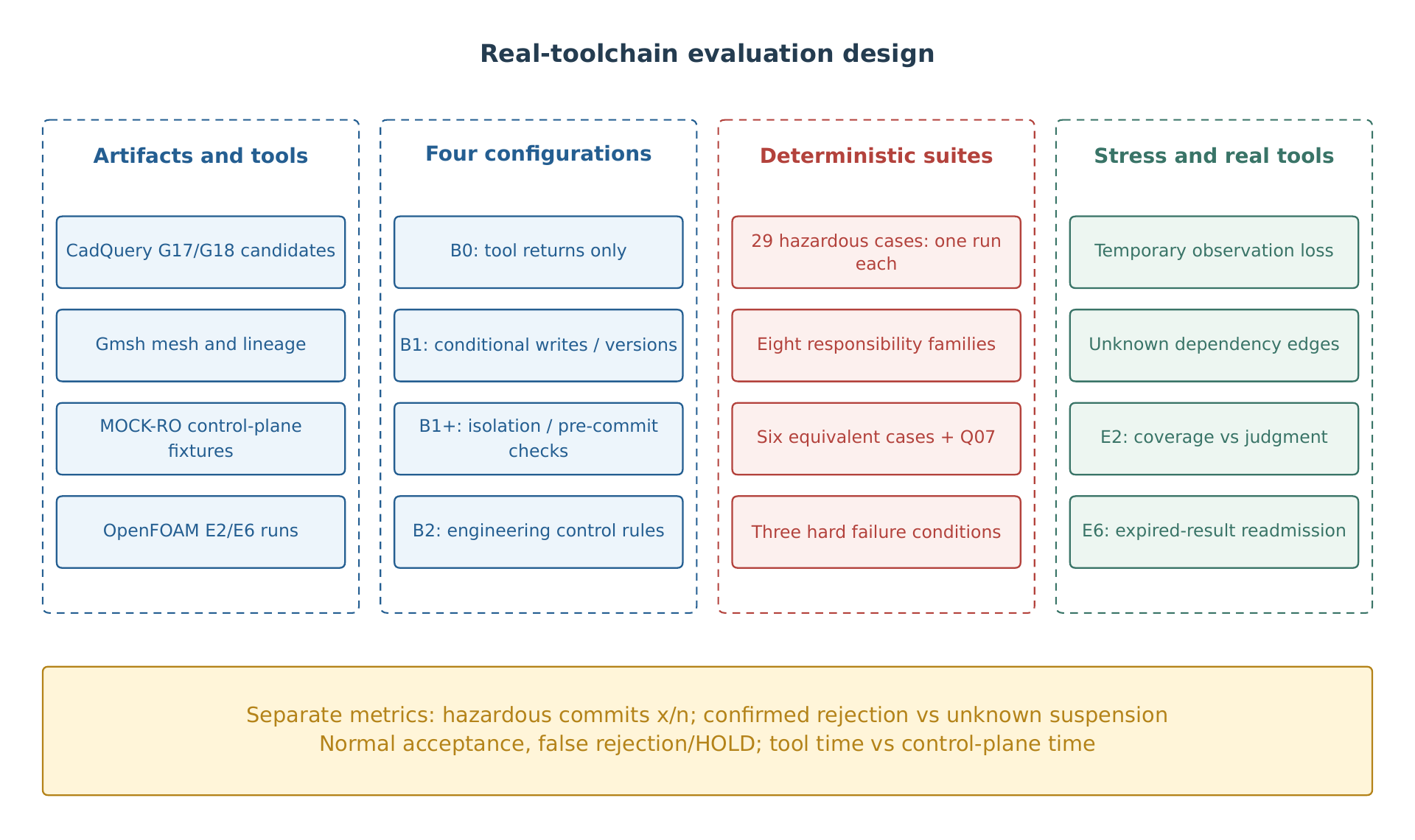}
\caption{Real-toolchain evaluation design and separation metrics.}\label{fig:10}
\end{figure}

\subsection{Four Comparison Configurations, Fault Suite, and Metrics}

B0 relies on tool success and report labels. B1 adds conditional writes, object versions, idempotency, and tool state. B1+ adds candidate isolation, preactivation revision checks, and expired-result isolation. B2 enables this prototype's engineering rules. Section~11.2 introduces the additional B-Std component abstraction; the common interpretation in Section~9.1 applies to every configuration.

The deterministic suite has 29 hazardous cases: three each for T1–T3 and four each for T4–T8. Each case receives one valid run per configuration; reruns occur only after tool or injection failures. Results are reported as hazardous-commit counts over the relevant family size, without repeated sampling of deterministic rules or confidence intervals. The compliant suite contains six equivalent variants that should be accepted directly and one temporarily missing-observation trajectory that should HOLD and then recover. Random repetitions are used only for temporary observation loss, correlated failures, concurrent interleavings, and unknown dependency edges.

\subsection{Deterministic Hazardous Suite, Terminal States, and Hard Failure Conditions}

Figure 11 reports hazardous commits in the 29 frozen cases: 29, 26, 22, and 0 for B0, B1, B1+, and B2, respectively. B1 handles three cases; B1+ handles four more; B2's assembled rules cover the remaining 22. This describes agreement between the predefined specification and its implementation. B2 yields 22 confirmed blocks, four unknown-isolation outcomes, and three REBASE outcomes. The 22 include three DENIED, seven FAIL, and twelve REJECTED. Isolating UNKNOWN is not identifying a confirmed hazard, and 0/29 is neither an unseen-fault detection rate nor a safety probability.

\begin{figure}[H]\centering
\includegraphics[width=\linewidth,height=0.77\textheight,keepaspectratio]{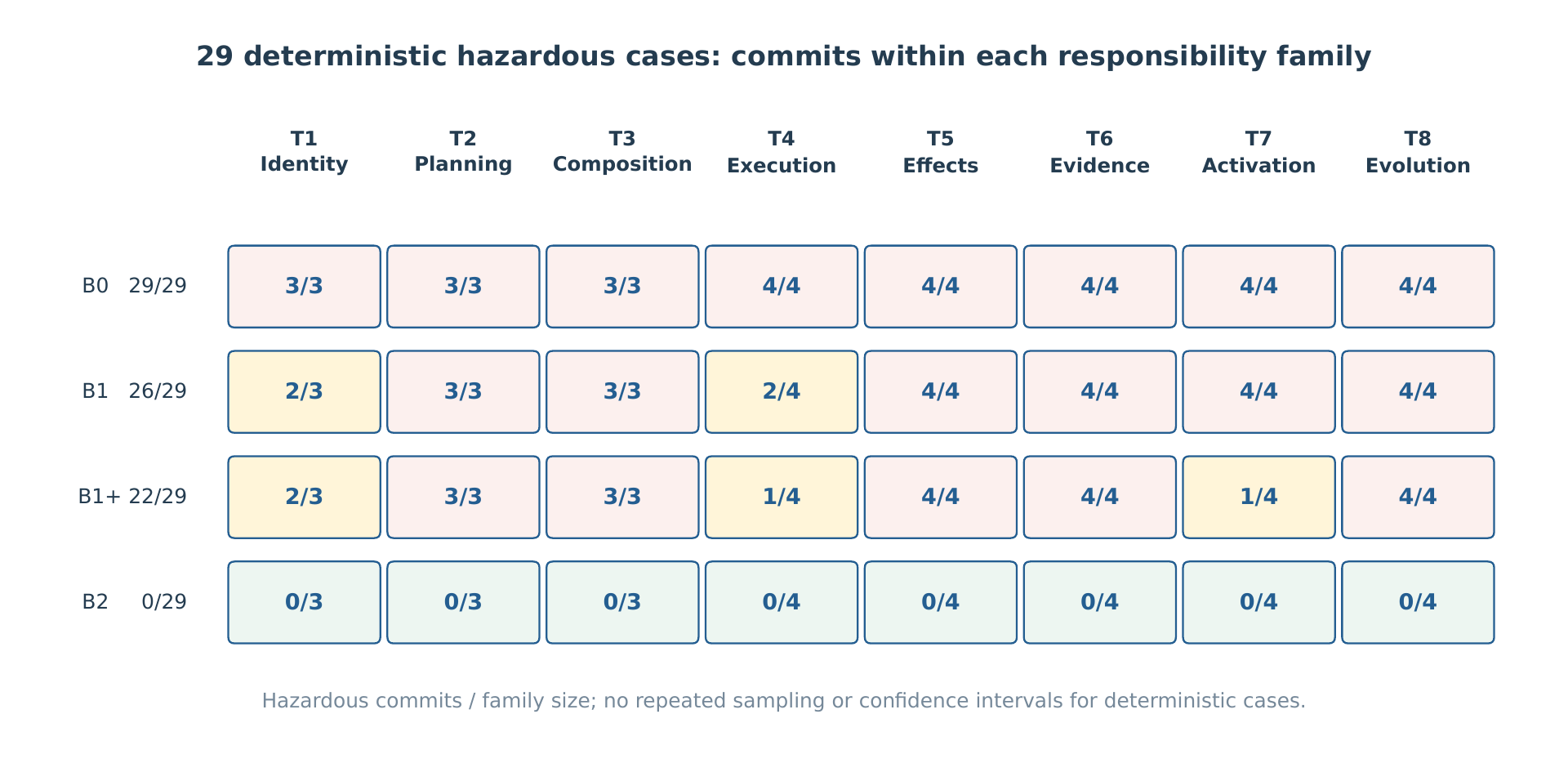}
\caption{Per-family results of four comparison configurations on 29 deterministic hazardous cases.}\label{fig:11}
\end{figure}

The E6 token condition also includes three long-job scenarios: B0/B1 yield three hazardous commits, whereas B1+/B2 isolate all three pending current-context readmission. None triggers a registered E1/E5/E6 assertion failure in the frozen suite.

\subsection{Compliant Equivalence, Temporary Observation Loss, and Positive Controls}

B2 accepts all six directly admissible trajectories, with 0/6 false rejections. B1 falsely rejects Q02 by treating a unit-representation difference as an engineering difference, yielding 1/6 false rejections. Q07 temporarily removes a mandatory evidence source; B2 returns HOLD with the correct reason code, then accepts after the observation recovers without context drift. Preserving unknowns therefore need not mean permanent rejection: the decision record must retain evidence-completion conditions and recovery paths.

Observation-loss probabilities are 1\%, 5\%, 10\%, and 20\%, with 200 runs per combination. Table~\ref{tab:t10} reports HOLD rates for compliant candidates; hazardous ALLOW is zero in all twelve combinations. Missing observations on hazardous cases largely convert some confirmed rejections into HOLD, directly checking three-state implementation rather than empirical detection performance. The more informative result is the availability cost for compliant candidates. Independently removing four mandatory observations—mesh digest, role binding, lineage query, and baseline query—at a 20\% per-observation probability produces 55.5\% false HOLD. Correlated object-service or lineage-service failures remove fields in groups and change both the statistical structure and HOLD rate. All HOLD cases recover after observations return, and reason codes are correct in every case.

\FloatBarrier
\begingroup
\small
\setlength{\parskip}{0pt}\setlength{\LTpre}{8pt}\setlength{\LTpost}{10pt}
\begin{longtable}{@{}P{65.96pt}P{83.55pt}P{96.75pt}P{96.75pt}P{96.75pt}@{}}
\caption{Compliant-candidate HOLD rates and hazardous permission under temporary observation loss.}\label{tab:t10}\\
\toprule
\textbf{Per-item loss probability} & \textbf{Independent-loss HOLD} & \textbf{Object-service-correlated HOLD} & \textbf{Lineage-service-correlated HOLD} & \textbf{Hazardous ALLOW} \\
\midrule\endfirsthead
\multicolumn{5}{l}{\small\itshape Table \thetable\ (continued)}\\
\toprule
\textbf{Per-item loss probability} & \textbf{Independent-loss HOLD} & \textbf{Object-service-correlated HOLD} & \textbf{Lineage-service-correlated HOLD} & \textbf{Hazardous ALLOW} \\
\midrule\endhead
\midrule\multicolumn{5}{r}{\small Continued on next page}\\\endfoot
\bottomrule\endlastfoot
1\% & 5.0\% & 1.5\% & 1.5\% & 0/600 \\
5\% & 22.5\% & 3.0\% & 4.5\% & 0/600 \\
10\% & 35.0\% & 11.5\% & 8.0\% & 0/600 \\
20\% & 55.5\% & 22.0\% & 17.0\% & 0/600 \\
\end{longtable}
\endgroup

A positive control intentionally collapses UNKNOWN to PASS to check that the stress test is discriminative. At the four loss levels, it permits hazardous execution 2, 13, 23, and 48 times out of 200 under independent loss; 1, 10, 12, and 35 times under object-service correlation; and 2, 11, 10, and 39 times under lineage-service correlation. These failures distinguish absence of an unknown trigger from correct preservation of unknowns.

\subsection{Mechanism Ablation by Responsibility Family}

Figure~12 reports hazardous commits restored within each responsibility family: 6/6 after removing plan/composition checks, 4/4 for actual-effect observation, 1/4 for common-cause checks, 3/4 for atomic activation, and 2/4 for dependency propagation. In A5, F26/F27 escape but recovery and evidence checks still block F28/F29. Only F19 in the common-cause family requires that particular mechanism. These denominators are family sizes, not detection-rate samples; the residual blocking demonstrates overlapping controls.

\begin{figure}[H]\centering
\includegraphics[width=\linewidth,height=0.77\textheight,keepaspectratio]{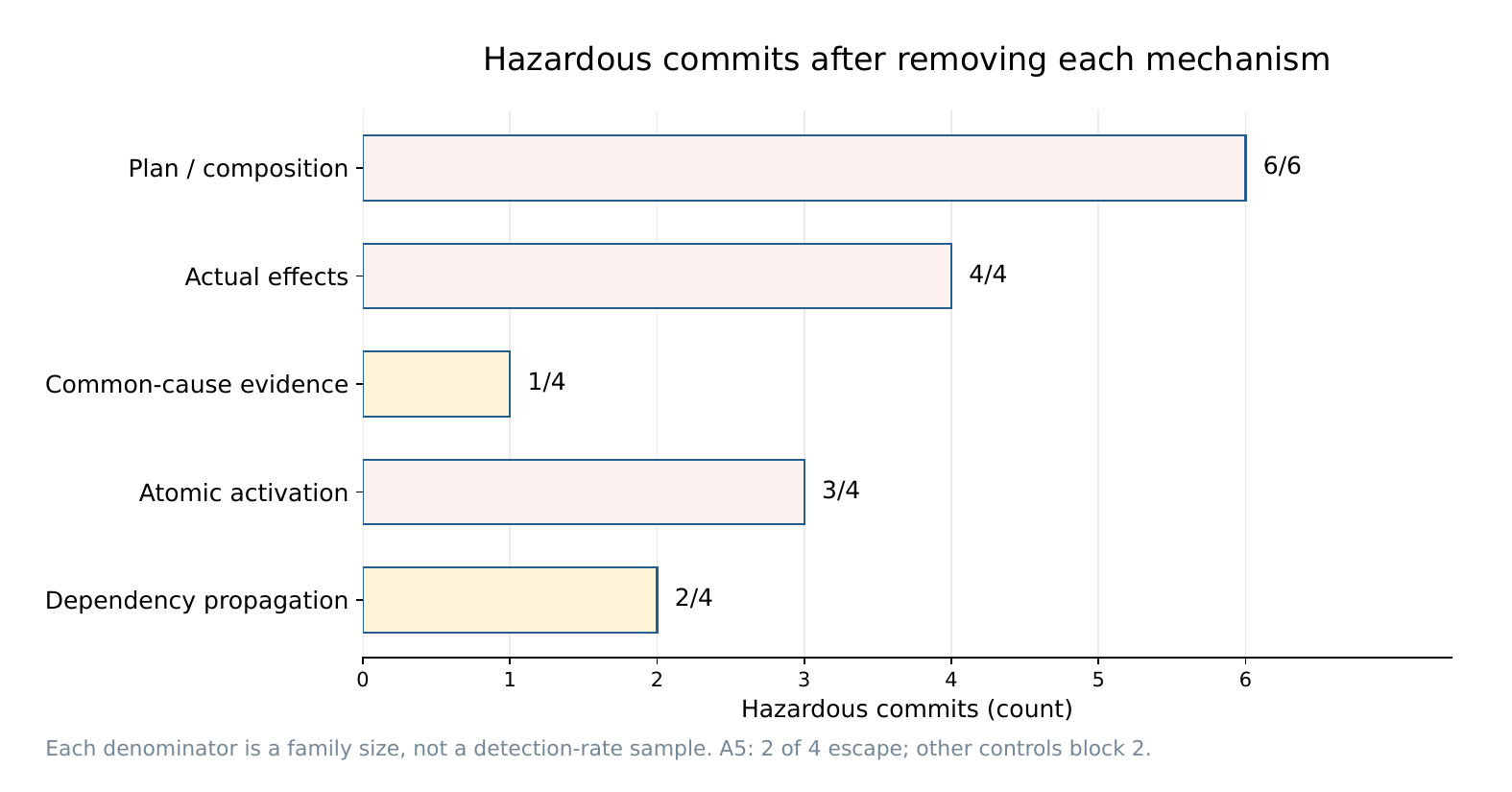}
\caption{Mechanism ablation reported within each responsibility family.}\label{fig:12}
\end{figure}

\subsection{Unknown Dependency-Edge Stress}

We run 200 seeds at edge-hiding levels of 0\%, 5\%, 10\%, 20\%, and 40\%. Treating unknown edges as absent produces no confirmed false inclusions but omission rates of 0, 5.08\%, 13.23\%, 38.44\%, and 61.73\%. Treating them as present eliminates omissions but yields mean false inclusions of 0, 0.015, 0.025, 0.065, and 0.085 nodes per run. Three-state propagation has neither confirmed omissions nor confirmed false inclusions, at the cost of mean UNRESOLVED counts of 0, 0.230, 0.570, 1.770, and 3.045 nodes per run. Figure 13 therefore presents safety, scope, and unresolved cost together rather than only a favorable metric.

\begin{figure}[H]\centering
\includegraphics[width=\linewidth,height=0.77\textheight,keepaspectratio]{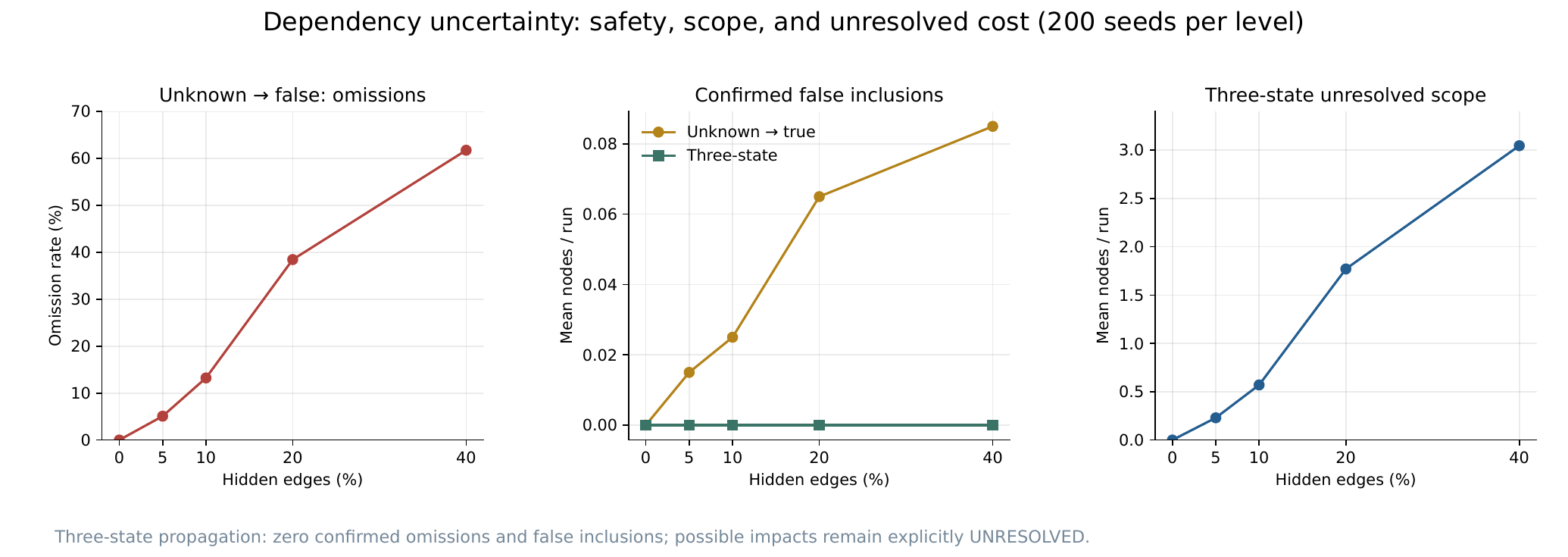}
\caption{Safety, scope, and unresolved-cost trade-offs under unknown dependency edges.}\label{fig:13}
\end{figure}

\subsection{Concurrent Semantics and Control-Plane Overhead}

Concurrency stress uses real thread interleavings and a SQLite CAS baseline service, with 200 runs per scenario. When baseline drift precedes activation, all 200 runs return REBASE without COMMITTED; when activation precedes drift, all 200 commit without REBASE. With two to eight threads, 200 logical idempotent operations produce exactly 200 tool side effects. In token-timing tests, 199 runs actually expire and all are isolated, with zero direct reuse without readmission. The remaining run completes before expiry and commits normally. This is not a 99.5\% isolation success rate: one run did not require isolation.

Microbenchmarks use 2,000 warm-up iterations, ten processes, and 10,000 iterations per operation. P50 times for object references, V/U aggregation, effect comparison, and lineage queries are 0.9, 0.7, 0.4, and 1.9 \ensuremath{\mu}s. Dependency closure and B/C/A checking take 26.0 and 61.2 \ensuremath{\mu}s. The complete service path includes stage \ensuremath{\rightarrow} evaluate \ensuremath{\rightarrow} activate and SQLite persistence, with substantially higher tail latency than kernel computation.

The test machine has an Intel Core Ultra 9 285H, 16 cores, 95.5 GB memory, and NVMe storage. At 32 clients, P95 is approximately 1.19 s, mainly due to serialized SQLite writes and in-process lock convoying. Non-monotonic P50 with concurrency also reflects batch scheduling and lock contention, not a scalability advantage. Figure 14 separates microsecond kernel work from millisecond-to-second service paths so that storage bottlenecks are not mistaken for control-semantic costs.

\begin{figure}[H]\centering
\includegraphics[width=\linewidth,height=0.77\textheight,keepaspectratio]{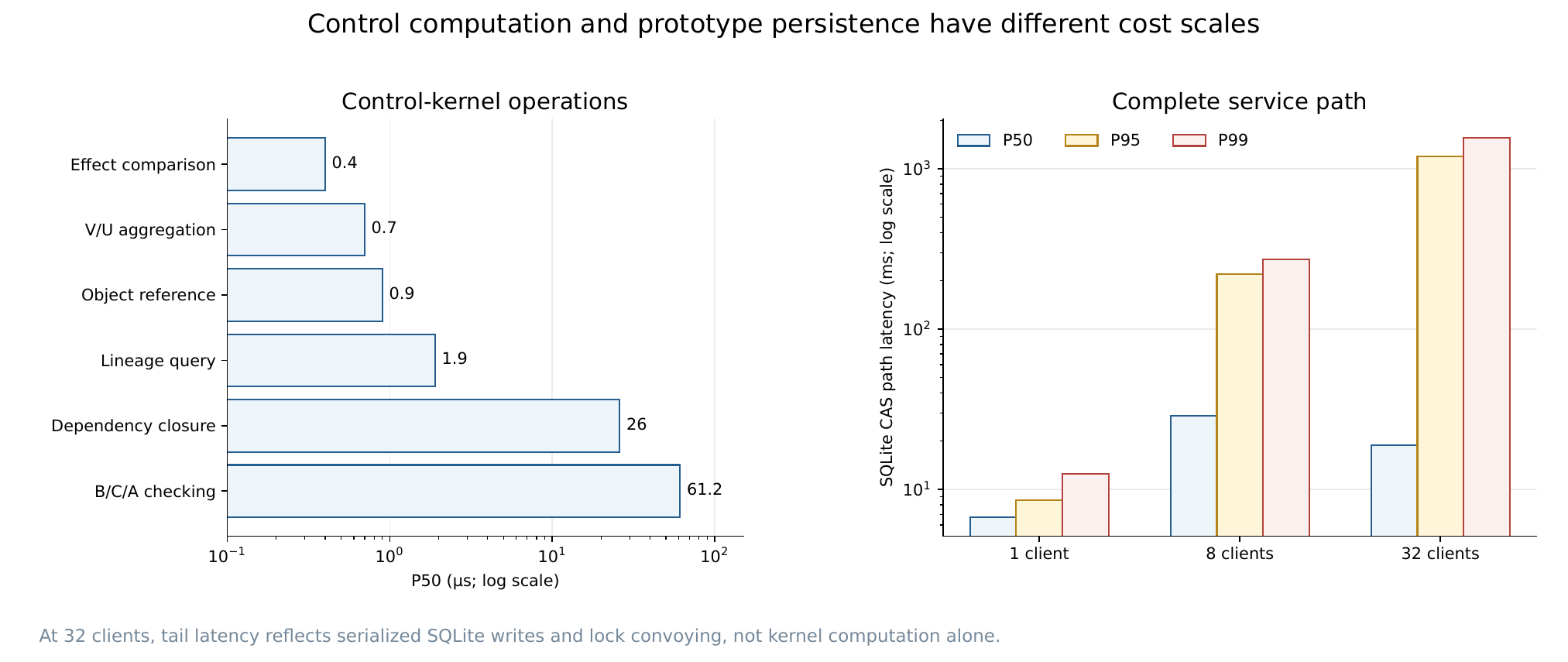}
\caption{Overhead decomposition for the control kernel and prototype storage service.}\label{fig:14}
\end{figure}

\subsection{Connection to the Tool and Interface Studies}
Sections~10--11 move from prescribed control-plane inputs to solver jobs, Agent calls, artifact records, and persistence. Their role is to test whether facts survive the interfaces assumed by the specification. The decisive counterexamples concern missing checks and changed fact meanings, rather than merely incorrect terminal labels.

\section{State-Separation Cases in a Real OpenFOAM Toolchain}\label{sec:section_10}

These real solver runs are semantic-separation fixtures: E2 exercises the distinction between evidence coverage and a criterion judgment; E6 exercises the distinction between job completion and permission to consume its result. The reported numerical fields establish the inputs to these decisions, not validated cold-plate performance. The reduced-order values in Section~2 describe a separate motivating scenario and are not calibration targets for these runs.

\subsection{Solver Pipeline, Per-Case Calibration, and Evidence Boundaries}

E2 and E6 use real solver execution. Gmsh generates a 200 \ensuremath{\times} 16 \ensuremath{\times} 24 hexahedral single-channel mesh. Scripts generate OpenFOAM v1912 dictionaries and run gmshToFoam, decomposePar, eight-process \mbox{buoyant}\allowbreak{}\mbox{Boussinesq}\allowbreak{}\mbox{Simple}\allowbreak{}\mbox{Foam}, reconstructPar, and independent result extraction through WSL2. Post-processing rereads Tmax, outlet volume-average temperature, and inlet–\allowbreak{}outlet \mbox{p\_}\allowbreak{}\mbox{rgh} differences from field files, converting pressure drop using density. Solver return codes and summaries are not themselves engineering judgments.

Wall thermal boundaries use fixedGradient, calibrated per case against Q/(\ensuremath{\dot{m}}\ensuremath{\cdot}cp) to keep outlet temperature-rise error below 5\%. E6 uses 404.94 K/m and achieves +0.01\% temperature-rise error. This per-case calibration constructs control-experiment fixtures satisfying the stated energy check; it does not validate the predictive capability of a common boundary model across operating conditions. Runtime control, numerical checks, and physical-model validity are therefore evaluated separately.

\subsection{E2: Complete Coverage Does Not Imply Engineering Acceptance}

All four cases in Table~\ref{tab:t12} retain logs and re-extractable T, \mbox{p\_}\allowbreak{}\mbox{rgh}, and U fields. Against the frozen criteria Tmax \ensuremath{\leq} 40 °C and \ensuremath{\Delta}p \ensuremath{\leq} 45 kPa, nominal G17/G18 yield 29.82 °C/17.56 kPa and 29.74 °C/7.04 kPa, respectively, and PASS. Low-flow G17 yields 74.85 °C/0.49 kPa and FAIL. Its Coverage remains COMPLETE because the records needed to answer the specified questions are present. This demonstrates non-substitutability of the two states, not validated thermal-fluid performance of a real product.

\FloatBarrier
\begingroup
\fontsize{8.3}{10.4}\selectfont
\setlength{\parskip}{0pt}\setlength{\LTpre}{8pt}\setlength{\LTpost}{10pt}
\begin{longtable}{@{}P{54.83pt}P{80.13pt}P{37.96pt}P{42.18pt}P{37.96pt}P{42.18pt}P{63.26pt}P{63.26pt}@{}}
\caption{OpenFOAM hard-requirement cases, energy checks, and engineering judgments.}\label{tab:t12}\\
\toprule
\textbf{Case} & \textbf{Conditions} & \textbf{Solve (s)} & \textbf{Tmax (°C)} & \textbf{\ensuremath{\Delta}p (kPa)} & \textbf{Rise error} & \textbf{Coverage} & \textbf{Judgment} \\
\midrule\endfirsthead
\multicolumn{8}{l}{\small\itshape Table \thetable\ (continued)}\\
\toprule
\textbf{Case} & \textbf{Conditions} & \textbf{Solve (s)} & \textbf{Tmax (°C)} & \textbf{\ensuremath{\Delta}p (kPa)} & \textbf{Rise error} & \textbf{Coverage} & \textbf{Judgment} \\
\midrule\endhead
\midrule\multicolumn{8}{r}{\small Continued on next page}\\\endfoot
\bottomrule\endlastfoot
G17 S1 & Nominal 8 L/min & 46.4 & 29.82 & 17.56 & +0.01\% & COMPLETE & PASS \\
G18 S1 & Nominal 8 L/min & 37.8 & 29.74 & 7.04 & \ensuremath{-}0.04\% & COMPLETE & PASS \\
G17 S1 low flow & Pump degradation, 1 L/min; Re \ensuremath{\approx} 1188 & 30.4 & 74.85 & 0.49 & \ensuremath{-}2.55\% & COMPLETE & FAIL \\
G17 S1 E6 & 8 L/min; endTime 1000 & 91.0 & 29.03 & 17.57 & +0.01\% & COMPLETE & PASS \\
\end{longtable}
\endgroup

\subsection{E6: Passing Physical Checks Does Not Imply Lifecycle Admission}

When E6's approximately 91-second real solver run completes, its initial 60-second token has expired. A +0.01\% temperature-rise error and \mbox{temperature/}\allowbreak{}\mbox{pressure} values within the case thresholds establish only the listed numerical checks, not current-context admission. B0/B1 directly use the expired result. B1+/B2 isolate it and reread B/C/A, recording \mbox{READMITTED\_}\allowbreak{}\mbox{COMMITTED/}\allowbreak{}\mbox{ACCEPT} without drift or \mbox{REBASE\_}\allowbreak{}\mbox{REQUIRED} with drift. This illustrates separation of numerical judgment, job completion, and permission to use results; it does not establish physical-model sufficiency or novelty of token mechanisms.

\begin{figure}[H]\centering
\includegraphics[width=\linewidth,height=0.77\textheight,keepaspectratio]{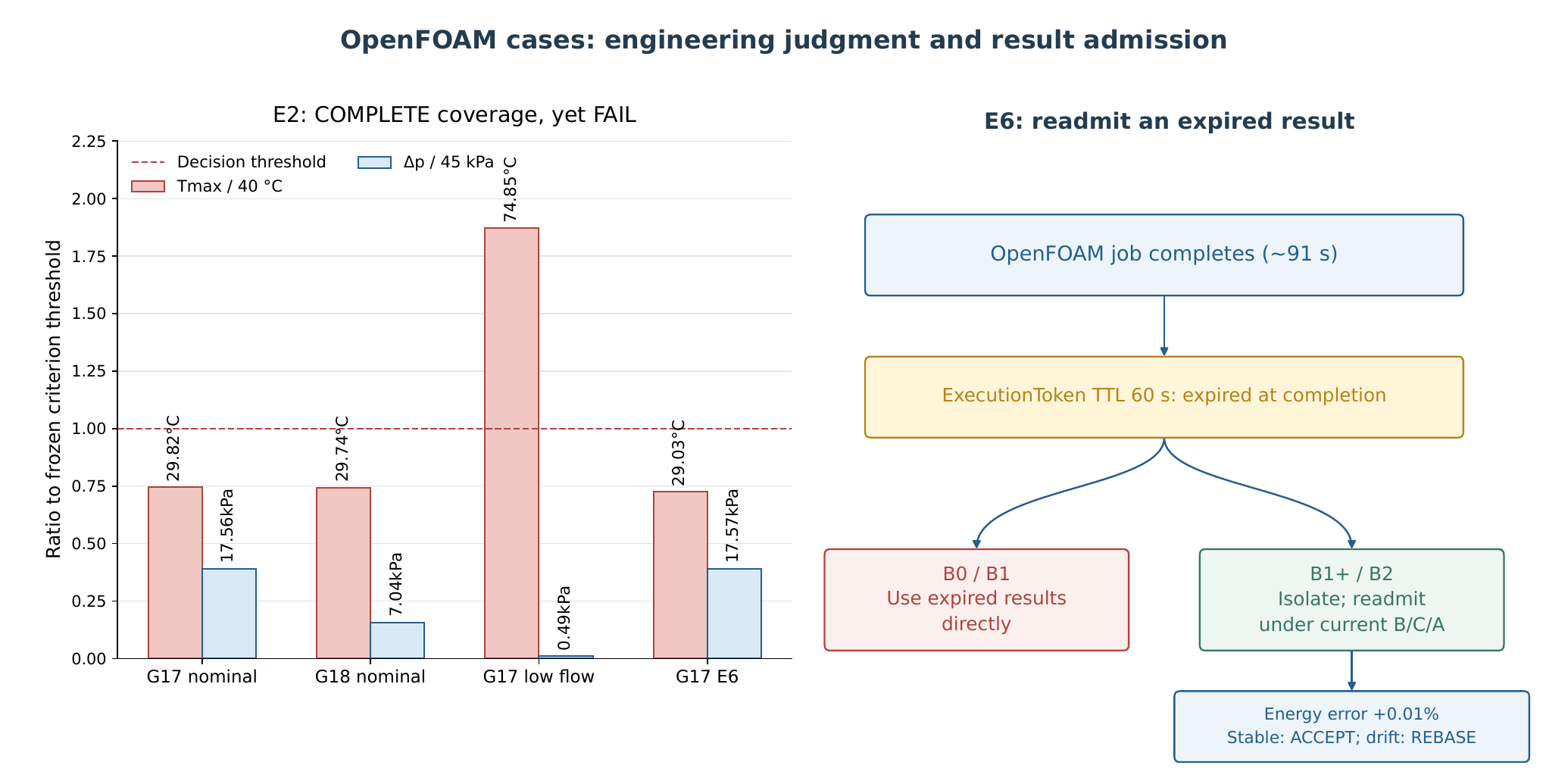}
\caption{End-to-end outcomes of OpenFOAM E2 and E6.}\label{fig:15}
\end{figure}

\subsection{Physical-Model Limitations}

OpenFOAM results test control semantics, not cold-plate certification. Independently calibrated fixedGradient values do not validate a common boundary model. The low-flow Reynolds number of approximately 1,188 is laminar, yet the case uses RAS kEpsilon; it is therefore a FAIL fixture, not evidence of low-flow modeling accuracy. The proxy scenario in Section 2 and the real solver also use different physical calibrations: nominal G17's proxy pressure drop of 98.15 kPa cannot be combined with OpenFOAM's approximately 17.56 kPa as design-improvement evidence. Real fields and logs are used only to demonstrate separation of Coverage from Decision, physical judgment from Admission, and job completion from current applicability.

The deliberately short E6 lease tests readmission at an expiry boundary. Deployment policy should use the separate job-lifetime and decision-authority semantics of Section~7.3; the fixture does not evaluate automatic lease sizing.

\section{Extended Evaluation, Actual Tool Invocation, and Commit Checking}\label{sec:section_11}

This section evaluates internal specification cases, agent execution, representation transformations, a second engineering domain, and local persistence. These experiments answer different questions and are not combined into whole-framework validation. WP2 includes actual CAD, meshing, and structural-solver calls. Task-completion and commit-checking scripts are counted separately; the latter contains an error in hole-position semantics. Tool execution, script outputs, and engineering compliance are therefore reported as three evidential layers.

\subsection{Internal Frozen Cases and Specification Regression Protocol}

Internal cases store public tasks, declarative worlds, expected states, and sealed Oracles in separate directories; runners and expected-state readers use separate paths. Experiment A contains 48 cases—24 hazardous and 24 compliant—executed by five configurations. Four-tuple deduplication finds no exact collision with F01–F29 and flags four near-duplicates. This is separation of files and read paths, not independence of case semantics, rule knowledge, or reviewers. File organization, hashes, and deduplication support traceability, not generalization to unknown faults.

The internal evaluation uses file isolation and terminal-state comparison, but lacks the semantic and personnel independence needed for an unseen-fault holdout. The v3 authoring materials include rules and triggers. Initial exact terminal agreement is 28/40; v3.1 reaches 40/40 through joint regression of rules, Oracle, and checking conditions on the same cases. These results show that specifications and checking protocols can themselves be defective. The later count is not unseen-fault detection evidence. Versioned artifacts and terminal records make the corresponding conditions traceable.

\subsubsection*{ECE 1: Consistency of Decision and Activation States}

ECE 1 contains 48 cases, 24 per cold-plate and bracket domain; each domain has eight compliant, eight hazardous, four mandatory-unknown, and four composite cases. The execution side does not read expected fields, but Oracle derivation and review are performed by the same AI without personnel separation. This is an internal frozen-case evaluation, not an externally independent holdout.

ECE 1 checks decision and activation states as separate expected fields. The single-field convention has 16/48 terminal mismatches, mainly because the Oracle conflates these states. Under the two-field ECE 1.1 convention, agreement is 48/48 (Table~\ref{tab:t13}). The two-field convention separates the evaluator result from activation; the rows are same-set regression results.

\FloatBarrier
\begingroup
\small
\setlength{\parskip}{0pt}\setlength{\LTpre}{8pt}\setlength{\LTpost}{10pt}
\begin{longtable}{@{}P{112.94pt}P{167.15pt}P{171.67pt}@{}}
\caption{Two-field specification terminal-state agreement in ECE 1.1.}\label{tab:t13}\\
\toprule
\textbf{Measure} & \textbf{Count} & \textbf{Interpretation} \\
\midrule\endfirsthead
\multicolumn{3}{l}{\small\itshape Table \thetable\ (continued)}\\
\toprule
\textbf{Measure} & \textbf{Count} & \textbf{Interpretation} \\
\midrule\endhead
\midrule\multicolumn{3}{r}{\small Continued on next page}\\\endfoot
\bottomrule\endlastfoot
ECE 1.1 exact terminal agreement & 48/48 & Separate decision and activation fields; no artifact-confirmed hazardous commit \\
ECE 1.1 hazardous commits & 0 & Sixteen hazardous and eight composite cases blocked; sixteen compliant cases COMMITTED \\
B2 terminal distribution & 16 compliant COMMITTED; 16 hazardous blocked (14 REJECTED + 2 DENIED); 8 composite REJECTED; 8 mandatory-unknown HOLD & Unknowns are not counted as passes \\
Weak-baseline comparison & B0/B1/B1+/B-Std permit all cases & 125 \mbox{hazardous/}\allowbreak{}\mbox{composite/}\allowbreak{}\mbox{unknown} variant rows labeled \mbox{unsafe\_}\allowbreak{}\mbox{commit}; corresponding mechanisms are intentionally disabled \\
ECE 1 run 1: defect discovery & 16/48 terminal mismatches & Attributed to Oracle vocabulary rather than system code in this protocol; disclosed separately \\
\end{longtable}
\endgroup

\subsubsection*{ECE 2: Fixture Validity and Multiconfiguration Regression}

ECE 2 contains 48 cases and five system variants, producing 240 evaluation rows. Fixture checks identify seven inconsistencies between generated facts and expectations: five arise from recursive merging of declaration dictionaries, and two action-order perturbations exist only in task packages rather than \mbox{hidden\_}\allowbreak{}\mbox{facts.} Under those fixture conditions, B2's exact terminal agreement is 41/48. Although the fixture labels ST2-12 hazardous because of an action-order violation, its actual hidden action sequence lacks that violation; the label alone cannot establish an artifact-confirmed hazardous commit. Table~\ref{tab:t14} separates fixture conditions from agreement results.

\FloatBarrier
\begingroup
\small
\setlength{\parskip}{0pt}\setlength{\LTpre}{8pt}\setlength{\LTpost}{10pt}
\begin{longtable}{@{}P{112.94pt}P{167.15pt}P{171.67pt}@{}}
\caption{ECE 2 fixture conditions and terminal agreement; 48 cases and 240 rows are counted separately.}\label{tab:t14}\\
\toprule
\textbf{Fixture/rule condition} & \textbf{Case/row count} & \textbf{Interpretation} \\
\midrule\endfirsthead
\multicolumn{3}{l}{\small\itshape Table \thetable\ (continued)}\\
\toprule
\textbf{Fixture/rule condition} & \textbf{Case/row count} & \textbf{Interpretation} \\
\midrule\endhead
\midrule\multicolumn{3}{r}{\small Continued on next page}\\\endfoot
\bottomrule\endlastfoot
v1 fixtures / R1 & B2 terminal agreement 41/48 & Seven generated facts disagree with Oracle \\
Aligned run2 fixtures / R1 & B2 48/48; zero mismatches in 240 five-variant rows & Same-set regression with aligned \mbox{facts/}\allowbreak{}\mbox{expectations} \\
R2.1 type-guard regression & Zero mismatches in 240 five-variant rows & Compatibility with existing frozen facts; no new LLM calls \\
\end{longtable}
\endgroup

With generated facts and expectations aligned in run2, B2 has sixteen ALLOWED+COMMITTED, eighteen REJECTED, three DENIED, three FAIL, and eight HOLD outcomes. R2.1 agrees with expectations on all 240 rows of the same 48-case, five-configuration evaluation. These counts support compatibility of the listed fixtures and rules, not coverage of newly agent-generated parameters or adapter fact transformations, and do not resolve the semantic gap in Section 11.3.

\subsection{Experiment A: Internally Constructed Cases and Rule-Extension Regression}

\FloatBarrier
\begingroup
\small
\setlength{\parskip}{0pt}\setlength{\LTpre}{8pt}\setlength{\LTpost}{10pt}
\begin{longtable}{@{}P{72.29pt}P{72.29pt}P{72.29pt}P{72.29pt}P{72.29pt}P{72.29pt}@{}}
\caption{Five configurations on 24 hazardous and 24 compliant internal cases.}\label{tab:t15}\\
\toprule
\textbf{Variant} & \textbf{Hazardous commits} & \textbf{False reject} & \textbf{False HOLD} & \textbf{Correct hazardous terminal} & \textbf{Correct compliant terminal} \\
\midrule\endfirsthead
\multicolumn{6}{l}{\small\itshape Table \thetable\ (continued)}\\
\toprule
\textbf{Variant} & \textbf{Hazardous commits} & \textbf{False reject} & \textbf{False HOLD} & \textbf{Correct hazardous terminal} & \textbf{Correct compliant terminal} \\
\midrule\endhead
\midrule\multicolumn{6}{r}{\small Continued on next page}\\\endfoot
\bottomrule\endlastfoot
B0 & 24/24 & 0/24 & 0/24 & 0/24 & 24/24 \\
B1 & 22/24 & 0/24 & 0/24 & 2/24 & 24/24 \\
B1+ & 20/24 & 0/24 & 0/24 & 4/24 & 24/24 \\
B-Std & 20/24 & 0/24 & 0/24 & 4/24 & 24/24 \\
B2 & 0/24 & 0/24 & 0/24 & 24/24 & 24/24 \\
\end{longtable}
\endgroup

All 24 fault points in Experiment A receive their specified terminal states under the expanded B2 rules; fifteen correspond to newly added rules, so this is not zero-shot generalization. B1+ retains gateway checks, versioned references, candidate isolation, and preactivation revision checks. B-Std additionally assembles policy decisions, telemetry, PROV provenance, conditional writes, and human approval, but neither configuration implements the proposed actual-effect observation, evidence-support evaluation, or applicability propagation. Their tie provides no additional discrimination on this suite; it does not establish that these added infrastructure components lack value. The residual cases exercise engineering predicates absent from both configurations. Relative to B0, B1 additionally blocks two \mbox{schema/}\allowbreak{}\mbox{idempotency} faults.

The v3 cases were written by a contributor not involved in rule coding, but using guidance containing rule names, routing labels, and trigger facts. Forty cases include 32 \mbox{hazardous/}\allowbreak{}\mbox{mandatory-}\allowbreak{}\mbox{unknown} and eight compliant cases, yielding 200 rows across five configurations. Four-tuple deduplication finds no exact development-set duplicates, but different wording does not establish independent engineering semantics or triggers. Sealing, deduplication, and before/after hashes support procedural traceability. Table~\ref{tab:t16} presents specification terminal states, not unseen-fault holdout results.

\FloatBarrier
\begingroup
\small
\setlength{\parskip}{0pt}\setlength{\LTpre}{8pt}\setlength{\LTpost}{10pt}
\begin{longtable}{@{}P{112.94pt}P{167.15pt}P{171.67pt}@{}}
\caption{Terminal-state distribution of the internal v3.1 specification set.}\label{tab:t16}\\
\toprule
\textbf{Case category} & \textbf{Count} & \textbf{B2 terminal states; v3.1 exact agreement 40/40} \\
\midrule\endfirsthead
\multicolumn{3}{l}{\small\itshape Table \thetable\ (continued)}\\
\toprule
\textbf{Case category} & \textbf{Count} & \textbf{B2 terminal states; v3.1 exact agreement 40/40} \\
\midrule\endhead
\midrule\multicolumn{3}{r}{\small Continued on next page}\\\endfoot
\bottomrule\endlastfoot
VIOLATION: confirmed violation & 24 & All blocked: 19 REJECTED, 1 DENIED, 4 \mbox{REBASE\_}\allowbreak{}\mbox{REQUIRED}; no activation \\
\mbox{REQUIRED\_}\allowbreak{}\mbox{UNKNOWN} & 8 & All isolated: 8 HOLD \\
COMPLIANT & 8 & 8/8 accepted and correctly committed; no false rejection \\
\end{longtable}
\endgroup

Among the 32 hazardous or mandatory-unknown v3.1 cases, hazardous activations are 32 for B0, 28 for B1, 23 for B1+, 23 for B-Std, and zero for B2. The final 40/40 agreement follows same-set revision of rules, fixtures, and expectations.

Terminal checking must compare both decision and activation states, distinguish RUNNING from confirmed violations, and retain unknowns when declared effects lack factual support. These requirements concern rules, Oracles, and checking assertions; a generic program-pass result is insufficient. Confirmed violations take precedence over mandatory unknowns. HOLD for insufficient evidence demonstrates isolation of unknowns, not discovery of a new hazard class.

\subsection{Experiment B: Agent Interaction, Tool Execution, and Artifact Checking}

The typed-interaction evaluation follows requirements reading, intent generation, typed feedback, and plan adjustment. It has 24 templates—eight each for normal, ambiguous, and drifting cases—with three seeds and two control configurations, totaling 144 episodes. The LLM uses GPUStack's qwen3.6-35b-a3b-fp8 service; a deterministic harness executes tool behavior, without autonomous CAD, mesh, or solver calls. This batch measures use of typed feedback and recovery behavior. The separate \mbox{pair\_}\allowbreak{}\mbox{b1} batch includes actual tool calls, and its execution scope and metrics are reported separately.

\FloatBarrier
\begingroup
\fontsize{8.3}{10.4}\selectfont
\setlength{\parskip}{0pt}\setlength{\LTpre}{8pt}\setlength{\LTpost}{10pt}
\begin{longtable}{@{}P{54.83pt}P{59.05pt}P{50.61pt}P{42.18pt}P{54.83pt}P{54.83pt}P{50.61pt}P{54.83pt}@{}}
\caption{agent-loop metrics for the deterministic arm; 144 episodes.}\label{tab:t17}\\
\toprule
\textbf{Config.} & \textbf{Scenario} & \textbf{Unsafe attempts} & \textbf{Blocked} & \textbf{Hazardous commits} & \textbf{Complete} & \textbf{HOLD recoveries} & \textbf{Human takeovers} \\
\midrule\endfirsthead
\multicolumn{8}{l}{\small\itshape Table \thetable\ (continued)}\\
\toprule
\textbf{Config.} & \textbf{Scenario} & \textbf{Unsafe attempts} & \textbf{Blocked} & \textbf{Hazardous commits} & \textbf{Complete} & \textbf{HOLD recoveries} & \textbf{Human takeovers} \\
\midrule\endhead
\midrule\multicolumn{8}{r}{\small Continued on next page}\\\endfoot
\bottomrule\endlastfoot
B1+ & Normal (24) & 0 & 0 & 0 & 24/24 & 0 & 0 \\
B1+ & Ambiguous (24) & 21 & 6 & 15 & 9/24 & 0 & 0 \\
B1+ & Drift (24) & 24 & 9 & 15 & 9/24 & 0 & 0 \\
B2 & Normal (24) & 0 & 0 & 0 & 24/24 & 0 & 0 \\
B2 & Ambiguous (24) & 21 & 21 & 0 & 24/24 & 0 & 0 \\
B2 & Drift (24) & 24 & 24 & 0 & 21/24 & 12 & 3 \\
\end{longtable}
\endgroup

\FloatBarrier
\begingroup
\fontsize{8.3}{10.4}\selectfont
\setlength{\parskip}{0pt}\setlength{\LTpre}{8pt}\setlength{\LTpost}{10pt}
\begin{longtable}{@{}P{62.36pt}P{49.89pt}P{41.58pt}P{33.26pt}P{45.73pt}P{45.73pt}P{41.58pt}P{45.73pt}P{49.89pt}@{}}
\caption{LLM-arm metrics; qwen3.6-35b-a3b-fp8, thinking off, 144 episodes.}\label{tab:t18}\\
\toprule
\textbf{Config.} & \textbf{Scenario} & \textbf{Unsafe attempts} & \textbf{Blocks} & \textbf{Hazardous commits} & \textbf{Complete} & \textbf{HOLD recoveries} & \textbf{Human takeovers} & \textbf{LLM tokens} \\
\midrule\endfirsthead
\multicolumn{9}{l}{\small\itshape Table \thetable\ (continued)}\\
\toprule
\textbf{Config.} & \textbf{Scenario} & \textbf{Unsafe attempts} & \textbf{Blocks} & \textbf{Hazardous commits} & \textbf{Complete} & \textbf{HOLD recoveries} & \textbf{Human takeovers} & \textbf{LLM tokens} \\
\midrule\endhead
\midrule\multicolumn{9}{r}{\small Continued on next page}\\\endfoot
\bottomrule\endlastfoot
B2-LLM & Normal (24) & 0 & 0 & 0 & 24/24 & 0 & 0 & 15,487 \\
B2-LLM & Ambiguous (24) & 37 & 37 & 0 & 13/24 & 0 & 11 & 69,461 \\
B2-LLM & Drift (24) & 27 & 27 & 0 & 17/24 & 0 & 7 & 44,553 \\
B1+-LLM & Normal (24) & 0 & 0 & 0 & 24/24 & 0 & 0 & 15,282 \\
B1+-LLM & Ambiguous (24) & 13 & 3 & 10 & 11/24 & 0 & 3 & 18,065 \\
B1+-LLM & Drift (24) & 15 & 9 & 6 & 17/24 & 0 & 1 & 23,559 \\
\end{longtable}
\endgroup

In typed interaction, deterministic B2 completes 69/72 episodes—24 normal, 24 ambiguous, and 21 drifting—versus 42/72 for B1+. LLM B2 completes 54/72—24, 13, and 17—versus 52/72 for B1+. B2 blocks all 64 registered unsafe LLM intents, a result defined by the harness and its rules rather than independent safety checking of actual artifacts. The LLM's eighteen unfinished episodes reach the six-turn limit and transfer to humans; subsequent human activity is not evaluated. The deterministic arm includes twelve HOLD recoveries. The LLM often supplies related facts in its first intent, so this batch has no naturally triggered HOLD-recovery count. Tables~\ref{tab:t17} and~\ref{tab:t18} give the details.

Additional forced-HOLD experiment A3 injects first-turn HOLD (\mbox{force\_}\allowbreak{}\mbox{first}) into eight episodes H01–H08, covering job freshness, checkpoint order, capability revalidation, and related defects. Eight episodes \ensuremath{\times} two configurations \ensuremath{\times} two seeds yield 32 LLM runs. Under B2, all fourteen H01–H07 runs recover within two turns through HOLD \ensuremath{\rightarrow} evidence \mbox{completion/}\allowbreak{}\mbox{revalidation} \ensuremath{\rightarrow} COMMITTED. Both H08 runs, involving a capability upgrade that cannot self-recover, correctly transfer to human control; no hazardous commits occur. Under B1+, all sixteen runs are \mbox{unsafe\_}\allowbreak{}\mbox{committed.} The same forced trajectories expose the escape surface under the two configurations. Records are in \mbox{agent\_}\allowbreak{}\mbox{records\_}\allowbreak{}\mbox{llm\_}\allowbreak{}\mbox{hold\_}\allowbreak{}\mbox{v1.}\allowbreak{}\mbox{jsonl.}

\subsubsection*{WP2: Actual Tool Execution and Commit-Linked Artifact Review}

After qualifying the cold-plate CAD–\allowbreak{}mesh and bracket CAD–\allowbreak{}mesh–\allowbreak{}CalculiX environments, \mbox{pair\_}\allowbreak{}\mbox{b1} runs thinking-off and thinking-on configurations of the same qwen3.6-35b-a3b-fp8 service. Both use temperature = 0.3, \mbox{max\_}\allowbreak{}\mbox{tokens} = 32768, identical tasks and seeds: N1–N4 normal, A1–A4 ambiguous, D1–D4 drifting, and U1–U4 unknown-recovery tasks, each with three seeds, totaling 96 task instances. The LLM proposes tools and parameters; adapters actually call CadQuery, Gmsh, and the bracket solver. Cold-plate tasks stop at CAD/meshing and do not include autonomous OpenFOAM calls. Recorded model identifiers are consistent in all 48 cases of each arm.

\FloatBarrier
\begingroup
\small
\setlength{\parskip}{0pt}\setlength{\LTpre}{8pt}\setlength{\LTpost}{10pt}
\begin{longtable}{@{}P{75.78pt}P{124.81pt}P{115.90pt}P{129.27pt}@{}}
\caption{\mbox{pair\_}\allowbreak{}\mbox{b1} task-completion script counts; twelve cases per category per arm, with asymmetric execution configurations.}\label{tab:t19}\\
\toprule
\textbf{Task category} & \textbf{Thinking-off strict completion} & \textbf{Thinking-on strict completion} & \textbf{Tasks per arm} \\
\midrule\endfirsthead
\multicolumn{4}{l}{\small\itshape Table \thetable\ (continued)}\\
\toprule
\textbf{Task category} & \textbf{Thinking-off strict completion} & \textbf{Thinking-on strict completion} & \textbf{Tasks per arm} \\
\midrule\endhead
\midrule\multicolumn{4}{r}{\small Continued on next page}\\\endfoot
\bottomrule\endlastfoot
Normal & 12 & 4 & 12 \\
Requirement ambiguity & 9 & 0 & 12 \\
\mbox{Activation/}\allowbreak{}\mbox{state} drift & 12 & 3 & 12 \\
Unknown recovery & 9 & 0 & 12 \\
Total & 42 & 7 & 48 \\
\end{longtable}
\endgroup

The completion script requires task completion registration, successful execution of every tool in \mbox{tools\_}\allowbreak{}\mbox{all}, and readback of preset postconditions. It reports 42/48 and 7/48. The artifact label strict denotes this conjunction, not every engineering invariant, and is not a safe-completion metric. Table~\ref{tab:t21} shows disagreement between commit checking and artifact facts. Category counts describe this execution batch, not a causal effect of enabling reasoning.

All six unfinished thinking-off cases reach the turn limit. The 41 unfinished thinking-on cases comprise 37 \mbox{parse\_}\allowbreak{}\mbox{fail}:\mbox{thinking\_}\allowbreak{}\mbox{only}, two exhausted shared worker budgets, and two turn limits; these are distinct causes. Response traces retain content, reasoning, \mbox{finish\_}\allowbreak{}\mbox{reason}, \mbox{response\_}\allowbreak{}\mbox{id}, and usage. Of 259 thinking-on calls, 149 have \mbox{finish\_}\allowbreak{}\mbox{reason} = length with empty content and reasoning only, which cannot be parsed into executable actions. All 304 thinking-off calls have \mbox{finish\_}\allowbreak{}\mbox{reason} = stop. This protocol exposure is directly inspectable for the given configuration, but does not establish that reasoning modes generally reduce intelligence or engineering capability.

Table~\ref{tab:t20} reports resources and execution exposure as event counts. Tool success is not task completion, and commit events are not independent tasks. Both runners self-report \mbox{dangerous\_}\allowbreak{}\mbox{commit} = 0, a flag reconciled separately with commit checking below. Rule-derived unsafe intents are blocked 60/60 and 15/15, respectively. These counts indicate blocking of triggered rules, not detection rates against an independent hazard Oracle.

\FloatBarrier
\begingroup
\small
\setlength{\parskip}{0pt}\setlength{\LTpre}{8pt}\setlength{\LTpost}{10pt}
\begin{longtable}{@{}P{65.96pt}P{83.55pt}P{96.75pt}P{96.75pt}P{96.75pt}@{}}
\caption{\mbox{pair\_}\allowbreak{}\mbox{b1} resources and execution events; 48 cases per arm, selected trajectories only.}\label{tab:t20}\\
\toprule
\textbf{Config.} & \textbf{LLM calls} & \textbf{Tool requests / successful returns} & \textbf{Commit events} & \textbf{\mbox{total\_}\allowbreak{}\mbox{tokens}} \\
\midrule\endfirsthead
\multicolumn{5}{l}{\small\itshape Table \thetable\ (continued)}\\
\toprule
\textbf{Config.} & \textbf{LLM calls} & \textbf{Tool requests / successful returns} & \textbf{Commit events} & \textbf{\mbox{total\_}\allowbreak{}\mbox{tokens}} \\
\midrule\endhead
\midrule\multicolumn{5}{r}{\small Continued on next page}\\\endfoot
\bottomrule\endlastfoot
Thinking off & 304 & 163 / 162 & 105 & 804,139 \\
Thinking on & 259 & 86 / 82 & 22 & 5,622,673 \\
\end{longtable}
\endgroup

The usage.\mbox{total\_}\allowbreak{}\mbox{tokens} field includes context inputs and generated outputs, not just reasoning tokens or monetary cost. The thinking-on total is approximately seven times the thinking-off total, but completion counts and failure paths differ, so this is not the extra cost of equal completed work. Selected trajectories contain one and four tool failures, respectively. Attempt-level logs additionally contain one engine exception and three API transport failures. Unique task–\allowbreak{}seed keys select trajectories; attempt events are not added again to task terminal counts.

This is not a controlled causal comparison of the reasoning switch. Thinking off uses one worker and thinking on three concurrent workers. The 9,000-second budget is shared across a batch, not assigned per task. The \mbox{wall\_}\allowbreak{}\mbox{s} field accumulates from worker startup and cannot represent individual task duration. The thinking-on w4 runner loads the completed-task set differently from other workers. Service contention, remaining budget, and retry policies provide alternative explanations for completion differences. Results are therefore descriptive counts for this service and execution configuration.

The 49 formal COMMIT events comprise 42 thinking-off and seven thinking-on events. The \mbox{commit\_}\allowbreak{}\mbox{audit} script independently reads digests, parameters, lineage, and sandbox boundaries rather than runner danger labels, reporting 48 passes and one failure. However, its formula treats relative hole translations as absolute distances from plate edges, so it cannot establish 48 engineering-compliant commits. All eighteen commit-linked bracket generation records contain \mbox{mounting\_}\allowbreak{}\mbox{interface\_}\allowbreak{}\mbox{preserved} = false, including seventeen labeled as passing by that script. All eighteen STEP digests match the existing files. This bracket-record check does not establish complete geometric compliance for other commits.

\FloatBarrier
\begingroup
\small
\setlength{\parskip}{0pt}\setlength{\LTpre}{8pt}\setlength{\LTpost}{10pt}
\begin{longtable}{@{}P{75.78pt}P{124.81pt}P{115.90pt}P{129.27pt}@{}}
\caption{Layered comparison of commit records and checking grounds; no aggregate compliance rate is calculated.}\label{tab:t21}\\
\toprule
\textbf{Evidence layer} & \textbf{Existing records} & \textbf{Supported interpretation} & \textbf{Not established} \\
\midrule\endfirsthead
\multicolumn{4}{l}{\small\itshape Table \thetable\ (continued)}\\
\toprule
\textbf{Evidence layer} & \textbf{Existing records} & \textbf{Supported interpretation} & \textbf{Not established} \\
\midrule\endhead
\midrule\multicolumn{4}{r}{\small Continued on next page}\\\endfoot
\bottomrule\endlastfoot
Formal commits & 49 (42 + 7) & Actual baseline-entry records in this batch & All engineering conditions satisfied \\
Commit-audit outputs & 48 pass; 1 fail & Script outputs with incorrect translation semantics & 48/49 engineering compliance \\
Bracket associations & 18 with interface preservation = false & Eighteen STEP digests match; seventeen script passes & Risk rate from independent full-geometry checking \\
U4 counterexample & Hole x \ensuremath{\approx} 120.5; plate length \ensuremath{\approx} 120 & Record-derived right clearance \ensuremath{\approx} \ensuremath{-}3.7 mm & Compliance of every other commit \\
\end{longtable}
\endgroup

\begin{figure}[H]\centering
\includegraphics[width=\linewidth,height=0.77\textheight,keepaspectratio]{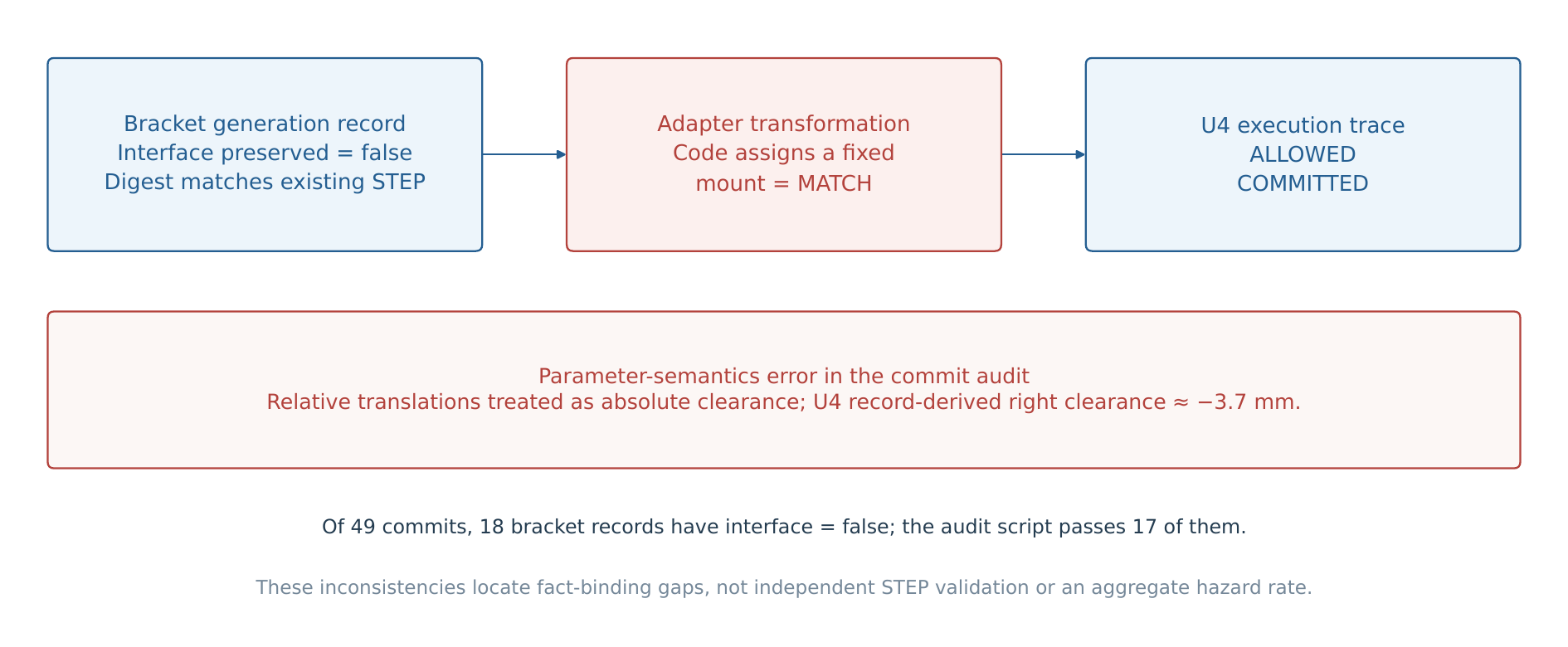}
\caption{Semantic breaks in bracket artifact adaptation and checking, reconstructed from existing code and records.}\label{fig:16}
\end{figure}

\mbox{U4\_}\allowbreak{}\mbox{\_}\allowbreak{}\mbox{s101} commits despite violating interface preservation. The generator adds \mbox{hole\_}\allowbreak{}\mbox{shift\_}\allowbreak{}\mbox{x/}\allowbreak{}\mbox{y} to nominal hole positions; a zero y translation does not place the hole center on an edge. Instead, the artifact's x translation of 15.5 mm moves a nominal right-hole center from 105 to 120.5 mm. With a plate length of approximately 120 mm and radius 3.2 mm, the record-derived signed right-edge clearance is 120 \ensuremath{-} 120.5 \ensuremath{-} 3.2 \ensuremath{\approx} \ensuremath{-}3.7 mm, violating the minimum 4 mm requirement. The audit formula min(15.5, 0) \ensuremath{-} 3.2 = \ensuremath{-}3.2 mm incorrectly treats translations as absolute distances. Although it also yields a negative value, it does not respect the generator's parameter definitions and cannot support passing judgments for other commits.

The gap spans two interfaces. U4's final plan actually contains \mbox{hole\_}\allowbreak{}\mbox{shift\_}\allowbreak{}\mbox{x/}\allowbreak{}\mbox{y}, so the issue is not simply missing agent hole parameters. Those translations are not converted into the hole table, radius, and plate-extent inputs required by the clearance predicate. Meanwhile, tool output already records \mbox{mounting\_}\allowbreak{}\mbox{interface\_}\allowbreak{}\mbox{preserved} = false, but the stored \mbox{cad\_}\allowbreak{}\mbox{bracket} adapter fixes mount = MATCH rather than forwarding that negative fact to difference evaluation. Tool parameters, rule inputs, and artifact facts are not consistently mapped. The eighteen records contain generation metadata, not independently parsed final STEP geometry. They contradict the consistency of the original audit's grounds, but do not justify new \mbox{hazardous/}\allowbreak{}\mbox{compliant} rates of 18/49 or 31/49.

BFM-1.1 replaces the historical fixed mount = MATCH path with mapping from existing artifact-role metadata, frozen geometric requirements, and provenance binding. The adapter and offline replay call the same pure mapping function. Its outputs enter existing rule interfaces through \mbox{binding\_}\allowbreak{}\mbox{changes}, unknowns, and controlled \mbox{measured\_}\allowbreak{}\mbox{facts.} The evaluation below tests fact transformation and local gates without rerunning original agent tasks. The mapping binds relative hole-shift semantics, artifact digests, source revisions, the four-hole requirement, minimum edge clearance, and three-state interface status before invoking the existing gate.

R2.1 type and empty-parameter guards address input form, not correspondence between geometric parameters and role predicates. BFM-1.1 adds required-hole-count checking, record-derived edge-clearance evaluation, and propagation of negative interface facts. Insufficient provenance or mandatory fields yield unknowns; confirmed violations are not swallowed by other unknowns. Rule compatibility, mapping contracts, and end-to-end tool execution remain different evaluation levels.

\paragraph{Evidence for the remaining commits.}
The two archived commit-audit files contain 49 events: 18 bracket events and 31 cold-plate events. The latter were not outside the audit: the script conditionally checks artifact digests, CAD--mesh lineage, envelope metadata, selected channel-count readback, and a wall-width calculation using actual call arguments. However, checks guarded by field or file availability can be skipped, and the per-event report records issues rather than a complete executed-check manifest. The available reports therefore do not establish that all 31 events underwent every listed check, independent geometric inspection, or exhaustive requirement coverage. This is a coverage limitation of the historical audit, not evidence that those 31 artifacts violate an engineering requirement.

\subsubsection*{Deterministic Local Regression of Bracket Fact Mapping}

BFM-1.1 reuses eighteen historically committed bracket records and one existing nominal control. Replay locates artifact JSON, STEP digests, and commit associations through an input manifest, calls the adapter's pure function, and passes its outputs through \mbox{r\_}\allowbreak{}\mbox{effect\_}\allowbreak{}\mbox{binding} and \mbox{r\_}\allowbreak{}\mbox{struct\_}\allowbreak{}\mbox{min\_}\allowbreak{}\mbox{edge\_}\allowbreak{}\mbox{distance} into the effect and composition rule families. Baseline activation is evaluated only in memory. No LLM is called, no geometry is generated, and no mesh or solver is run. Original task trajectories and commit records remain unchanged.

All eighteen abnormal records map to MISMATCH; local gates return FAIL without requesting activation. Fifteen have record-derived minimum clearance of \ensuremath{-}3.2 mm, and three U4 records have \ensuremath{-}3.7 mm. The nominal control has 11.8 mm, maps to MATCH, and passes the local gate. Provenance binding passes for all nineteen cases, with identical input hashes before and after execution. Clearances derive from generation-role metadata and frozen plate extents, not independent STEP measurement. COMMITTED for the nominal control denotes only the in-memory minimal gate result.

\FloatBarrier
\begingroup
\small
\setlength{\parskip}{0pt}\setlength{\LTpre}{8pt}\setlength{\LTpost}{10pt}
\begin{longtable}{@{}P{75.78pt}P{124.81pt}P{115.90pt}P{129.27pt}@{}}
\caption{Local regression and interface-contract results for bracket fact mapping.}\label{tab:t22}\\
\toprule
\textbf{Evaluation object} & \textbf{Size} & \textbf{Result} & \textbf{Interpretation} \\
\midrule\endfirsthead
\multicolumn{4}{l}{\small\itshape Table \thetable\ (continued)}\\
\toprule
\textbf{Evaluation object} & \textbf{Size} & \textbf{Result} & \textbf{Interpretation} \\
\midrule\endhead
\midrule\multicolumn{4}{r}{\small Continued on next page}\\\endfoot
\bottomrule\endlastfoot
Historical abnormal records & 18 & All MISMATCH; no activation eligibility & Existing negative facts reach local rules; historical commits unchanged \\
Existing nominal control & 1 & MATCH; 11.8 mm clearance; local gate allows & Mounting-interface subcheck, not full-toolchain compliance \\
Fixed interface contracts & C01–C12 & All pass; formal failure list empty & Required-hole count and coexisting V/U; no new engineering artifacts \\
Simulated bad mappings & 3 classes & All contract violations detected & Fixed MATCH; omitted hole-count check; invalid fields swallowing violations \\
\end{longtable}
\endgroup

Contract tests cover conflict between negative facts and plan declarations, \mbox{missing/}\allowbreak{}\mbox{invalid} fields, \mbox{provenance/}\allowbreak{}\mbox{version} mismatch, threshold boundaries, determinism, required hole count, and coexisting V/U. Negative self-tests simulate three incorrect mapping outputs in memory and check their contract violations; they are neither new agent trials nor independent geometry experiments. Replay acceptance checks semantic decisions, provenance binding, and input hashes together. Formal and negative-self-test reports are stored separately. The evidence supports fact mapping and local gating for specified artifacts and contract conditions, not completeness of mandatory checks for arbitrary objects.

\subsection{Experiment C: Semantics-Preserving Transformations and Nearby Counterexamples}

Experiment C constructs 100 variants across ten representation transformations. Authors review them against semantic criteria independent of runner outputs, but without third-party personnel separation. An implementation lacking action-order dependencies and evidence of equivalence accepts V-013: moving geometry modification after solving leaves a parseable sequence but changes execution semantics. Five further variants are accepted before equivalence is established. Same-set regression after adding these conditions yields 92 ACCEPT, five HOLD, and three REJECTED outcomes (V-013/023/083). HOLD denotes unknown equivalence; the three rejections are confirmed execution-order violations. This tests rule coverage of identified problems, not generalization to unseen variants.

\FloatBarrier
\begingroup
\small
\setlength{\parskip}{0pt}\setlength{\LTpre}{8pt}\setlength{\LTpost}{10pt}
\begin{longtable}{@{}P{112.94pt}P{167.15pt}P{171.67pt}@{}}
\caption{Experiment C transformation categories and terminal states; ten tasks \ensuremath{\times} ten transformations, separating confirmed equivalence, unknown equivalence, and execution-order violations.}\label{tab:t23}\\
\toprule
\textbf{Transformation} & \textbf{ACCEPT} & \textbf{REJECTED / HOLD / REBASE} \\
\midrule\endfirsthead
\multicolumn{3}{l}{\small\itshape Table \thetable\ (continued)}\\
\toprule
\textbf{Transformation} & \textbf{ACCEPT} & \textbf{REJECTED / HOLD / REBASE} \\
\midrule\endhead
\midrule\multicolumn{3}{r}{\small Continued on next page}\\\endfoot
\bottomrule\endlastfoot
Equivalent units (mm/m) & 10/10 & 0 / 0 / 0 \\
Renaming & 10/10 & 0 / 0 / 0 \\
CAD modification execution order & 7/10; three order violations rejected & 3 / 0 / 0 \\
Equivalent role identifiers & 10/10 & 0 / 0 / 0 \\
Action reordering & 10/10 & 0 / 0 / 0 \\
Equivalent mesh metadata & 10/10 & 0 / 0 / 0 \\
Lineage wrapper nodes & 10/10 & 0 / 0 / 0 \\
Patch-version upgrade & 8/10; two equivalence-evidence gaps HOLD & 0 / 2 / 0 \\
Field order & 10/10 & 0 / 0 / 0 \\
Commutative post-processing & 7/10; three equivalence-evidence gaps HOLD & 0 / 3 / 0 \\
\end{longtable}
\endgroup

\subsection{Experiment D: Transfer to Mechanical-Bracket Structural Analysis}

The second domain reconnects a full toolchain: a CadQuery parametric bracket with a 120 \ensuremath{\times} 80 mm plate, four Ø6.4 mounting holes, and a 24 \ensuremath{\times} 24 mm load-bearing boss; a Gmsh C3D10 tetrahedral mesh; CalculiX ccx 2.21 static analysis, installed locally in WSL2 without root; and independent extraction of six criteria: von Mises stress below yield, displacement within limits, mass budget, mounting-interface predicates, Mount \mbox{reaction/}\allowbreak{}\mbox{external-}\allowbreak{}\mbox{load} balance within 2\%, and evidence digest lineage. Frozen requirements are LC1 at 2,000 N with displacement \ensuremath{\leq} 0.5 mm, LC2 at 3,000 N with displacement \ensuremath{\leq} 0.8 mm, a 900 g mass budget, and STEEL/AL6061 materials. The harness runtime and blind-run protocol are unchanged, corroborated by timestamps and import-only access. Two interoperability corrections are fixed: the last two midside nodes are swapped when exporting Gmsh tet10 as ccx C3D10, established through 17,280 element-variant comparisons; and FRD reactions use the FORC block, whose all-node sum is zero, requiring Mount-set partitioning for balance checks.

\FloatBarrier
\begingroup
\fontsize{8.3}{10.4}\selectfont
\setlength{\parskip}{0pt}\setlength{\LTpre}{8pt}\setlength{\LTpost}{10pt}
\begin{longtable}{@{}P{64.16pt}P{55.61pt}P{55.61pt}P{64.16pt}P{59.89pt}P{51.33pt}P{77.00pt}@{}}
\caption{Experiment D reference toolchain cases and six-criterion judgments; \mbox{records/}\allowbreak{}\mbox{gold\_}\allowbreak{}\mbox{records\_}\allowbreak{}\mbox{d.}\allowbreak{}\mbox{jsonl.}}\label{tab:t24}\\
\toprule
\textbf{Record} & \textbf{Change} & \textbf{von Mises (MPa)} & \textbf{Displacement (mm)} & \textbf{Mass (g)} & \textbf{Reaction imbalance} & \textbf{Judgment} \\
\midrule\endfirsthead
\multicolumn{7}{l}{\small\itshape Table \thetable\ (continued)}\\
\toprule
\textbf{Record} & \textbf{Change} & \textbf{von Mises (MPa)} & \textbf{Displacement (mm)} & \textbf{Mass (g)} & \textbf{Reaction imbalance} & \textbf{Judgment} \\
\midrule\endhead
\midrule\multicolumn{7}{r}{\small Continued on next page}\\\endfoot
\bottomrule\endlastfoot
G1 nominal steel t8 & None & 21.17 & 6.6e-5 & 603.8 & 0.0\% & PASS: all six criteria \\
G2 thickness t10 & Increased thickness & 21.22 & 6.4e-5 & 752.5 & 0.0\% & PASS \\
G3 opening 40 \ensuremath{\times} 30 & Opening & 81.31 & 1.1e-3 & 547.3 & 0.0\% & PASS \\
G4 AL6061 & Material & 21.07; yield 240 & 2.2e-4; \ensuremath{\approx}3.3\ensuremath{\times} & 207.7; \ensuremath{\approx}1/2.9 & 0.0\% & PASS \\
G5 stale mesh reuse & Injected fault & — & — & — & — & Rejected: lineage digest mismatch \\
G6 t12 over budget & Negative case & 21.19 & 6.1e-5 & 901.2 > 900 & 0.0\% & Rejected: mass constraint \\
\end{longtable}
\endgroup

The blind-run suite reuses evaluate() and TierPolicy and adds six domain predicates. It contains 48 cases, six per T1–T8: 25 hazardous and 23 compliant. T7 has four hazardous and two compliant cases, including use DENIED and B/C/A-drift injections; every other family has three hazardous and three compliant. Five variants yield 240 rows, all satisfying their assertions. Hazardous commits are B0 25/25, B1 23/25, B1+ 20/25, B-Std 20/25, and B2 0/25. The layered reuse audit (Table~\ref{tab:t25}, \mbox{reuse\_}\allowbreak{}\mbox{audit\_}\allowbreak{}\mbox{v2}) counts fifteen distinct reused rules—five core and ten blind-run—versus six new predicates, with 33 reused and thirteen new case-level rule references. Deployable new-domain code totals 748 lines (606 toolchain, ninety domain rules, fifty-two decision graph), versus 1,440 reused runtime lines (1,300 harness, 140 blind-run rules), a ratio of 51.9\%. Including 1,145 lines of new-domain scaffolding for generation, blind runs, verification, reference paths, and audit yields 1,893 lines, an upper-bound ratio of 131.5\%. No state-machine, rule-engine, or evidence-chain code is added; T1/T4/T5/T7 add no rules.

\FloatBarrier
\begingroup
\small
\setlength{\parskip}{0pt}\setlength{\LTpre}{8pt}\setlength{\LTpost}{10pt}
\begin{longtable}{@{}P{65.96pt}P{83.55pt}P{96.75pt}P{96.75pt}P{96.75pt}@{}}
\caption{Mechanism-reuse audit by T1–T8 family; \mbox{records\_}\allowbreak{}\mbox{d/}\allowbreak{}\mbox{reuse\_}\allowbreak{}\mbox{audit\_}\allowbreak{}\mbox{v2.}\allowbreak{}\mbox{json.}}\label{tab:t25}\\
\toprule
\textbf{Family} & \textbf{Cases (hazardous)} & \textbf{Reused rule references} & \textbf{New rule references} & \textbf{Extension} \\
\midrule\endfirsthead
\multicolumn{5}{l}{\small\itshape Table \thetable\ (continued)}\\
\toprule
\textbf{Family} & \textbf{Cases (hazardous)} & \textbf{Reused rule references} & \textbf{New rule references} & \textbf{Extension} \\
\midrule\endhead
\midrule\multicolumn{5}{r}{\small Continued on next page}\\\endfoot
\bottomrule\endlastfoot
T1 identity & 6 (3) & 6 & 0 & Direct \mbox{core/}\allowbreak{}\mbox{blind-}\allowbreak{}\mbox{rule} reuse \\
T2 planning & 6 (3) & 2 & 6 & Structural-\mbox{obligation/}\allowbreak{}\mbox{invariant} schema constants; reused scope rules \\
T3 composition & 6 (3) & 6 & 1 & One physical invariant: hole-edge clearance \ensuremath{\geq} 4 mm \\
T4 conditions & 6 (3) & 6 & 0 & Reused condition-\mbox{label/}\allowbreak{}\mbox{unit-}\allowbreak{}\mbox{schema} rules \\
T5 binding & 6 (3) & 6 & 0 & Reused binding logic; geometric role predicates rewritten \\
T6 evidence & 6 (3) & 4 & 2 & One domain predicate: stress-evidence digest lineage \\
T7 use & 6 (4) & 1 & 0 & Reused use DENIED and drift gates \\
T8 recovery & 6 (3) & 2 & 4 & Two predicates: checkpoint three-state propagation and recomputation scope \\
Total & 48 (25) & 33 & 13 & Fifteen reused vs six new rules; deployable-domain/runtime ratio 51.9\%, or 131.5\% including scaffolding \\
\end{longtable}
\endgroup

A4 runs CalculiX on 5.0/3.0/1.8 mm meshes, obtaining p99 stresses of 4.82/3.70/2.62 MPa. The six listed criteria retain their judgments across this range, but the medium–\allowbreak{}fine p99 difference is approximately 41\%, and Richardson extrapolation near 1.0 MPa differs substantially from the 2.62 MPa fine result. Neither mesh convergence nor controlled error is established. The Mount reaction of approximately 1,999.99998 N balances the 2,000 N load, supporting that balance check only. Very small displacement and local stress remain sensitive to mounting constraints, load application, and geometric singularities. Agreement with a broad hand-calculated interval is not sufficient V\&V. Kernel versions and frozen artifacts document reuse, not physical validity.

\subsubsection*{WP3: Bracket Artifact Assembly and Fact-Patch Re-evaluation}

WP3 maps fifteen predefined bracket cases to CadQuery–\allowbreak{}Gmsh–\allowbreak{}CalculiX, using six geometries and two load cases for twelve solves. At t0, frozen plans, uses, and recovery conditions are assembled with real tool digests and other records; all fifteen decisions match frozen expectations. ST-11 reports 2.8 mm hole-edge clearance below 4 mm under its bound role predicate. This supports conformance for fixed inputs and bound predicates, not coverage of every autonomously generated agent parameter. At t1, \mbox{apply\_}\allowbreak{}\mbox{patch} directly changes input facts and invokes the same evaluator; no CAD, meshing, or solver execution occurs between t0 and t1. Table~\ref{tab:t26} preserves original decisions; Table~\ref{tab:t27} traces patches and their evidential level.

\FloatBarrier
\begingroup
\fontsize{8.3}{10.4}\selectfont
\setlength{\parskip}{0pt}\setlength{\LTpre}{8pt}\setlength{\LTpost}{10pt}
\begin{longtable}{@{}P{42.18pt}P{42.18pt}P{54.83pt}P{71.70pt}P{59.05pt}P{59.05pt}P{46.39pt}P{46.39pt}@{}}
\caption{Existing WP3 decision records for fifteen tasks mapped from ECE 1 bracket cases: t0 follows tool-record assembly; t1 re-evaluates patched facts and does not establish completed physical repair.}\label{tab:t26}\\
\toprule
\textbf{WP3} & \textbf{ST source} & \textbf{Category} & \textbf{Physical scenario} & \textbf{t0 decision} & \textbf{t1 input re-evaluation} & \textbf{t0 expected match} & \textbf{Script checklist} \\
\midrule\endfirsthead
\multicolumn{8}{l}{\small\itshape Table \thetable\ (continued)}\\
\toprule
\textbf{WP3} & \textbf{ST source} & \textbf{Category} & \textbf{Physical scenario} & \textbf{t0 decision} & \textbf{t1 input re-evaluation} & \textbf{t0 expected match} & \textbf{Script checklist} \\
\midrule\endhead
\midrule\multicolumn{8}{r}{\small Continued on next page}\\\endfoot
\bottomrule\endlastfoot
W3-01 & ST-01 & Compliant & Nominal t8 steel & ALLOWED & ALLOWED & Yes & Yes \\
W3-02 & ST-02 & Compliant & AL6061 & ALLOWED & ALLOWED & Yes & Yes \\
W3-03 & ST-04 & Compliant & 40 \ensuremath{\times} 30 opening & ALLOWED & ALLOWED & Yes & Yes \\
W3-04 & ST-06 & Compliant & Thickness t10 & ALLOWED & ALLOWED & Yes & Yes \\
W3-05 & ST-09 & Hazardous & Nominal & REJECTED & ALLOWED & Yes & Yes \\
W3-06 & ST-10 & Hazardous & Hole translation & REJECTED & ALLOWED & Yes & Yes \\
W3-07 & ST-11 & Hazardous & Hole translation & REJECTED & ALLOWED & Yes & Yes \\
W3-08 & ST-12 & Hazardous & Nominal & REJECTED & ALLOWED & Yes & Yes \\
W3-09 & ST-13 & Hazardous & Nominal + stale mesh & REJECTED & ALLOWED & Yes & Yes \\
W3-10 & ST-14 & Hazardous & Aluminum & REJECTED & ALLOWED & Yes & Yes \\
W3-11 & ST-16 & Hazardous & Nominal & DENIED & ALLOWED & Yes & Yes \\
W3-12 & ST-17 & Mandatory unknown & Nominal + old digest & HOLD & ALLOWED & Yes & Yes \\
W3-13 & ST-18 & Mandatory unknown & Nominal & HOLD & ALLOWED & Yes & Yes \\
W3-14 & ST-21 & Composite & Nominal + old digest & REJECTED & ALLOWED & Yes & Yes \\
W3-15 & ST-22 & Composite & Hole translation + aluminum & REJECTED & ALLOWED & Yes & Yes \\
\end{longtable}
\endgroup

\FloatBarrier
\begingroup
\small
\setlength{\parskip}{0pt}\setlength{\LTpre}{8pt}\setlength{\LTpost}{10pt}
\begin{longtable}{@{}P{75.78pt}P{124.81pt}P{115.90pt}P{129.27pt}@{}}
\caption{WP3 t1 input changes and the scope of their evidential interpretation.}\label{tab:t27}\\
\toprule
\textbf{Task} & \textbf{t1 input change} & \textbf{Artifact-update grounds} & \textbf{Interpretation of state change} \\
\midrule\endfirsthead
\multicolumn{4}{l}{\small\itshape Table \thetable\ (continued)}\\
\toprule
\textbf{Task} & \textbf{t1 input change} & \textbf{Artifact-update grounds} & \textbf{Interpretation of state change} \\
\midrule\endhead
\midrule\multicolumn{4}{r}{\small Continued on next page}\\\endfoot
\bottomrule\endlastfoot
W3-01–04 & Empty patch; same facts re-evaluated & Existing tool artifacts reused & Four ALLOWED outcomes retained; same-input stability only \\
W3-05 & Register \mbox{LC2\_}\allowbreak{}\mbox{proof}:\mbox{mass\_}\allowbreak{}\mbox{g} obligation & No extra solve/evidence acquisition & Conditional decision after plan declaration completion, not completed evidence \\
W3-06,07 & Register \mbox{min\_}\allowbreak{}\mbox{edge\_}\allowbreak{}\mbox{distance}; replace planned holes with nominal positions & Physical records remain 2.8 mm and interface false; no repaired geometry & ALLOWED evaluates substituted inputs, not repaired original artifacts \\
W3-08 & Rewrite action order as CAD \ensuremath{\rightarrow} mesh \ensuremath{\rightarrow} LC1 solve \ensuremath{\rightarrow} post-process & Sequence not re-executed & Checks updated plan, not historical execution order \\
W3-09,10 & Set checkpoint to mesh:B1; mark all required recomputation items & No new checkpoint, mesh, or solve & Changed declared recovery premises/scope, not completed recovery \\
W3-11 & Narrow requested use from \mbox{fatigue\_}\allowbreak{}\mbox{LCF} to \mbox{LC2\_}\allowbreak{}\mbox{proof} & Existing artifact and allowed-use scope retained & New request can remove original use violation without expanding asset permissions \\
W3-12,13 & Set field digest to current mesh digest; or job to COMPLETED with \mbox{freshness\_}\allowbreak{}\mbox{verified} = true & No new field-provenance check or job query & Conditional \mbox{lineage/}\allowbreak{}\mbox{freshness} re-evaluation, not new tool-side proof \\
W3-14 & Change field digest and add obligation & No new solve/evidence & Composite declaration patch, not artifact-side evidence closure \\
W3-15 & Mark recomputation complete; substitute nominal planned holes & Translated aluminum artifact still has 2.8 mm clearance and interface false & Original \mbox{ALLOWED/}\allowbreak{}\mbox{COMMITTED} retained, not interpreted as actual \mbox{repair/}\allowbreak{}\mbox{admission} success \\
\end{longtable}
\endgroup

All fifteen t1 outcomes are ALLOWED: four re-evaluations use empty patches and eleven use nonempty patches. This is an existing script output, not fifteen successful physical repairs. In W3-06, W3-07, and W3-15, physical records still contain 2.8 mm edge clearance and \mbox{mounting\_}\allowbreak{}\mbox{interface\_}\allowbreak{}\mbox{preserved} = false, while \mbox{apply\_}\allowbreak{}\mbox{patch} substitutes nominal hole positions in the plan. The evaluator's facts no longer correspond to the original artifact's invariants. This exposes a recovery provenance requirement: substituting a favorable declaration cannot discharge an existing confirmed violation in an unchanged artifact.

COMMITTED and \mbox{unsafe\_}\allowbreak{}\mbox{commit} at t1 are script outputs. The \mbox{unsafe\_}\allowbreak{}\mbox{commit} flag combines the task's Oracle category with baseline entry; the category is not redefined after changing a use request, so these flags cannot all be counted as a new engineering hazard rate. For hole-position tasks, unchanged physical negative records explicitly prevent interpreting declaration patches as actual recovery. Passing a script checklist is not artifact-repair verification.

Actual recovery requires rebinding changed engineering objects, tool execution records, new evidence, and their provenance to the current snapshot. Completed plan obligations or narrowed uses may constitute new request facts, but geometric repair, completed recomputation, corrected lineage, and job freshness require corresponding artifacts or authoritative observations. WP3 t0 instances and t1 conditional re-evaluations are reported separately; the latter do not establish an implemented T8 physical-repair loop or external validity.

\subsection{Experiment E: Persistence Replacement and Three-Service Decomposition}

SQLite is replaced by PostgreSQL 18.6, installed without root in WSL2. Policy evaluation, provenance queries, and object storage become three services communicating through JSON over HTTP. Policy evaluation reuses \mbox{r\_}\allowbreak{}\mbox{use\_}\allowbreak{}\mbox{control} and \mbox{effect\_}\allowbreak{}\mbox{family\_}\allowbreak{}\mbox{check} from \mbox{rules\_}\allowbreak{}\mbox{control.} Object storage implements baseline CAS using a single CTE statement for B/C/A guards, state transition, and audit logging, with row locks scoped to that statement; it also maintains content-addressed objects, the idempotent tool gateway, and lineage edges. Loads use 1/8/32/64 clients with 1,000 logical operations each, totaling 105,000 operations. The concurrency benchmark explicitly uses \mbox{synchronous\_}\allowbreak{}\mbox{commit} = off. The following paragraph reports the principal latency, throughput, retry, and decomposition measurements; Table~\ref{tab:t29} reports semantic assertions.

Throughput rises from 309 ops/s at one client to 518 at eight, then saturates near 270–300 at 32/64. Limits arise from a single-writer baseline CAS and optimistic retries on hot counters. P99 reaches 1.4/3.2 s at 32/64 clients, with retry convoying averaging thirteen to fifteen attempts per operation. At 64 clients, mean decomposition is 219.7 ms control-kernel time, 210.8 ms database time, and 5.8 ms network time. Arithmetic means are dominated by retry convoys and greatly exceed P50. Five-millisecond lock-wait sampling reports zero at every level: submillisecond statement locks turn contention into optimistic retries rather than sampled waits. These results locate scalability limits rather than establish a throughput claim.

\FloatBarrier
\begingroup
\small
\setlength{\parskip}{0pt}\setlength{\LTpre}{8pt}\setlength{\LTpost}{10pt}
\begin{longtable}{@{}P{112.94pt}P{167.15pt}P{171.67pt}@{}}
\caption{Semantic assertions for Experiment E controlled endpoints under the listed loads.}\label{tab:t29}\\
\toprule
\textbf{Assertion} & \textbf{Size} & \textbf{Result} \\
\midrule\endfirsthead
\multicolumn{3}{l}{\small\itshape Table \thetable\ (continued)}\\
\toprule
\textbf{Assertion} & \textbf{Size} & \textbf{Result} \\
\midrule\endhead
\midrule\multicolumn{3}{r}{\small Continued on next page}\\\endfoot
\bottomrule\endlastfoot
B/C/A drift before activation prevents COMMITTED & 16,481 cases & All \mbox{REBASE\_}\allowbreak{}\mbox{REQUIRED}; zero COMMITTED \\
Activation before drift avoids unnecessary REBASE & 20,954 cases & All terminal COMMITTED; 14,035 transient-conflict retries counted separately \\
One tool side effect per idempotency key & 6,220 keys & 6,220 unique side-effect sequence numbers, including transparent transport retries \\
No lost optimistic-counter updates & 21,059 operations & Final value equals operation count \\
Out-of-scope use DENY / lineage matching & 12,477 / 20,954 cases & All DENY / all match \\
\end{longtable}
\endgroup

Additional persistence records compare sync on/off. With synchronous commit, P50 across the four loads is 4.06/9.66/26.37/28.16 ms and P99 is 27.7/208.8/2,094.3/4,438.8 ms. In one sync-on trial combining 8,000 operations with immediate shutdown, acknowledged counts match final database values, and eight affected operations recover within two retries. This covers only the controlled service path. Sync off produces duplicate acknowledgments numbered 620–629, ten duplicate EXECUTED idempotency records, and twelve commit-state disagreements, demonstrating possible loss of acknowledged records within the durability window. Synchronous commit is one necessary configuration for resisting this database acknowledgment-loss failure in this implementation, not a sufficient condition for exactly-once effects in arbitrary external tools. Added latency, brief service outages, and legitimate rebasing likewise evaluate only their local endpoints.

Scope and limitations: \mbox{synchronous\_}\allowbreak{}\mbox{commit} = off is used only for the concurrency benchmark, and its crash-durability window is measured by the B2 crash trial above. PostgreSQL is a single WSL2 instance; multinode sharding and replication are untested. Drivers and services share one host with localhost forwarding; wide-area latency is untested.

\subsubsection*{WP4: Persistence Fault Injection with Synchronous Commit}

WP4 uses \mbox{synchronous\_}\allowbreak{}\mbox{commit} = on. At 1/8/32/64 clients and 200 operations per client, it executes 21,000 operations. Listed baseline assertions report no semantic errors. The 4,257 counter operations have no lost updates, and 3,724 idempotency keys retain single-execution records within the controlled endpoint model. P50 across the four loads is 4.06/9.66/26.37/28.16 ms and P99 is 27.7/208.8/2,094.3/4,438.8 ms. Table~\ref{tab:t31} reports reconciliation and acknowledged-state readback for five fault classes. F1/F2 recover in approximately 4.5/6.2 seconds, with \mbox{sixteen/}\allowbreak{}\mbox{twenty-}\allowbreak{}\mbox{four} transport errors. Timeout retries apply only to idempotency-safe endpoints; non-idempotent operations cannot be blindly replayed. F5 retains forty expired-result isolation and ten completion-race records. These results do not cover independent-node partitions, arbitrary external CAD/PLM effects, or global exactly-once execution; one local sync-on experiment is not a universal durability guarantee.

\FloatBarrier
\begingroup
\fontsize{8.3}{10.4}\selectfont
\setlength{\parskip}{0pt}\setlength{\LTpre}{8pt}\setlength{\LTpost}{10pt}
\begin{longtable}{@{}P{25.31pt}P{118.09pt}P{37.96pt}P{42.18pt}P{42.18pt}P{42.18pt}P{54.83pt}P{59.05pt}@{}}
\caption{Five WP4 persistence fault injections under sync on; readback counts individually successful rereads of client-acknowledged COMMITTED records.}\label{tab:t31}\\
\toprule
\textbf{Case} & \textbf{Injection} & \textbf{Recover (s)} & \textbf{I/O errors} & \textbf{Semantic errors} & \textbf{Reconc. gaps} & \textbf{Durable readback} & \textbf{Counter (acknowledged)} \\
\midrule\endfirsthead
\multicolumn{8}{l}{\small\itshape Table \thetable\ (continued)}\\
\toprule
\textbf{Case} & \textbf{Injection} & \textbf{Recover (s)} & \textbf{I/O errors} & \textbf{Semantic errors} & \textbf{Reconc. gaps} & \textbf{Durable readback} & \textbf{Counter (acknowledged)} \\
\midrule\endhead
\midrule\multicolumn{8}{r}{\small Continued on next page}\\\endfoot
\bottomrule\endlastfoot
F1 & Kill/restart store process & 4.502 & 16 & 0 & 0 & OK (283) & 303/303 \\
F2 & Immediate DB \mbox{shutdown/}\allowbreak{}\mbox{restart} & 6.156 & 24 & 0 & 0 & OK (282) & 301/301 \\
F3 & Client 50 ms timeout/retry; idempotent endpoints only & — & 0 & 0 & 0 & OK (139) & 144/144 \\
F4 & Duplicate \mbox{tool\_}\allowbreak{}\mbox{once/}\allowbreak{}\mbox{activate} requests & — & 0 & 0 & 0 & OK (298) & 303/303 \\
F5 & Token-\mbox{expiry/}\allowbreak{}\mbox{job-}\allowbreak{}\mbox{completion} race & — & 0 & 0 & 0 & OK (10) & 0/0 \\
\end{longtable}
\endgroup

\subsection{Fact Binding Across Actual Execution and Recovery}\label{sec:execution_binding}

This additional evaluation asks whether the fact-binding requirements of Section 7.13 survive actual tool execution and cross-module consumption. It uses the separate \mbox{exec\_}\allowbreak{}\mbox{chain\_}\allowbreak{}\mbox{v1} prototype and does not replace historical WP2 or WP3 records. Three experiment families distinguish actual CAD invocation and test-baseline activation, recovery re-evaluation, and adapter transfer. A contract-based challenge set subsequently probes the same paths. No new thermal or structural validation is claimed by these experiments.

\paragraph{Implementation identity and evidence endpoints.}
The implementations below are separate evidence sources, not a single before/after system upgrade. A positive result in the additional prototype does not establish what a rerun of historical pair\_b1 would do.
\begin{center}
\small
\begin{tabularx}{\linewidth}{@{}P{83pt}YYY@{}}
\toprule
Implementation & Execution input & Observed endpoint & Relation to other evidence \\
\midrule
Historical WP2/WP3 & Agent/tool trajectories; fixed artifacts and patched evaluation inputs & Recorded calls, historical commits, or re-evaluation decisions, according to the experiment & Immutable historical observations; WP3 t1 is not artifact reconstruction \\
BFM-1.1 & Existing bracket metadata and targeted contracts & Fact mapping and local gating & Offline regression; neither agent rerun nor STEP validation \\
\mbox{exec\_}\allowbreak{}\mbox{chain\_}\allowbreak{}\mbox{v1} & Actual CAD calls and selected fact-boundary interventions & Test-baseline readback on submission paths; registry decisions on recovery paths & Separate prototype; not pair\_b1 with a replacement adapter \\
V4 & Thirty sealed contract tasks against the additional execution paths & Test activation or recovery permission, separately recorded & First run and same-set repair regression remain distinct \\
\bottomrule
\end{tabularx}
\end{center}

The execution environment is Python 3.12.14, CadQuery 2.8.0, and the associated OCCT geometry kernel. The bracket and cold-plate paths build STEP artifacts through tool adapters, obtain fact mappings, evaluate gates, and, where specified, call an in-memory \mbox{Baseline}\allowbreak{}\mbox{Service}. Activation observations are test-service readbacks, not commercial PLM commits or crash-durable transactions. Recovery experiments instead use an in-memory artifact registry and a \mbox{Recovery}\allowbreak{}\mbox{Gate}. They generate or register changed artifacts but do not invoke the baseline activation interface. A recovery permission is therefore reported as a decision only.

\subsubsection{Actual CAD Calls and Adapter-Fact Transmission}

The deterministic bracket experiment contains twelve cases: two nominal, two geometric violations, two unavailable-fact cases, two hole-count cases, two coexisting violation/unknown cases, and two declaration/artifact conflicts. Every case begins with actual CAD construction. Eight cases additionally alter the fact-reading boundary or submitted claims; they are boundary-injection tests rather than eight independently malformed geometries. The recorded outcomes are two ALLOWED/COMMITTED, seven FAIL without activation, and three HOLD without activation. Missing observations do not imply invalid geometry: an independently read STEP may satisfy the geometric checks while the controlled path remains on HOLD because its required observations are unavailable.

A parameter-matched old/new mapping comparison uses two violating settings. Both old fixed-MATCH runs enter their isolated test baselines; both repaired-mapping runs return FAIL without activation. The recorded artifact digests differ between comparison runs, so the comparison is not presented as byte-identical replay of one artifact. It supports a local behavioral contrast under matching construction parameters and the same downstream gate, not complete control of all possible tool-side differences.

The separate STEP checker imports the exported file and reads cylindrical faces without consulting the generator's interface-preservation flag or calling the mapping function under test. It evaluates a four-hole requirement and clearance relative to the specified plate boundary. This is a separate inspection path using the same CadQuery/OCCT kernel, not an independent geometry implementation or product-certification procedure. Its reported distances are not pooled with metadata-derived distances: partial cylindrical faces and holes crossing the plate boundary can yield different geometric and analytic descriptions. The two valid cases use different construction parameters: E1-K1-a has zero additional hole translation and a checker-reported clearance of 11.8 mm; E1-K1-b has \mbox{hole\_}\allowbreak{}\mbox{shift\_}\allowbreak{}\mbox{x=}\allowbreak{}\mbox{1} mm and \mbox{hole\_}\allowbreak{}\mbox{shift\_}\allowbreak{}\mbox{y=}\allowbreak{}\mbox{0.}\allowbreak{}\mbox{5} mm and a clearance of 10.8 mm. These are different geometries, not conflicting measurements of the same artifact. The violating cases have negative checker-reported clearances. Agreement on rejection does not establish accuracy of every reported distance outside the intended hole family.

The separate agent experiment uses qwen3.6-35b-a3b-fp8 with thinking disabled, temperature 0.3, seed 20261007, a 2,048-token response budget, and one worker. The model actually invokes \mbox{cad\_}\allowbreak{}\mbox{bracket}; submission uses the same gate as the deterministic path. Of eight task records, four have \mbox{fulfilled=}\allowbreak{}\mbox{true}: EA-01, EA-02, EA-07, and EA-08. The other four terminate at the turn limit: two repeatedly attempt violating submissions and two lack the required observations. No tool errors are recorded. Thus task completion is 4/8, not a safety-success rate. EA-07 supplies a positive sequence: an initial missing-fact HOLD, a violating-artifact FAIL, actual reconstruction, and a test-baseline COMMIT. The records contain no flagged hazardous commits in this small batch; this is not a statistically supported bound on agent risk or a reasoning-mode comparison.

\subsubsection{Recovery Decisions and Adapter Transfer}

The recovery module tests six scenario families with two instances each: plan-only changes, favorable declarations, substituted metadata, actual reconstruction, unavailable new facts, and artifact change between evaluation stages. Requests cannot replace the registry's current facts. The twelve recorded outcomes are eight REJECTED, two HOLD, and two ALLOWED. Both positive cases reconstruct an artifact and retain the earlier violation history. Artifact identity is a composite of object reference, revision, and digest: some fixtures reuse the local revision label B1 while the content digest changes. These records do not demonstrate enterprise-wide monotonic revision allocation.

A four-configuration comparison examines the original consumer, mapping repair alone, recovery binding alone, and both together on three selected cases. The original configuration permits all three incorrect recoveries. Mapping repair alone blocks plan and declaration replacement but not the substituted-metadata case. The recovery-binding configurations reject all three. These are localized implementation variants; the binding-only path has its own access to registered facts, so the experiment does not prove that either layer can generally compensate for an incorrect upstream adapter.

The cold-plate adapter experiment contains twelve cases covering nominal geometry, geometric violations, missing facts, favorable claims about a violating artifact, mismatched digests, and reconstruction. The two reconstruction cases change FAIL/no activation into ALLOWED/test-baseline COMMITTED after rebuilding. Three deterministic handoff fixtures emulate transfer from a design role to a checking role: a current valid reference is admitted, a favorable summary of a violating artifact is rejected by the fact checks, and a stale reference is rejected by the version check. These fixtures test the handoff interface; they are not a run of multiple collaborating LLM agents. In particular, the capability to check all published cold-plate requirements is not established: mounting-hole diameter remains outside the implemented checks.

\begin{table}[H]
\centering\small
\caption{Additional actual-execution evidence. Units and result levels are kept separate.}\label{tab:execution_chain}
\begin{tabularx}{\linewidth}{@{}P{83pt}P{62pt}YY@{}}
\toprule
Experiment & Units & Recorded observation & Evidence limit \\
\midrule
Bracket deterministic & 12 cases & 2 commits; 7 FAIL; 3 HOLD & Eight cases include fact/claim-boundary interventions \\
Old/new mapping & 2 parameter settings, 4 runs & Old: 2 commits; repaired: 2 FAIL & Parameter-matched, not identical artifact bytes \\
Actual agent & 8 tasks & 4 completed; 4 turn limits; 0 tool errors & One model/configuration; no population safety estimate \\
Recovery module & 12 cases & 8 REJECTED; 2 HOLD; 2 ALLOWED & Registry and decision tests; no activation readback \\
Cold-plate transfer & 12 cases & Local gating and two rebuild-to-commit paths & Selected geometric predicates only \\
Role handoff & 3 fixtures & Current valid reference admitted; conflict and stale reference blocked & Deterministic role-interface tests, not multi-LLM coordination \\
\bottomrule
\end{tabularx}
\end{table}

\subsubsection{Contract-Based Challenges and Artifact Adjudication}\label{sec:v4_challenges}

The V4 set contains thirty tasks: fourteen bracket, six cold-plate, and ten recovery cases. The authoring material supplies engineering requirements, parameter semantics, interface states, and permissible boundary transformations. Expected outcomes are sealed with the tasks; execution does not use those outcomes to select control decisions. Task construction is reported separately from implementation, but personnel nonparticipation has not been independently documented in the supplied package. We therefore call this a contract-based challenge, not an externally independent blind evaluation. A matching seal authenticates the recorded task payload; it does not establish personnel independence or an unseen-task distribution.

The challenge has two forms of mismatch. A controller may violate an engineering requirement, or an expected outcome may be wrong or use a different interface convention. Disagreement therefore triggers examination of artifact facts and interface semantics rather than automatic acceptance of the implementation or the expected label. Two counterexamples are particularly informative.

\paragraph{Omitted mandatory check: V4-017.}
The published cold-plate requirements specify twelve channels of width 1.6 mm. A challenge builds twelve channels of width 2.4 mm. The initial engine checks channel count, wall thickness, and mounting-hole count but not channel width, and returns MATCH, ALLOWED, and COMMITTED with a changed test baseline. This is an observed erroneous activation, not merely a favorable decision label. Adding the missing frozen width requirement and evaluating STEP-read per-channel widths with a 0.001 mm tolerance changes the result to MISMATCH/FAIL without activation. The case demonstrates the consequence of a mismatch between the declared checking scope and the implemented predicate set. It does not show that all remaining requirements are covered. Section 7.13 specifies that an unimplemented mandatory predicate must contribute a closure unknown. The current experiments do not implement a general reconciliation of every published obligation with the executed-check inventory. Their positive decisions are consequently relative to the installed and bound predicates, not proof of satisfaction of the full published contract. In particular, the missing mounting-hole-diameter check is an implementation gap, not an optional requirement.

\paragraph{Incorrect fact consumption: V4-028.}
The expected outcome labels a reconstructed bracket with a 10 mm hole translation as a valid repair. Its artifact record places the far-side hole centers at 115 mm in a 120 mm plate, with radius 3.2 mm. The record-derived clearance is consequently
\[
d_{\mathrm{edge}}=120-115-3.2=1.8\ \mathrm{mm}<4.0\ \mathrm{mm}.
\]
The record also states \mbox{mounting\_}\allowbreak{}\mbox{interface\_}\allowbreak{}\mbox{preserved=}\allowbreak{}\mbox{false}. Nevertheless, the old recovery path returns ALLOWED. The producer sends \mbox{measured\_}\allowbreak{}\mbox{facts.}\allowbreak{}\mbox{mount\_}\allowbreak{}\mbox{interface.}\allowbreak{}\mbox{binding\_}\allowbreak{}\mbox{status}, while the consumer reads a top-level \mbox{binding\_}\allowbreak{}\mbox{status}. Failure to find that top-level field is treated as absence of violations. The newly reconstructed content has no previously registered violation under its new identity, so neither mechanism prevents the erroneous permission. The expected label and implementation thus agree while both are wrong about the new artifact.

The repaired consumer recognizes the supplied nested and flat representations and maps missing required status or UNRESOLVED to HOLD. An existing-artifact recheck gives ALLOWED for V4-027 (nominal positions, record-derived clearance 11.8 mm), REJECTED for V4-028 (1.8 mm), and HOLD for missing-field copies. These rechecks neither generate new geometry nor constitute independent STEP measurement. The defect concerns recovery permission: earlier COMMITTED and baseline-change fields in this path were derived from permission, not read from an activation service, and are not used as evidence of a commit.

\begin{table}[H]
\centering\small
\caption{V4 label agreement, recorded checker output, and artifact-grounded findings. The columns measure different quantities; none is a population safety rate.}\label{tab:v4_evidence}
\begin{tabularx}{\linewidth}{@{}P{65pt}P{63pt}P{94pt}Y@{}}
\toprule
Stage & Exact original-label matches & Recorded checker / adjudication output & Artifact-grounded finding \\
\midrule
Initial execution & 24/30 & Original checker: 29/30 & V4-017 commits incorrectly. Review also identifies V4-028's incorrect recovery permission, which the checker missed. \\
Width-check repair & 25/30 & Checker: 30/30, still missing the recovery error & V4-017 blocked; V4-028 still allowed. Its agreement with an incorrect expected label is not correctness. \\
Schema repair & 24/30 & All 30 cases accounted for under current checking and adjudication & V4-017 blocked; V4-028 rejected; V4-027 allowed. Six original-label differences remain explicit. \\
Targeted recovery recheck & Not a label-agreement score & Two original artifacts and missing-field copies & Valid artifact allowed; invalid artifact rejected; missing status held. No activation-service readback. \\
\bottomrule
\end{tabularx}
\end{table}

\FloatBarrier
\begingroup\small
\setlength{\parskip}{0pt}
\begin{longtable}{@{}P{47pt}P{82pt}P{99pt}P{127pt}P{84pt}@{}}
\caption{Case-level interpretation of V4 counts. Original label means the sealed expectation; artifact truth means the available artifact-grounded finding, not exhaustive independent geometric validation.}\label{tab:v4_case_adjudication}\\
\toprule
\textbf{Case ID} & \textbf{Original label} & \textbf{Implementation output} & \textbf{Artifact truth / grounds} & \textbf{Final category} \\
\midrule\endfirsthead
\multicolumn{5}{l}{\small\itshape Table \thetable\ (continued)}\\
\toprule
\textbf{Case ID} & \textbf{Original label} & \textbf{Implementation output} & \textbf{Artifact truth / grounds} & \textbf{Final category} \\
\midrule\endhead
\bottomrule\endlastfoot
V4-017 & FAIL; no activation & ALLOWED + COMMITTED $\to$ FAIL $\to$ FAIL & Channel width is 2.4 mm, not the required 1.6 mm. & Missing-check defect; local repair adds one label match. \\
V4-028 & ALLOWED & ALLOWED $\to$ ALLOWED $\to$ REJECTED & New artifact's recorded clearance is 1.8 mm $<4.0$ mm; interface preservation is false. & Wrong label and consumer defect; correct repair removes one label match. \\
V4-027 & ALLOWED & ALLOWED throughout & Nominal reconstruction: recorded clearance 11.8 mm. & Valid recovery control; permission only, no activation readback. \\
V4-001 / V4-015 & ALLOWED; no activation & ALLOWED + COMMITTED throughout & Installed checks pass; the submit interface includes test-baseline activation. & Interface-convention difference, not an extra unauthorized activation. \\
V4-002 & ALLOWED; no activation & FAIL throughout & STEP inspection finds two surviving holes, not the required four. & Incorrect geometric expectation; implementation blocks. \\
V4-020 & REJECTED & HOLD throughout & Swapped digest prevents source binding at the mapping stage. & Intervention-stage difference; no activation. \\
V4-030 & HOLD & REJECTED throughout & Artifact digest changes between evaluation stages; consumer detects mismatch. & Intervention-stage difference; no verified activation. \\
\end{longtable}
\endgroup

Arrows in Table~\ref{tab:v4_case_adjudication} follow initial execution, width-check repair, and Schema repair. Recovery rows report decisions only; historical fields derived from permission are not evidence of activation. The count change is therefore
\[
24\;\xrightarrow{\text{V4-017: +1 correct label match}}\;25
\;\xrightarrow{\text{V4-028: -1 match to a wrong label}}\;24.
\]
These are counts out of thirty unchanged original labels, not accuracy estimates. The final six differences comprise two interface conventions (001/015), two geometric expectations (002/028), and two intervention-stage expectations (020/030); all remain in the accounting. The twenty-two cases not listed in the table remain label-matched throughout.

The two counterexamples expose complementary prerequisites. Checking-scope completeness determines which obligations must be evaluated; fact-consumption consistency determines whether their results survive module boundaries. Repairs establish the observed behavior for the specified paths and counterexamples. General mandatory-check closure, all malformed-state handling, and a complete recovery-to-commit protocol remain outside this evidence. First-run outputs, repair regressions, decision labels, and post hoc adjudication remain separately versioned in the experimental archive.

\section{Discussion}\label{sec:section_12}

\subsection{Relationship Between the Technical Framework and Local Empirical Evidence}

The technical framework assigns non-substitutable responsibilities to engineering actions, evidence, and lifecycle change. Table~\ref{tab:t3} summarizes its mechanisms, Figure~6 connects the three chains, and Figure~9 organizes T6's normal and diagnostic paths. The evidence is strongest where it exposes an otherwise hidden boundary between these responsibilities.

The prototype supplies observable positive and negative instances. E1 and T5 ablation examine differences between tool success and registered effect observations. E2 has COMPLETE coverage but FAIL judgment. E6 and unknown-edge tests examine readmission and re-evaluation after changes. WP3 t0 supplies specification-evaluation instances for fixed artifacts and inputs, while t1 records conditional evaluation after fact patches. Table~\ref{tab:t32} consolidates each experiment's role and uncovered scope.

WP2 and WP3 expose different cross-module engineering-fact binding gaps. The historical WP2 adapter loses an existing negative interface-preservation fact as artifacts enter the control plane. WP3 substitutes nominal plan-hole positions without changing the artifact, yet re-evaluation returns ALLOWED. Locally correct rules and transitions therefore do not establish consistency of provenance, actual object versions, configurations, and transformations. BFM-1.1 provides local repair evidence for the former gap on existing artifacts; it does not validate the historical recovery path. The separate \mbox{exec\_}\allowbreak{}\mbox{chain\_}\allowbreak{}\mbox{v1} implementation in Section~\ref{sec:execution_binding} adds actual CAD execution and recovery-consumer tests without changing those historical records.

The additional counterexamples make the interface requirements more precise. V4-017 extracts geometric facts but omits a published width requirement from the check inventory, resulting in an erroneous test-baseline commit. V4-028 constructs a new artifact with an explicit negative interface fact, but a consumer expecting a different field structure interprets it as absence of violations and permits recovery. These failures distinguish checking-scope completeness from fact-consumption consistency. They must not be conflated with the historical U4 diagnosis: its plan already contained hole-translation parameters, while the adapter discarded a negative fact and the later audit misinterpreted translation semantics. The three cases locate different breaks in the responsibility for engineering facts. Neither provenance hashing nor a three-state decision function alone enforces either property. A controller must establish which checks are mandatory and ensure that producers and consumers preserve the same object-bound interpretation of their results.

The width-check repair blocks V4-017 in the actual adapter/test-baseline path. The Schema repair rejects the invalid V4-028 artifact, permits the valid V4-027 control, and holds specified missing-status copies. These observations support local repairs and their discriminating behavior, not closure of all instances of the two failure classes. The published cold-plate hole-diameter requirement still lacks an implementation; complete malformed-state and same-version violation-priority handling are also outside the observed repair evidence. Restoring applicability requires current artifact facts, while formal activation additionally requires its own commit and readback evidence.

\subsection{Closed-Set Results, Internal Baselines, and Shared Assumptions}

The frozen suites share taxonomy, rule, and Oracle assumptions, while the internal baselines deliberately omit selected engineering checks. Their role is mechanism discrimination under the common scope of Section~9.1. External systems may provide specialized object plugins or approval procedures absent from these configurations; no product-level ranking follows.

Reproducible records differ from engineering ground truth beyond the specification. Internal cases expose inconsistency among tests, rules, and Oracles; WP2 shows that a separate checking script can still misinterpret tool parameters. The conflict between its 48/49 passes and eighteen negative bracket-generation records requires validity analysis of the checker itself. Not reading runner danger labels is insufficient to establish independent personnel or correct engineering semantics.

\FloatBarrier
\begingroup
\small
\setlength{\parskip}{0pt}\setlength{\LTpre}{8pt}\setlength{\LTpost}{10pt}
\begin{longtable}{@{}P{112.94pt}P{167.15pt}P{171.67pt}@{}}
\caption{Evidence levels and conclusion boundaries; batches are not combined into a safety probability.}\label{tab:t32}\\
\toprule
\textbf{Evidence} & \textbf{Supported local conclusion} & \textbf{Not supported} \\
\midrule\endfirsthead
\multicolumn{3}{l}{\small\itshape Table \thetable\ (continued)}\\
\toprule
\textbf{Evidence} & \textbf{Supported local conclusion} & \textbf{Not supported} \\
\midrule\endhead
\midrule\multicolumn{3}{r}{\small Continued on next page}\\\endfoot
\bottomrule\endlastfoot
\mbox{Taxonomy/}\allowbreak{}\mbox{four-}\allowbreak{}\mbox{plane} design & Systematic objects, control moments, modules, interfaces & Taxonomy completeness or deployment of every module \\
Frozen suite/ablation & Specified-rule conformance and effects of mechanism removal & Unseen-fault detection or whole-toolchain safety \\
OpenFOAM E2/E6 & \mbox{Coverage/}\allowbreak{}\mbox{judgment} and \mbox{completion/}\allowbreak{}\mbox{admission} separations & Sufficient model V\&V or product certification \\
Internal cases/same-set regression & Treatment of identified rule, fixture, and Oracle defects & Personnel-independent holdout generalization \\
\mbox{pair\_}\allowbreak{}\mbox{b1} tool/task records & Real tool execution and preset completion counts & Pure reasoning-mode causal effect or safe-completion rate \\
Artifact/audit conflict & Eighteen negative interface records contradict seventeen audit passes & Original 48/49 compliance or a new aggregate risk rate \\
Local persistence & Conditional writes, idempotency, durability limits at specified endpoints & Arbitrary external exactly-once effects or multinode production capability \\
WP3 t0/t1 records & Fifteen t0 specification matches; t1 conditional fact-patch re-evaluation & Fifteen actual repairs or artifact compliance from declarations alone \\
BFM-1.1 local regression & Eighteen abnormal records blocked, one nominal passes; twelve contracts and three bad-mapping checks & Revised historical commits, full agent reruns, independent STEP validation, or unseen-fault detection \\
Additional execution-chain prototype & Actual bracket/cold-plate calls; four of eight agent tasks complete; local gating and separate STEP inspection path & Commercial PLM, independent geometry kernel, aggregate safe-completion rate \\
V4 challenge and repairs & Omitted-check commit and Schema-consumption permission localized; selected fixes pass same-set regression and artifact controls & External blind generalization, complete predicate coverage, recovery commits from decision labels \\
\end{longtable}
\endgroup

\subsection{Isolation and Availability Costs}

The V/U discipline preserves unknown mandatory facts as HOLD, preventing hazardous permission when observations are missing. Its cost is a 55.5\% compliant-path HOLD rate at 20\% independent per-observation loss. Correlated object-service and lineage-service failures yield 22.0\% and 17.0\%. Lower rates do not imply safer correlated failures; joint field loss and recovery change the event structure. Deployments cannot improve availability by changing UNKNOWN to PASS: the positive control produces hazardous permission under the same stress. Prefer versioned observation caches, heterogeneous probes for critical roles, asynchronous evidence completion, and executable recovery actions derived from HOLD reason codes.

Dependency propagation has a similar trade-off. Treating unknowns as false narrows re-evaluation but omits 61.73\% of safety-relevant impacts at 40\% hidden edges. Treating them as true avoids omission but unconditionally expands scope. Three-state propagation exposes possible impacts as an average of 3.045 UNRESOLVED nodes per run. The objective is not minimal re-evaluation at any cost, but identifiable entry points for review or additional observation without presenting unknowns as facts.

\subsection{Threats to Validity}

\paragraph{Internal validity.}
Rules, injectors, and Oracles share assumptions; same-set repair is subject to knowledge of failures. V4 sealed expectations are not used by the controller, but sealing alone does not establish independent personnel or semantic novelty. The WP2 arms also differ in worker count, a shared 9,000-second budget, and runner revision; their 149 length responses and 37 thinking-only terminations cannot identify a causal reasoning-mode effect.

\paragraph{Construct validity.}
Positive decisions are relative to installed checks and correctly bound observations. The known open interfaces are: an unimplemented cold-plate installation-hole-diameter predicate; incomplete malformed-status and same-version violation-priority handling in recovery consumers; historical WP3 fact patches without artifact updates; and recovery permission without activation readback. These are distinct from repaired width and Schema cases. The historical 48/49 audit score is invalid as an engineering-compliance measure, while the remaining cold-plate audit has conditional-check coverage rather than exhaustive inspection. Artifact hashes establish correspondence, not correctness. Shared OCCT geometry checks and generation-derived distances do not provide independent-kernel geometric truth; the CFD and structural models are control fixtures, not certified engineering predictions.

\paragraph{External validity.}
Evidence comes from fixed tasks, open-source tools, two domains, and local experimental prototypes. The eight-task \mbox{exec\_}\allowbreak{}\mbox{chain\_}\allowbreak{}\mbox{v1} Agent batch uses one model and configuration. Deterministic handoff fixtures test interfaces rather than multiple LLMs collaborating. In-memory test baselines and local SQLite/PostgreSQL services do not establish commercial PLM integration, independent-node fault tolerance, or exactly-once effects for arbitrary external tools. The study excludes cross-organizational governance and does not replace cybersecurity or functional-safety certification.

\paragraph{Statistical conclusion validity.}
Task--seed repetitions are dependent, case-family denominators are small, and the thinking-on arm has seven formal commits. Counts are descriptive; no overall safety interval or relative-safety estimate is calculated. Label agreement, engineering adjudication, and completion counts measure different constructs. Performance results characterize this prototype's storage bottlenecks: the SQLite service uses 200 operations at each client level, while PostgreSQL workloads have different concurrency and durability settings. Neither throughput nor tail latency is extrapolated to long-running industrial jobs or production deployments.

\subsection{Preconditions and Boundaries for Building the Complete Substrate}

Key prerequisites are that agents cannot arbitrarily remove obligation sources or mandatory checks; actual tool effects connect to invariants; artifact facts retain semantics when entering control objects; \mbox{evidence/}\allowbreak{}\mbox{version} services provide current grounds; and changes notify relevant consumers. Intra-enterprise multi-agent collaboration additionally requires handoff records to retain decision objects and scopes, with each agent obtaining task authority and assuming \mbox{observation/}\allowbreak{}\mbox{evidence} responsibilities. Section 6.13 provides common interfaces for these handoffs.

The framework retains the full scope of trustworthy construction, execution, asset governance, recovery, and evolution. Tables~\ref{tab:t2} and~\ref{tab:t32} and Section 12.4 consolidate current evidential boundaries. Existing-artifact repair evidence is complemented by selected actual execution, a separate STEP inspection path, adapter transfer, and new-artifact recovery decisions. General mandatory-check closure, independently validated arbitrary geometry, production activation, and complete lifecycle services remain unestablished. Future implementations can expand along the three chains, evaluating each extension through its engineering objects and fact interfaces.

\section{Conclusion}\label{sec:section_13}

This paper presents a research framework for trustworthy industrial-agent execution within a single enterprise governance domain. Eight control domains, a four-plane architecture, and three control chains connect engineering actions, evidence, and lifecycle evolution through typed objects and version-bound interfaces. The framework separates tool returns from actual effects, evidence coverage from engineering judgment, and historical lifecycle state from current applicability. Multi-agent handoffs extend the architecture; their evaluation is limited to deterministic interface tests.

Specification tests, ablations, uncertainty stress, CAD/CAE execution, and local-persistence experiments support selected mechanisms across distinct prototypes. OpenFOAM cases demonstrate why complete coverage need not imply engineering acceptance and completed jobs still require current-context admission. Artifact analysis and additional executions reveal complementary failure modes: requirements omitted from checks, negative facts lost in adaptation, and inconsistent fact consumption during recovery. Targeted repairs provide local regression evidence, not comprehensive protection or retrospective correction of historical submissions.

The central implication is that trustworthy execution requires joint control of checking scope, fact transformation, and decision consumption. Correct rules alone are insufficient when required checks are absent or engineering facts lose their meaning across interfaces. The taxonomy and architecture identify these responsibilities; the implementation and evidence matrices delimit what has been demonstrated. General checking completeness, end-to-end recovery, and broader industrial validation remain open. The framework provides a common basis for pursuing these problems without treating local success as a guarantee for the whole execution chain.

The resulting claim is intentionally layered: the paper establishes a complete systems research framework and a vocabulary for developing industrial-agent trust controls, while the experiments establish only selected mechanisms and failure boundaries on the implemented paths. This is sufficient to make the architecture a reusable starting point for subsequent methods, adapters, and independent evaluations; it is not a claim that every industrial toolchain or unknown failure is already covered.

\section*{Data and Code Availability}
The experimental code and data do not yet have a publicly accessible download repository. Source inventories, execution configurations, counting rules, parameter-semantics analyses, and artifact fingerprints are retained in an internal experimental archive. Internal paths and SHA-256 identifiers establish artifact correspondence, not public availability, independent reproduction, or engineering validity. The article reports the experimental protocols, principal parameters, results, prototype boundaries, and limitations needed to interpret its claims without access to that archive.

\section*{Terminology and State Guide}
\FloatBarrier
\begingroup\small
\setlength{\parskip}{0pt}
\begin{longtable}{@{}P{141pt}P{316pt}@{}}
\caption{Glossary and compact reading guide.}\label{tab:glossary}\\
\toprule
\textbf{Term} & \textbf{Meaning / scope} \\
\midrule\endfirsthead
\multicolumn{2}{l}{\small\itshape Table \thetable\ (continued)}\\
\toprule
\textbf{Term} & \textbf{Meaning / scope} \\
\midrule\endhead
\bottomrule\endlastfoot
T1--T8 & Authority and capability; intent and planning; admission and composition; tool execution; actual effects; evidence; baseline and use; recovery and evolution. \\
V / U & Base sets of confirmed violations and mandatory unknowns, with grounded object references. Optional unknowns do not discharge mandatory conditions. \\
$V^*/U^*$ & Final sets including closure-evaluation grounds. A confirmed applicable violation has priority over unknowns. \\
ActualEngineeringDelta & Version-bound before/after differences, bindings, downstream references, and observation gaps; sources remain distinguished. \\
EvaluationContextSnapshot & Frozen references to object, baseline, rule, capability, policy, and observation scope; invalidation blocks unsupported reuse. \\
ExecutionToken / Lease & Execution-bound credential / time-limited task authority; neither is a capability or credibility Envelope. \\
Envelope & Versioned capability, credibility, use, or disclosure scope, depending on object type. \\
ACCEPT\_CANDIDATE / ACCEPT & Eligibility for activation / completed authoritative model-baseline acceptance. \\
HOLD / REBASE\_REQUIRED & Await required grounds / reestablish drifted references; neither implies a completed repair. \\
COMPLETE / PARTIAL / UNRESOLVED & Evidence coverage states, not engineering pass or failure. \\
PASS / FAIL / UNKNOWN & Engineering criterion judgments from applicable evidence. \\
VALID / INVALID / UNRESOLVED & Current applicability of a historical effect or support object, separate from its lifecycle. \\
Gap / Obligation & Unmet or unresolved evidence/activity need / typed requirement that specifies what must be established. \\
B/C/A & Current baseline, candidate content, and acceptance-condition references, each compared to its own frozen counterpart. \\
Decision / activation / readback & Permission determination / authoritative state change / observation that confirms the resulting baseline state. \\
\end{longtable}
\endgroup

\clearpage
\setlength{\bibsep}{2.5pt}
\bibliographystyle{trust-unsrtnat}
\bibliography{references}

@misc{r55,
  author = {Wang, Zhongjie and Liu, Mingyi and Yu, Shuang and He, Xiang and Wang, Xiao and Xu, Xiaofei},
  title = {{Toward Services Computing 3.0: Delegated Action and Agentic Service Relationships}},
  year = {2026},
  howpublished = {Zenodo preprint},
  doi = {10.5281/zenodo.23227987},
  url = {https://zenodo.org/records/23227987},
}

@misc{r3,
  author = {{OpenAI}},
  title = {{Guardrails and Human Review}},
  howpublished = {Online documentation},
  year = {n.d.},
  url = {https://developers.openai.com/api/docs/guides/agents/guardrails-approvals},
}

@misc{r4,
  author = {{Microsoft}},
  title = {{Microsoft Agent Framework: Observability}},
  howpublished = {Online documentation},
  year = {n.d.},
  url = {https://learn.microsoft.com/en-us/agent-framework/agents/observability/},
}

@misc{r5,
  author = {{Amazon Web Services}},
  title = {{Amazon Bedrock AgentCore Policy: Core Concepts}},
  howpublished = {Online documentation},
  year = {n.d.},
  url = {https://docs.aws.amazon.com/bedrock-agentcore/latest/devguide/policy-core-concepts.html},
}

@misc{r6,
  author = {{Amazon Web Services}},
  title = {{Amazon Bedrock AgentCore: Observability}},
  howpublished = {Online documentation},
  year = {n.d.},
  url = {https://docs.aws.amazon.com/bedrock-agentcore/latest/devguide/observability.html},
}

@inproceedings{r7,
  author = {Zhan, Qiusi and Liang, Zhixiang and Ying, Zifan and Kang, Daniel},
  title = {{InjecAgent: Benchmarking Indirect Prompt Injections in Tool-Integrated Large Language Model Agents}},
  booktitle = {Findings of the Association for Computational Linguistics: ACL 2024},
  year = {2024},
  pages = {10471--10506},
  publisher = {Association for Computational Linguistics},
  doi = {10.18653/v1/2024.findings-acl.624},
  url = {https://aclanthology.org/2024.findings-acl.624/},
}

@inproceedings{r8,
  author = {Debenedetti, Edoardo and Zhang, Jie and Balunovi{\'c}, Mislav and Beurer-Kellner, Luca and Fischer, Marc and Tram{\`e}r, Florian},
  title = {{AgentDojo: A Dynamic Environment to Evaluate Prompt Injection Attacks and Defenses for LLM Agents}},
  booktitle = {Advances in Neural Information Processing Systems},
  year = {2024},
  volume = {37},
  url = {https://proceedings.neurips.cc/paper_files/paper/2024/file/97091a5177d8dc64b1da8bf3e1f6fb54-Paper-Datasets_and_Benchmarks_Track.pdf},
}

@misc{r9,
  author = {{OPC Foundation}},
  title = {{OPC Unified Architecture---Part 2: Security Model}},
  howpublished = {Online documentation},
  year = {n.d.},
  url = {https://reference.opcfoundation.org/specs/OPC-10000-2/full},
}

@misc{r10,
  author = {{OPC Foundation}},
  title = {{OPC Unified Architecture---Part 1: Overview and Concepts}},
  howpublished = {Online documentation},
  year = {n.d.},
  url = {https://reference.opcfoundation.org/specs/OPC-10000-1/4.4.1},
}

@misc{r12,
  author = {{Siemens}},
  title = {{Siemens unveils industrial AI and digital twin innovations at CES 2025}},
  year = {2025},
  url = {https://press.siemens.com/global/en/pressrelease/siemens-unveils-breakthrough-innovations-industrial-ai-and-digital-twin-technology-ces},
}

@misc{r13,
  author = {{Siemens}},
  title = {{The new AI agent doesn't just make suggestions---it gets the job done}},
  year = {2026},
  url = {https://news.siemens.com/en-gb/the-new-ai-agent-doesnt-just-make-suggestions/},
}

@misc{r14,
  author = {{NASA}},
  title = {{NASA-STD-7009B: Standard for Models and Simulations}},
  year = {2024},
  url = {https://standards.nasa.gov/standard/NASA/NASA-STD-7009},
}

@misc{r20,
  author = {{Alibaba Cloud}},
  title = {{Model Studio: Security Overview}},
  howpublished = {Online documentation},
  year = {n.d.},
  url = {https://help.aliyun.com/zh/model-studio/security/overview},
}

@misc{r21,
  author = {{Alibaba Cloud}},
  title = {{Agent Studio: Lifecycle Observability, Evaluation, and Optimization}},
  howpublished = {Online documentation},
  year = {n.d.},
  url = {https://docs.agent.bailian.aliyun.com/zh/loop},
}

@misc{r22,
  author = {{Huawei}},
  title = {{Agentic AI Products for Secure Autonomous agent Environments and the AgentArts Platform}},
  year = {2026},
  url = {https://www.huawei.com/cn/news/2026/6/inspire-agenticera-agenticinfra},
}

@misc{r23,
  author = {{W3C}},
  title = {{PROV-O: The PROV Ontology, W3C Recommendation}},
  year = {2013},
  url = {https://www.w3.org/TR/prov-o/},
}

@misc{r24,
  author = {{JSON Schema}},
  title = {{JSON Schema Draft 2020-12}},
  year = {2022},
  url = {https://json-schema.org/draft/2020-12},
}

@misc{r25,
  author = {{OpenTelemetry}},
  title = {{OpenTelemetry Specification}},
  howpublished = {Online documentation},
  year = {n.d.},
  url = {https://opentelemetry.io/docs/specs/otel/},
}

@misc{r26,
  author = {{Cloud Native Computing Foundation}},
  title = {{CloudEvents Specification}},
  howpublished = {Online documentation},
  year = {n.d.},
  url = {https://cloudevents.io/},
}

@misc{r27,
  author = {{IETF}},
  title = {{HTTP Semantics}},
  year = {2022},
  url = {https://www.rfc-editor.org/rfc/rfc9110},
  howpublished = {RFC 9110},
}

@misc{r28,
  author = {{Open Policy Agent}},
  title = {{Open Policy Agent: Management APIs and Deployment}},
  howpublished = {Online documentation},
  year = {n.d.},
  url = {https://www.openpolicyagent.org/docs/management-introduction},
}

@misc{r29,
  author = {{Model Context Protocol}},
  title = {{Model Context Protocol: Server Primitives and Tools (Draft)}},
  howpublished = {Online documentation},
  year = {n.d.},
  url = {https://modelcontextprotocol.io/specification/draft/server/index},
}

@misc{r30,
  author = {{ASME}},
  title = {{VVUQ 50.1---Guide to a Model Life Cycle Approach That Incorporates Verification, Validation, and Uncertainty Quantification}},
  year = {2025},
  url = {https://www.asme.org/codes-standards/find-codes-standards/vvuq501-vvuq-50-1-guide-to-a-model-life-cycle-approach-that-incorporates-verification-validation-and-uncertainty-quantification},
}

@misc{r31,
  author = {{Confidential Computing Consortium}},
  title = {{Common Terminology for Confidential Computing}},
  year = {2023},
  url = {https://confidentialcomputing.io/wp-content/uploads/sites/10/2023/03/Common-Terminology-for-Confidential-Computing.pdf},
}

@article{r33,
  author = {Leucker, Martin and Schallhart, Christian},
  title = {{A brief account of runtime verification}},
  journal = {The Journal of Logic and Algebraic Programming},
  year = {2009},
  volume = {78},
  number = {5},
  pages = {293--303},
  doi = {10.1016/j.jlap.2008.08.004},
  url = {https://doi.org/10.1016/j.jlap.2008.08.004},
}

@misc{r34,
  author = {{ASME}},
  title = {{V\&V 40---Assessing Credibility of Computational Modeling through Verification and Validation: Application to Medical Devices}},
  year = {2018},
  url = {https://www.asme.org/codes-standards/publications-information/verification-validation-uncertainty},
}

@inproceedings{r35,
  author = {Ruan, Yangjun and Dong, Honghua and Wang, Andrew and Pitis, Silviu and Zhou, Yongchao and Ba, Jimmy and Dubois, Yann and Maddison, Chris J. and Hashimoto, Tatsunori},
  title = {{Identifying the Risks of LM Agents with an LM-Emulated Sandbox}},
  booktitle = {International Conference on Learning Representations},
  year = {2024},
  url = {https://openreview.net/pdf?id=GEcwtMk1uA},
}

@inproceedings{r36,
  author = {Yao, Shunyu and Shinn, Noah and Razavi, Pedram and Narasimhan, Karthik},
  title = {{$\tau$-bench: A Benchmark for Tool-Agent-User Interaction in Real-World Domains}},
  booktitle = {International Conference on Learning Representations},
  year = {2025},
  url = {https://openreview.net/pdf?id=roNSXZpUDN},
}

@inproceedings{r37,
  author = {Xia, Hongfei and Wang, Hongru and Liu, Zeming and Yu, Qian and Guo, Yuhang and Wang, Haifeng},
  title = {{SafeToolBench: Pioneering a Prospective Benchmark to Evaluating Tool Utilization Safety in LLMs}},
  booktitle = {Findings of the Association for Computational Linguistics: EMNLP 2025},
  year = {2025},
  pages = {17643--17660},
  publisher = {Association for Computational Linguistics},
  doi = {10.18653/v1/2025.findings-emnlp.958},
  url = {https://aclanthology.org/2025.findings-emnlp.958/},
}

@article{r38,
  author = {Falcone, Yli{\`e}s and Mounier, Laurent and Fernandez, Jean-Claude and Richier, Jean-Luc},
  title = {{Runtime Enforcement Monitors: Composition, Synthesis, and Enforcement Abilities}},
  journal = {Formal Methods in System Design},
  year = {2011},
  volume = {38},
  number = {3},
  pages = {223--262},
}

@misc{r39,
  author = {Mehmood, Usama and Sheikhi, Sanaz and Bak, Stanley and Smolka, Scott A. and Stoller, Scott D.},
  title = {{The Black-Box Simplex Architecture for Runtime Assurance of Autonomous CPS}},
  year = {2021},
  howpublished = {arXiv:2102.12981},
  url = {https://arxiv.org/abs/2102.12981},
}

@incollection{r40,
  author = {Hull, Richard and Damaggio, Elio and Fournier, Fabiana and Gupta, Manmohan and Heath, Fenno (Terry) and Hobson, Stacy and Linehan, Mark and Maradugu, Sridhar and Nigam, Anil and Sukaviriya, Piyawadee and Vaculin, Roman},
  title = {{Introducing the Guard-Stage-Milestone Approach for Specifying Business Entity Lifecycles}},
  booktitle = {Lecture Notes in Computer Science},
  year = {2011},
  pages = {1--24},
  doi = {10.1007/978-3-642-19589-1_1},
  publisher = {Springer Berlin Heidelberg},
  url = {https://doi.org/10.1007/978-3-642-19589-1_1},
}

@article{r41,
  author = {Damaggio, Elio and Hull, Richard and Vacul{\'i}n, Roman},
  title = {{On the Equivalence of Incremental and Fixpoint Semantics for Business Artifacts with Guard-Stage-Milestone Lifecycles}},
  journal = {Information Systems},
  year = {2013},
  volume = {38},
  number = {4},
  pages = {561--584},
  doi = {10.1016/j.is.2012.09.002},
  url = {https://doi.org/10.1016/j.is.2012.09.002},
}

@incollection{r42,
  author = {Butt, Anila Sahar and Fitch, Peter},
  title = {{ProvONE+: A Provenance Model for Scientific Workflows}},
  booktitle = {Lecture Notes in Computer Science},
  year = {2020},
  pages = {431--444},
  doi = {10.1007/978-3-030-62008-0_30},
  publisher = {Springer International Publishing},
  url = {https://doi.org/10.1007/978-3-030-62008-0_30},
}

@misc{r43,
  author = {Leo, Simone and Crusoe, Michael R. and Rodr{\'i}guez-Navas, Laura and Sirvent, Ra{\"u}l and Kanitz, Alexander and De Geest, Paul and Wittner, Rudolf and Pireddu, Luca and Garijo, Daniel and Fern{\'a}ndez, Jos{\'e} M. and Colonnelli, Iacopo and Gallo, Matej and Ohta, Tazro and Suetake, Hirotaka and Capella-Gutierrez, Salvador and de Wit, Renske and Kinoshita, Bruno P. and Soiland-Reyes, Stian},
  title = {{Recording provenance of workflow runs with RO-Crate}},
  year = {2023},
  howpublished = {arXiv:2312.07852},
  url = {https://arxiv.org/abs/2312.07852},
}

@article{r44,
  author = {Dhruv, Akash and Dubey, Anshu},
  title = {{Managing Software Provenance to Enhance Reproducibility in Computational Research}},
  journal = {Computing in Science \& Engineering},
  year = {2023},
  volume = {25},
  number = {3},
  pages = {60--65},
  doi = {10.1109/mcse.2023.3314288},
  url = {https://doi.org/10.1109/MCSE.2023.3314288},
}

@article{r45,
  author = {Cederbladh, Johan and Cicchetti, Antonio and Suryadevara, Jagadish},
  title = {{Early Validation and Verification of System Behaviour in Model-based Systems Engineering: A Systematic Literature Review}},
  journal = {ACM Transactions on Software Engineering and Methodology},
  year = {2024},
  volume = {33},
  number = {3},
  pages = {1--67},
  doi = {10.1145/3631976},
  url = {https://doi.org/10.1145/3631976},
}

@article{r46,
  author = {Brahma, Arindam and Wynn, David C.},
  title = {{Concepts of change propagation analysis in engineering design}},
  journal = {Research in Engineering Design},
  year = {2023},
  volume = {34},
  number = {1},
  pages = {117--151},
  doi = {10.1007/s00163-022-00395-y},
  url = {https://doi.org/10.1007/s00163-022-00395-y},
}

@article{r47,
  author = {Wang, Haoqi and Thomson, Vincent and Tang, Chengtong},
  title = {{Change propagation analysis for system modeling using Semantic Web technology}},
  journal = {Advanced Engineering Informatics},
  year = {2018},
  volume = {35},
  pages = {17--29},
  doi = {10.1016/j.aei.2017.11.004},
  url = {https://doi.org/10.1016/j.aei.2017.11.004},
}

@article{r48,
  author = {Wei, Ran and Foster, Simon and Mei, Haitao and Yan, Fang and Yang, Ruizhe and Habli, Ibrahim and O'Halloran, Colin and Tudor, Nick and Kelly, Tim and Nemouchi, Yakoub},
  title = {{ACCESS: Assurance Case Centric Engineering of Safety-Critical Systems}},
  journal = {Journal of Systems and Software},
  year = {2024},
  volume = {213},
  pages = {112034},
  doi = {10.1016/j.jss.2024.112034},
  url = {https://doi.org/10.1016/j.jss.2024.112034},
}

@inproceedings{r49,
  author = {Liu, Yanming and Peng, Xinyue and Cao, Jiannan and Wang, Xinyi and Deng, Songhang and Chen, Jintao and Yin, Jianwei and Zhang, Xuhong},
  title = {{ToolGate: Contract-Grounded and Verified Tool Execution for LLMs}},
  booktitle = {Findings of the Association for Computational Linguistics: ACL 2026},
  year = {2026},
  pages = {9653--9684},
  publisher = {Association for Computational Linguistics},
  doi = {10.18653/v1/2026.findings-acl.470},
  url = {https://aclanthology.org/2026.findings-acl.470/},
}

@inproceedings{r50,
  author = {Zheng, Keyou and Su, Xuyang and Leng, Jiewu},
  title = {{SPADA: A Verifiable Test-Driven Agent for Controllable Parametric CAD Assembly Generation}},
  booktitle = {Proceedings of the 43rd International Conference on Machine Learning},
  series = {Proceedings of Machine Learning Research},
  volume = {306},
  publisher = {PMLR},
  year = {2026},
  pages = {163547--163572},
  url = {https://proceedings.mlr.press/v306/zheng26q.html},
}

@misc{r51,
  author = {Zheng, Peng and Dong, Xintong and Li, Chuanyang and Jing, Jiaxin and Han, Chuqi and Shen, Hailong and Song, Yanzhi and Yang, Zhouwang},
  title = {{ReliCAD: From Uncertain LLM Generation to Reliable Parametric CAD Modeling}},
  year = {2026},
  howpublished = {arXiv:2609.22325},
  url = {https://arxiv.org/abs/2609.22325},
}

@incollection{r52,
  author = {Xu, Xiaofei and Wang, Zhongjie},
  title = {{Value-Aware Service Model Driven Architecture and Methodology}},
  booktitle = {E-Government, ICT Professionalism and Competences Service Science},
  series = {IFIP International Federation for Information Processing},
  volume = {280},
  pages = {277--286},
  publisher = {Springer},
  year = {2008},
  doi = {10.1007/978-0-387-09712-1_29},
  url = {https://link.springer.com/chapter/10.1007/978-0-387-09712-1_29},
}

@inproceedings{r53,
  author = {He, Xiang and Tu, Zhiying and Liu, Lei and Xu, Xiaofei and Wang, Zhongjie},
  title = {{Optimal Evolution Planning and Execution for Multi-version Coexisting Microservice Systems}},
  booktitle = {Service-Oriented Computing (ICSOC 2020)},
  series = {Lecture Notes in Computer Science},
  volume = {12571},
  pages = {3--18},
  publisher = {Springer},
  year = {2020},
  doi = {10.1007/978-3-030-65310-1_1},
  url = {https://link.springer.com/chapter/10.1007/978-3-030-65310-1_1},
}

@article{r54,
  author = {He, Xiang and Tu, Zhiying and Xu, Xiaofei and Wang, Zhongjie},
  title = {{Programming framework and infrastructure for self-adaptation and optimized evolution method for microservice systems in cloud--edge environments}},
  journal = {Future Generation Computer Systems},
  volume = {118},
  pages = {263--281},
  year = {2021},
  doi = {10.1016/j.future.2021.01.008},
  url = {https://www.sciencedirect.com/science/article/pii/S0167739X21000170},
}

\end{document}